\documentclass[usenatbib]{mnras}
\usepackage{natbib,amssymb,amsmath,times,graphicx,url,aas_macros}
\usepackage{setspace}      
\usepackage{epstopdf}       
\usepackage{verbatim}       
\usepackage{bm}		  
\usepackage{fleqn}		  
\usepackage{Wurster_macros}	 
\newcommand\ttnow{$t = 1.45$~t$_\text{ff}$}  
\newcommand\tnow{$1.45$~t$_\text{ff}$}  
\newcommand\ttendIh{$t \approx1.45$~t$_\text{ff}$}  
\newcommand\tendIh{$1.45$~t$_\text{ff}$}  
\newcommand\ttendNh{$t \approx1.36$~t$_\text{ff}$}  
\newcommand\tendNh{$1.36$~t$_\text{ff}$}  
\newcommand\ttend{$t = t_\text{final}$}  
\newcommand\tend{$t_\text{final}$}

\title[Star cluster formation with non-ideal MHD]{There is no magnetic braking catastrophe: Low-mass star cluster and protostellar disc formation with non-ideal magnetohydrodynamics}

\author[Wurster, Bate \& Price]{James Wurster$^{1}$\thanks{J.Wurster@exeter.ac.uk}, Matthew R. Bate$^{1}$\thanks{M.R.Bate@exeter.ac.uk}, and Daniel J. Price$^{2}$ \\
$^{1}$School of Physics and Astronomy, University of Exeter, Stocker Rd, Exeter EX4 4QL, UK \\
$^{2}$School of Physics and Astronomy, Monash University, VIC 3800, Australia \\
}
\date{Submitted: Revised: Accepted: }
\pagerange{\pageref{firstpage}--\pageref{lastpage}} \pubyear{2019}

\begin{document}
\label{firstpage}
\bibliographystyle{mnras}
\maketitle

\begin{abstract}
We present results from the first radiation non-ideal magnetohydrodynamics (MHD) simulations of low-mass star cluster formation that resolve the fragmentation process down to the opacity limit.  We model 50~M$_\odot$ turbulent clouds initially threaded by a uniform magnetic field with strengths of 3, 5 10 and 20 times the critical mass-to-magnetic flux ratio, and at each strength, we model both an ideal and non-ideal (including Ohmic resistivity, ambipolar diffusion and the Hall effect) MHD cloud.  Turbulence and magnetic fields shape the large-scale structure of the cloud, and similar structures form regardless of whether ideal or non-ideal MHD is employed.  At high densities ($10^6 \lesssim n_{\rm H} \lesssim 10^{11}$~cm$^{-3}$), all models have a similar magnetic field strength versus density relation, suggesting that the field strength in dense cores is independent of the large-scale environment.  Albeit with limited statistics, we find no evidence for the dependence of the initial mass function on the initial magnetic field strength, however, the star formation rate decreases for models with increasing initial field strengths; the exception is the strongest field case where collapse occurs primarily along field lines.  Protostellar discs with radii $\gtrsim 20$~au form in all models, suggesting that disc formation is dependent on the gas turbulence rather than on magnetic field strength.  We find no evidence for the magnetic braking catastrophe, and find that magnetic fields do not hinder the formation of protostellar discs.
\end{abstract}


\section{Introduction}
\label{intro}

Stars are born and evolve in a dynamic environment, and are neither born nor evolve in isolation.  Therefore, how a star forms and evolves is directly influenced by its environment. Environmental influences in star forming molecular clouds include turbulence \citepeg{PadoanNordlund2002,MckeeOstriker2007}, radiative feedback \citepeg{Hatchell+2013,Siciliaaguilar+2013,FallKrumholzMatzner2010,SkinnerOstriker2015,KimKimOstriker2016},  magnetic fields \citepeg{HeilesCrutcher2005} and low ionisation fractions \citepeg{MestelSpitzer1956,NakanoUmebayashi1986,UmebayashiNakano1990}.

 Molecular clouds are large, structured complexes, with typical masses of $10^3-10^5$~\Msun, sizes of $2-15$~pc \citep[e.g.][and references therein]{BerginTafalla2007} and magnetic field strengths of a few times the critical mass-to-flux ratio \citep[e.g.][and references therein]{Crutcher2012}.  The clouds often contain a few, nearly parallel, large filaments, whose length may be comparable to the length of the cloud  \citepeg{Tachihara+2002,BurkertHartmann2004}.  Magnetic fields are observed to be approximately perpendicular to dense filaments \citep{Heyer+1987,AlvesFrancoGirart2008,FrancoAlvesGirart2010}, while in the less dense regions, magnetic fields are approximately parallel to the structure \citep[i.e. to striations; e.g.][]{Goldsmith+2008,Planck2016or,Planck2016role}.  Dense filaments, and especially the places where they intersect, are home to regions of intense star formation.  Individual young stars and their protostellar discs have been resolved in these regions \citepeg{Dunham+2011,Lindberg+2014,Tobin+2015,Gerin+2017}, and in many cases, the magnetic field geometry has been inferred in and near the discs \citepeg{Stephens+2014,Stephens+2017,Cox+2018}.

There have been numerous numerical studies of the formation and evolution of large ($M \ge 10^3$~\Msun) molecular clouds, that both exclude \citepeg{Walch+2012,DaleNgoumouErcolanoBonnell2014,Dale2017,LeeHennebelle2018a,LeeHennebelle2018b} and include \citepeg{GendelevKrumholz2012,Myers+2014,Geen+2015,Girichidis+2016,Geen+2018,LeeHennebelle2019} magnetic fields.  These studies allow for a direct comparison with observations on global (whole cloud) scales.  However, limited numerical resolution means that stellar properties must be inferred from sub-grid models since models on these scales currently do not resolve the fragmentation of individual stars. 

Simulating the formation and evolution of individual stars requires resolving the `opacity limit for fragmentation' --- the minimum Jeans mass as a function of density where radiation is trapped by dust opacity, leading to gas heating and a stable fragment \citep{Rees1976,LowLyndenbell1976}.   This minimum mass is several Jupiter masses and, when using smoothed particle hydrodynamics, must be resolved by 50-100 particles \citep{BateBurkert1997}. At our current computational capabilities, this restricts us to clouds of 50-500~\Msun while modelling the formation of individual stars without sub-resolution models.  

To date, many studies on this scale either ignored magnetic fields or neglected radiation by employing a barotropic equations of state (where temperature is simply prescribed as a function of density). Simulations of 50~\Msun{} clouds by \citet{BateBonnellBromm2003} using a barotropic equation of state produced as many brown dwarfs as stars --- hence over-producing the number of low-mass objects. Larger simulations of 500~{\Msun} clouds \citep{Bate2009sp}, or with better statistics from multiple realisations of 50~{\Msun} clouds \citep{Liptai+2017}, confirmed this discrepancy with the observed initial mass function (IMF). Once radiative feedback was included \citep{Bate2009rfb,Bate2012,Offner+2009}, the proportion of brown dwarfs decreased, bringing the mass function  into good agreement with the observed IMF.  Other low-mass cluster simulations have investigated the effect of different cloud density and velocity structure \citepeg{Bate2009ics,Liptai+2017,JonesBate2018ics}, metallicity \citepeg{Bate2014,Bate2019} and stellar radiation algorithms \citepeg{JonesBate2018fb}.  

Here, we focus on the effect of magnetic fields on low mass clouds \citep[as previously studied in, e.g.,][]{PriceBate2008,PriceBate2009,Hennebelle+2011,Myers+2013}; for reviews on the role of magnetic fields in molecular clouds, see \citet{Crutcher2012} and \citet{HennebelleInutsuka2019}.  These studies concluded that magnetic fields influenced the global evolution of the cloud, with magnetic pressure-supported voids and filamentary structures that followed the magnetic field lines forming in the strong field models.  The star formation rate was found to decrease with increasing initial magnetic field strength.  Our initial conditions most closely match that of \citet{PriceBate2008,PriceBate2009}, although they used Euler potentials to model the magnetic field, whereas we  directly evolve the induction equation.  Euler potentials enforce $\bm{\nabla}\cdot \bm{B} \equiv 0$, however, they cannot capture the full physics of magnetohydrodynamics (MHD) since helicity is identically zero \citepeg{Price2010,Price2012}; this means that complex magnetic fields structures, such as winding or combined poloidal/toroidal fields, can not be modelled with Euler potentials, but they can be with the direct evolution of the magnetic field (and divergence cleaning) that we use in this paper.

In this paper, we present results from low-mass radiation non-ideal magnetohydrodynamic star cluster simulations that resolve the fragmentation process down to the opacity limit.  In \secref{sec:methods}, we present our methods, and in \secref{sec:ic} we present our initial conditions.  We present our results in \secref{sec:results} and discuss them with respect to the literature in \secref{sec:discussion}.  Our conclusions are presented in \secref{sec:conclusion}.  Supplementary discussion of the dependence of the calculations on resolution and of the disc populations that are produced in the calculations are presented in Appendices \ref{app:resolution} and \ref{app:discs}, respectively.

\section{Methods}
\label{sec:methods}
\subsection{Radiation non-ideal magnetohydrodynamics}

We solve the equations of self-gravitating, radiation non-ideal magnetohydrodynamics in the form
\begin{eqnarray}
\frac{{\rm d}\rho}{{\rm d}t} & = & -\rho \nabla\cdot \bm{v}, \label{eq:cty} \\
\frac{{\rm d} \bm{v}}{\rm{d} t} & = & -\frac{1}{\rho}\nabla \cdot \left[\left(p+\frac{B^2}{2}\right)\mathbb{I} - \bm{B}\bm{B}\right] \notag \\
 &-& \nabla\Phi + \frac{\kappa \mbox{\boldmath$F$}}{c}, \label{eq:mom} \\
\rho \frac{\rm d}{{\rm d}t} \left(\frac{\bm{B} }{\rho} \right) & = & \left( \bm{B} \cdot \nabla \right) \bm{v}  - \bm{\nabla} \times \left\{ \eta_\text{OR} \bm{J} +  \eta_\text{HE}  \left(\bm{J}\times\bm{\hat{B}}\right) \right.\notag \\
                                                                                          & -&  \left.\left[ \eta_\text{AD}\left(\bm{J}\times\bm{\hat{B}}\right)\times\bm{\hat{B}}\right]\right\} \label{eq:ind}, \\
\rho \frac{\rm d}{{\rm d}t}\left( \frac{E}{\rho}\right) & = & -\nabla\cdot \bm{F} - \mbox{$\nabla \bm{v}${\bf :P}} \\
& + & 4\pi \rho \left( \kappa_{\rm d} B_\text{P,d} + \kappa_{\rm g} B_\text{P,g} \right) - c \rho (\kappa_{\rm d} + \kappa_{\rm g}) E, \label{eq:radiation} \\
\rho \frac{{\rm d}u}{{\rm d}t} & = & -p \nabla\cdot\bm{v} - 4 \pi \kappa_{\rm g} \rho B_\text{P,g} + c \rho \kappa_{\rm g} E\notag + \Gamma - \Lambda_\text{line}  \\
                                             & - &  \Lambda_\text{gd} + \eta_\text{OR}\bm{J}^2 +\eta_\text{AD}\left[ \bm{J}^2  - \left(\bm{J}\cdot\hat{\bm{B}}\right)^2 \right], \label{eq:matter} \\
\rho \Lambda_\text{ISR} & = & \kappa_{\rm d} \rho \left(4 \pi B_\text{P,d} - c E \right) - \Lambda_\text{gd}   \label{eq:dust}  \\
\nabla^{2}\Phi & = & 4\pi G\rho, \label{eq:grav}
\end{eqnarray}
where ${\rm d}/{{\rm d}t} \equiv \partial/\partial t  + \bm{v}\cdot \nabla$ is the Lagrangian derivative,  $\rho$ is the density, ${\bm  v}$ is the velocity, $p$ is the gas pressure, ${\bm B}$ is the magnetic field, $\bm{J} = \bm{\nabla}\times\bm{B}$ is the current density, $\Phi$ is the gravitational potential, $E$ is the radiation energy density, $\mbox{\boldmath $F$}$ is the radiative flux, {\bf P} is the radiation pressure tensor, $c$ is the speed of light, $G$ is the gravitational constant, and $\mathbb{I}$ is the identity matrix.  The non-ideal MHD coefficients for Ohmic resistivity, the Hall effect and ambipolar diffusion are given by $\eta_\text{OR}$, $\eta_\text{HE}$ and $\eta_\text{AD}$, respectively.  We assume units for the magnetic field such that the Alfv{\'e}n speed is $v_{\rm A} = \vert B\vert/\sqrt{\rho}$ \citep[see][]{PriceMonaghan2004a}.

For the treatment of the equation of state and radiative transfer, we use the combined radiative transfer and diffuse interstellar medium model of \citet{BateKeto2015}.  The underlying method of radiative transfer is grey flux-limited diffusion, using the implicit algorithms devised by \cite{WhitehouseBateMonaghan2005} and \cite{WhitehouseBate2006}.   \citeauthor{BateKeto2015} extended this method to treat gas and dust temperatures separately, and included various heating and cooling mechanisms that are particularly important at low densities.  In Eqns.~\ref{eq:radiation} to \ref{eq:dust}, $B_\text{P,d}$ is the frequency-integrated Planck function at the dust temperature, and $B_\text{P,g}$ is the frequency-integrated Planck function at the gas temperature, where the Planck function depends on temperature as  $B_\text{P} = (\sigma_\text{B}/\pi)T^4$ where $\sigma_\text{B}$ is the Stefan-Boltzmann constant.  The quantities $\kappa_\text{d}$ and $\kappa_\text{g}$ and the frequency-integrated opacities of the dust and gas, respectively.  The quantity $\Gamma$ includes cosmic ray, photoelectric, and molecular hydrogen formation heating terms.  The quantity $\Lambda_\text{line}$ provides gas cooling due to atomic and molecular line emission, and $\Lambda_\text{gd}$ treats thermal energy transfer between the gas and the dust.  Eqn.~\ref{eq:dust} sets the dust to be in thermal equilibrium with the total radiation field, taking into account the energy transfer between the gas and the dust.  The total radiation field is comprised of two components: the external interstellar radiation field ($\Lambda_\text{ISM}$ is the dust heating rate per unit volume due to the interstellar radiation field), and the radiation energy density of the flux-limited diffusion radiation field.

The equation of state is described by \cite{WhitehouseBate2006}. Briefly, we employ an ideal gas equation of state that assumes a 3:1 mix of ortho- and para-hydrogen (see \citealp{Boley+2007}) and treats the dissociation of molecular hydrogen and the ionisations of hydrogen and helium.  The mean molecular weight is taken to be $\mu_{\rm g} = 2.38$ at low temperatures, and we use dust opacity tables from \cite{PollackMckayChristofferson1985} and gas opacity tables from \cite{Ferguson+2005}.

\subsection{Smoothed particle radiation non-ideal magnetohydrodynamics}

We use the three-dimensional smoothed particle hydrodynamics (SPH) code \sphng \ that originated from \citet{Benz1990}, but has since been substantially extended to include individual particle timesteps and sink particles \citep{BateBonnellPrice1995}, variable smoothing lengths \citep{PriceMonaghan2004b,PriceMonaghan2007}, radiative transfer \citep{WhitehouseBateMonaghan2005,WhitehouseBate2006,BateKeto2015}, magnetohydrodynamics \citep{Price2012} and non-ideal MHD \citep{WursterPriceAyliffe2014,WursterPriceBate2016}.  The code is parallelised using both {\sc OpenMP} and MPI.

The density and smoothing length of each SPH particle are calculated iteratively by summing over its nearest neighbours and using $h = 1.2 (m/\rho)^{1/3}$ where $h$, $m$ and $\rho$ are the SPH particle's smoothing length, mass and density, respectively \citep{PriceMonaghan2004b,PriceMonaghan2007}.  The remainder of the terms, including our magnetic variable of $\bm{B}/\rho$ (Eqn.~\ref{eq:ind}), are solved using the standard smoothed particle magnetohydrodynamics (SPMHD) scheme \citep[for a review, see][]{Price2012}.  We use the artificial viscosity described in \citet{PriceMonaghan2005} to capture shocks.  For magnetic stability, we use the \cite{BorveOmangTrulsen2001} source-term approach and constrained hyperbolic divergence cleaning method of \citet{TriccoPriceBate2016}.  Artificial resistivity is included using the \citet{TriccoPrice2013} resistivity switch, which produces higher artificial resistivity \citep{Wurster+2017} and hence weaker magnetic fields strengths than the algorithm from \citet{Phantom2018} that we have used in our recent studies of isolated star formation \citep{WursterBatePrice2018sd,WursterBatePrice2018hd,WursterBatePrice2018ff}.  Our SPMHD algorithm allow us to capture the complex magnetic field structures, unlike in previous studies that used the Euler potential method \citep{PriceBate2008,PriceBate2009}.  To evolve the simulation, we employ a second-order Runge-Kutta-Fehlberg integrator \citep{Fehlberg1969}.

We use explicit timestepping for the Hall effect and ambipolar diffusion, with
\begin{equation}
\label{num:dtnimhd}
\text{d}t_\text{non-ideal} = C_\text{non-ideal}\frac{h^2}{\left|\eta\right|},
\end{equation}
where $C_\text{non-ideal}=1/2\pi$ is a constant equivalent to the Courant number \citep{WursterPriceAyliffe2014}.  Ohmic resistivity is calculated using an implicit solver \citep[see Appendix A of][]{WursterBatePrice2018sd}.
  
\section{Initial conditions}
\label{sec:ic} 

We initialise our clusters as spheres of cold, dense gas embedded in a warm medium.  The sphere has an initial radius of $R_0 =  0.1875$~pc and mass of 50~\Msun, yielding an initial uniform density of $1.22\times10^{-19}$~\gpercc; the characteristic timescale is the free-fall time of $t_\text{ff} = 1.9\times10^5$~yr.  The initial temperature structure of the sphere is set such that it is in thermal equilibrium with the interstellar radiation field; the initial temperature increases from $T\sim8.8$~K at the centre of the sphere to $\sim13$~K at its edge.  The sphere is embedded in a cube whose side-length is 0.75~pc and density is 30 times lower, with the warm and cold media in pressure equilibrium. We use boundary conditions such that the hydrodynamic forces are periodic but gravitational forces are not.  This `sphere-in-box' method prevents the need for magnetic boundary conditions at the edge of the initial sphere, which would immediately become deformed.

We impose an initial supersonic `turbulent' velocity field as described by \citet{OstrikerStoneGammie2001} and \citet{BateBonnellBromm2003}.  We generate a divergence-free random Gaussian velocity field with a power spectrum $P(k) \propto k^{-4}$, where $k$ is the wavenumber.  This is generated on a $128^3$ uniform grid, and the velocities of the particles are then interpolated from this grid. The velocity field is normalised so that the kinetic energy of the turbulence is equal to the magnitude of the gravitational potential energy of the initial sphere (i.e. excluding the warm medium).  This equates to an initial rms Mach number of the turbulence of $\mathcal{M} = 6.4$.  Each model uses the same initial velocity field, which uses the same random seed as in many of our previous simulations \citepeg{BateBonnellBromm2003,PriceBate2008,Bate2009rfb,Bate2014,Liptai+2017}.   We caution that a different random seed may yield different results \citepeg{Liptai+2017,Geen+2018}.  For example, in their study, \citet{Geen+2018} found the star formation efficiency to vary between 6 and 23 per cent simply by changing the random seed.  Ideally, we would perform many realisations of each model with different seeds, but this is beyond our current computing resources.  

We thread the entire domain with a uniform magnetic field initially parallel to the $z$-axis.  The non-ideal MHD coefficients of Ohmic resistivity, ambipolar diffusion and the Hall effect are computed using version 1.2.1 of the \textsc{Nicil} library \citep{Wurster2016}.  This includes a heavy ion represented by Magnesium \citep{Asplund+2009}, a light ion represented by the hydrogen and helium compounds, and a single grain species, $n_\text{g}$. We assume a single grain species with radius $a_\text{g} = 0.1~\mu$m and bulk density of $\rho_\text{b} = 3$~\gpercc \ \citep{Pollack+1994}, respectively, that includes three populations with charges $Z = -1, 0, +1$, respectively, where $n_\text{g} = n_\text{g}^- + n_\text{g}^0 + n_\text{g}^+$ to conserve grain density.  The grain number density is calculated from the local gas density, assuming a dust-to-gas ratio of 0.01. 

To model the long-term evolution of the cluster, we include sink particles \citep{BateBonnellPrice1995} with accretion radius $r_\text{acc} = 0.5$~au.  When a gas particle reaches $\rho = 10^{-5}$~\gpercc, near the end of the second collapse phase, it is replaced with a sink particle;  all particles within $0.25$~au of the centre of the sink particle are automatically accreted, and all particles within $0.5$~au are tested to determine if they are to be accreted.  Since we resolve the opacity limit, each sink particle represents one star or brown dwarf, and the properties of the sink reflect the properties of the star.  We merge sink particles that approach within 27~R$_\odot$ of each other. 

In this study, we model a 50~M$_\odot$ cloud in order to resolve the minimum local Jeans mass throughout the simulation, which is \appx0.0011~\Msun \ during the isothermal collapse phase \citep[][and discussion therein]{BateBurkert1997,BateBonnellBromm2003}.  Resolving this limit requires \sm75 particles per Jeans mass, or a minimum of $3.5 \times 10^6$ particles for a 50~\Msun{} cloud, as used in our previous studies.  By default, we use $5\times10^6$ equal mass SPH particles in the cold dense sphere, and an additional $2.3\times10^6$ particles in the warm medium.  To test the effect of resolution, we also perform two simulations using $35\times10^6$ equal mass SPH particles in the sphere; see Appendix~\ref{app:resolution}.

We test four magnetic field strengths: $B_0 = 6.48, 3.89, 1.94 \text{ and } 0.972 \times 10^{-5}$~G, which correspond to normalised mass-to-flux ratios of $\mu_0 = 3, 5, 10$ and 20.  The normalised mass-to-flux ratio is given by
\begin{equation}
\label{eq:mu0}
\mu_0 = \left(\frac{M}{\pi R_0^2B_0}\right)\left(\frac{M}{\Phi_\text{B}}\right)_\text{crit}^{-1},
\end{equation}
where
\begin{equation}
\label{eq:mucrit}
\left(\frac{M}{\Phi_\text{B}}\right)_\text{crit} = \frac{c_1}{3}\sqrt{\frac{5}{G}},
\end{equation}
is the critical mass-to-flux ratio in CGS units;  here, $\Phi_\text{B}$ is the magnetic flux threading the surface of the (spherical) cloud, and $c_1 \simeq 0.53$ is a parameter determined numerically by \citet{MouschoviasSpitzer1976}.

For each magnetic field strength at our fiducial resolution ($5\times10^6$ particles), we model both an ideal and non-ideal MHD cluster.  We also model a purely hydrodynamical model, for a total of nine simulations.  The magnetised models are labelled \emph{Ixx} and \emph{Nxx}, where \emph{xx} is the normalised mass-to-flux ratio.  We refer to the hydrodynamical model as `hydro' or \emph{Hyd}.

\section{Results}
\label{sec:results}

The different dynamics and physical processes included in each model led to variations in computational expense, so not all models had reached the same final time.  Our primary analysis is performed at \ttnow, the minimum time reached by all simulations.

\subsection{Large scale structure}
\label{sec:lss}

\subsubsection{Gas density evolution}

\begin{figure*}
\centering
\includegraphics[width=0.98\textwidth]{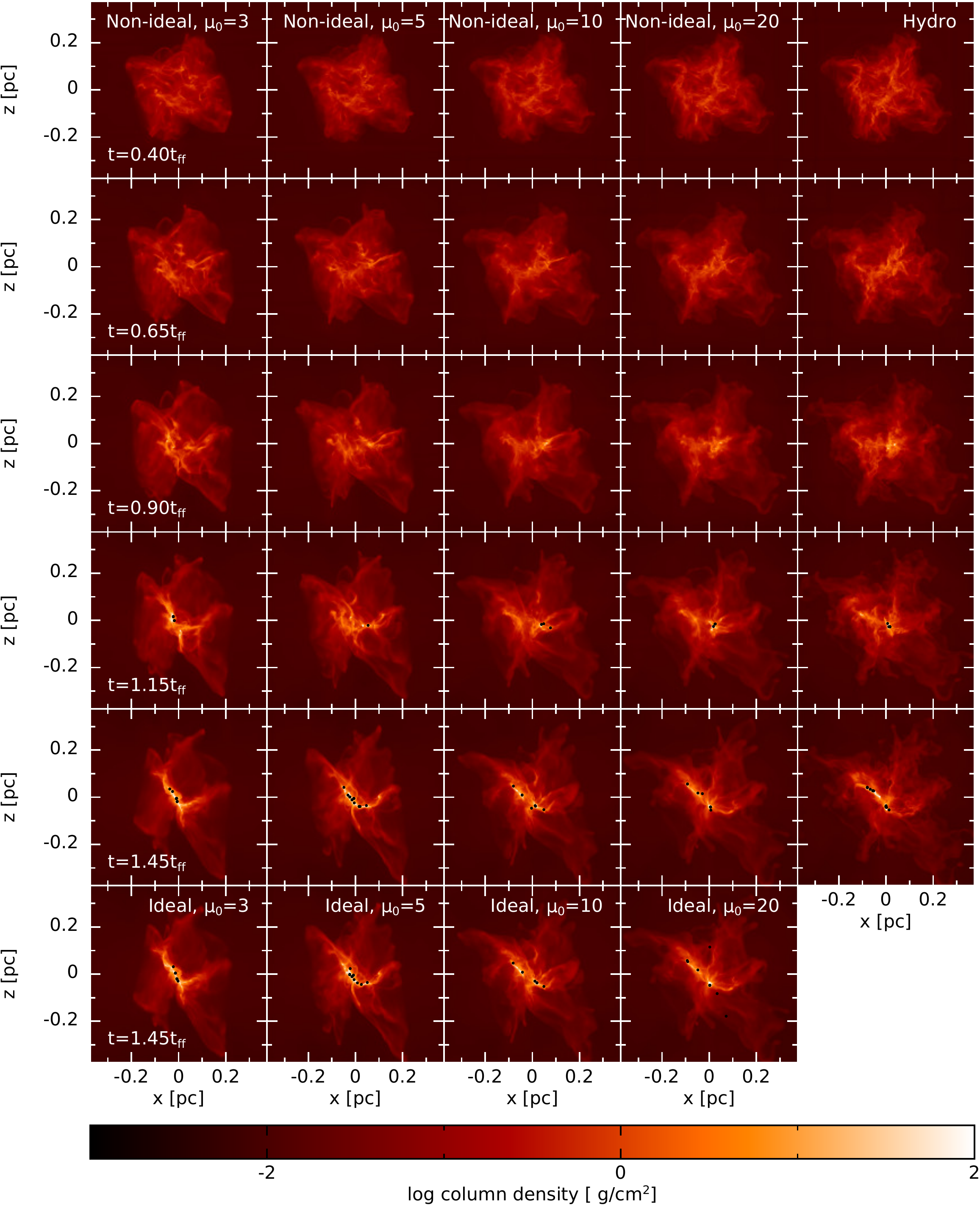}  
\caption{Top five rows: Evolution of the gas column density for the non-ideal MHD models and the hydro model.  Bottom two rows: Gas column density at $t = 2.76\times10^5$~kyr = 1.45~\tff \  for each model in our suite.  Each panel has a width of 0.75~pc, which is the initial width of our low-density box.  Each black dot represent the location of a sink particle and is not plotted to scale.  At strong initial magnetic field strengths, the initial evolution is approximately filamentary, whereas it is more clumpy for weaker magnetic fields.  By \ttnow, each model has formed a primary filament, and the weakly magnetised models have also formed several low-density filaments.   Even the weak magnetic field of \mueq{20} affects the evolution when compared to the purely hydrodynamical model.  Non-ideal MHD has a negligible effect on the global evolution of the cluster (comparing bottom two rows).}
\label{fig:columndensity:globalevol}
\end{figure*}

The top five rows of Fig.~\ref{fig:columndensity:globalevol} show the time evolution of the non-ideal MHD and hydro models (left to right, respectively). The bottom row shows the ideal MHD models at \ttnow, which may be compared to the row above showing the corresponding non-ideal models at the same time.

Magnetic fields have an immediate effect on the evolution of the large scale cloud structure, with visible differences in the large scale structure appearing as early as $t = 0.15$~\tff.  The weakly magnetised models become more filamentary at an earlier time, yielding slightly higher maximum densities.  Stronger magnetic fields, however, can be seen to smooth out the large-scale shock structure, as also found by \citet{PriceBate2008}; the effect is similar to increasing the gas pressure in hydrodynamical simulations \citep{BateBonnell2005}.  By $t =  0.40$~\tff, the evolution of the hydro model visibly differs from the magnetised models with \mueq{20}, showing that even a weak magnetic field can influence the large scale dynamics.

During the early evolution ($t \lesssim 0.40$~\tff), there are many transient features in the gas, with filaments forming, colliding and dispersing.  Persistent, dense clumps begin to form at $t = 0.40$~\tff \ (top row) in the hydro model first.  Due to small evolutionary changes already caused by the magnetic field, there is no obvious relation between the formation time of the first clumps and the initial magnetic field strength.  As the evolution continues, the clumps become denser and filamentary, which is more pronounced in the more strongly magnetised models.  Each model contains two primary groups of clumps/filaments, and these ultimately merge into a single, primary filament in the models with \mueq{3} and $5$.

At $t = 0.90$~\tff \ (third row), there is a weak qualitative trend regarding the dense gas near $x \sim 0.05$~pc and $z \sim 0$:  As the magnetic field strength is increased, this clump becomes more compact and denser.  However, this trend is broken by \emph{N03}, in which this clump does not exist.  Rather, the denser clump occurs at  $x \sim -0.05$~pc and $z \sim 0$ and is more elongated.   Although this qualitative difference may appear trivial, it demonstrates the effect of magnetic fields on the large-scale cloud structure.  This qualitative difference also suggests that the properties of \emph{N03} may not conform to the trend of the remaining three non-ideal models plus the hydro model.  This deviation from the trend will permeate most of our subsequent analysis.  This difference is a result of the clump preferentially forming along the magnetic field lines in \emph{N03}, thus becoming filamentary early.  In \emph{N05}, there are multiple regions where the gas has collapsed vertically into small clumps, and then the turbulent motion ultimately joins these clumps into a filament.  In the remaining models, filaments typically form without any preference to the $z$-axis (i.e., to the initial magnetic field direction).  

As the clusters evolve to  $t = 1.15$~\tff{} (fourth row), the dense regions in \emph{N05} to \emph{Hyd} collapse and host the birth of the first stars; thus the first stars form in these models in similar locations.  The dense clump in \emph{N03} also collapses to form stars, but this dense structure is already more elongated and denser than in the remaining models.

As the evolution continues, the filament becomes denser in \emph{N03}, while in the remaining models, the turbulent motion of the gas brings the various clumpy regions together to form the filament.  Thus, at \ttnow, each model contains one large scale (\sm0.2~pc) filament, but their formation mechanisms differ between the strong and weak field cases.

This primary filament is more vertical in the $yz$-plane than the $xz$-plane shown in Fig.~\ref{fig:columndensity:globalevol}, but this is simply a result of the initial turbulent velocity field.  In addition to the primary filament, there are many lower density filaments in all models, most of which emanate from the primary filament.  The low-density filaments are denser but shorter in the more strongly magnetised models, since only dense objects can overcome the magnetic field strength in these models.

The evolutionary sequence described above occurs in both our ideal and non-ideal MHD models; non-ideal effects have little effect on the large scale evolution.

\begin{figure}
\centering
\includegraphics[width=1.0\columnwidth]{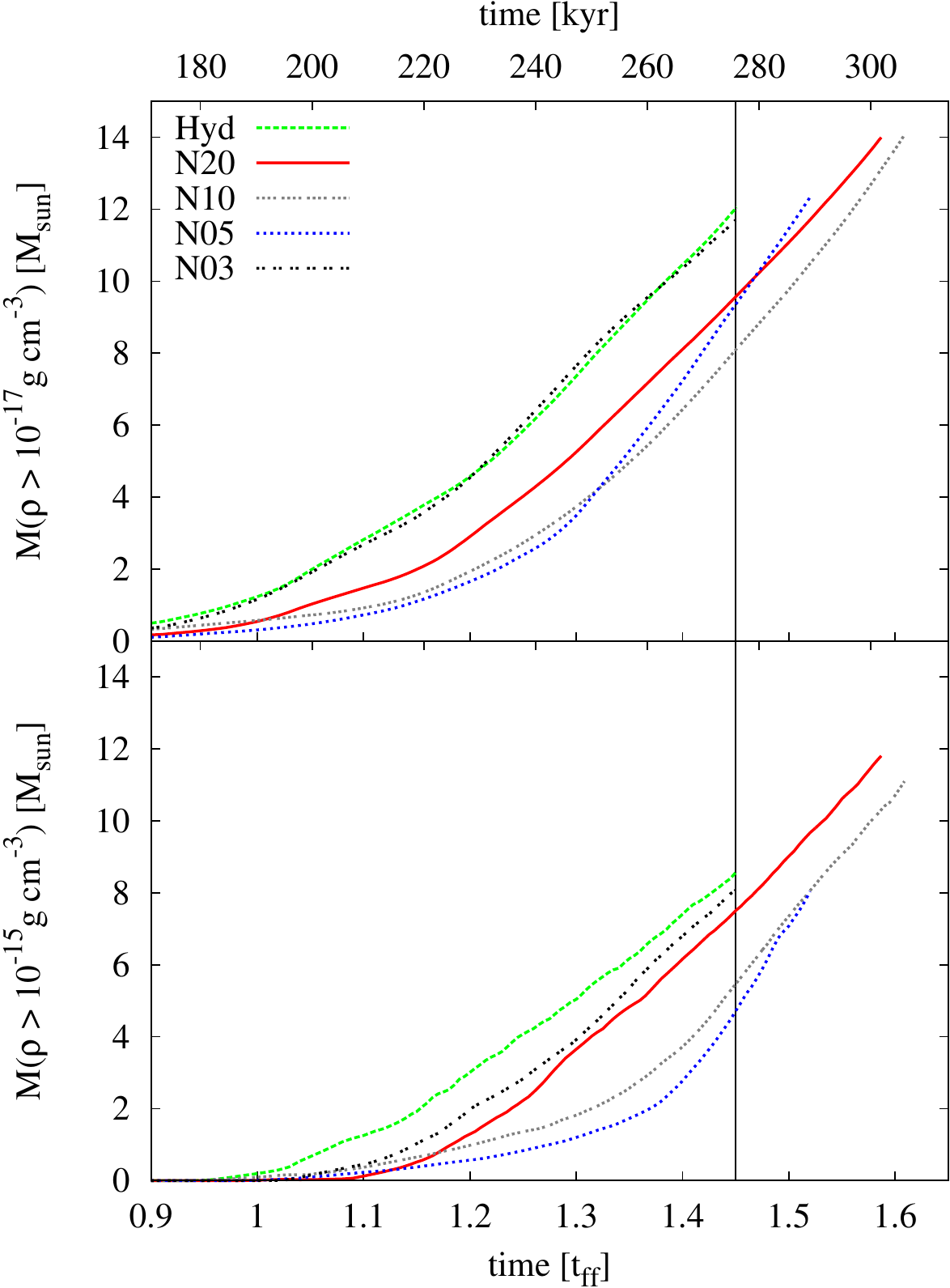}  
\caption{Mass above $\rho > 10^{-17}$ (top panel) and $10^{-15}$~\gpercc \ (lower panel)  for the hydro and non-ideal models; these masses include stellar masses.  The vertical line at \ttnow \ is the final time of the models that were evolved for the shortest amount of time.  At the higher threshold, there is less dense gas for models with increasing magnetic field strengths; model \emph{N03} is an outlier since the gas in this model quickly collapses along the initial magnetic field lines.  For all time and at each density threshold, the gas+stellar mass in an ideal model is slightly lower than its non-ideal counterpart.}
\label{fig:gasmass}
\end{figure}

 \fig{fig:gasmass} shows the gas+stellar mass above density thresholds of $\rho > 10^{-17}$ (top panel) and $10^{-15}$~\gpercc \ (lower panel) as a function of time for the hydro and non-ideal MHD models. The majority of the gas remains at low densities, with 76 to 87 per cent having \rhole{-17} (recall that \rhonaught{-19}).   Prior to \tapprox{1.32}, there is a trend of less gas with \rhogt{-17} for models with stronger magnetic field strengths (top panel).  This is reasonable since magnetic fields support gas from gravitationally collapsing, and was also found by \citet{PriceBate2008}.  The exception to this trend is \emph{N03}, which has the same amount of gas above $10^{-17}$~\gpercc{} as the hydro model.  This is a result of the rapid collapse of gas along the field lines to form the dense filament, as discussed above.

\subsubsection{Magnetic field evolution}

\begin{figure*}
\centering
\includegraphics[width=1.0\textwidth]{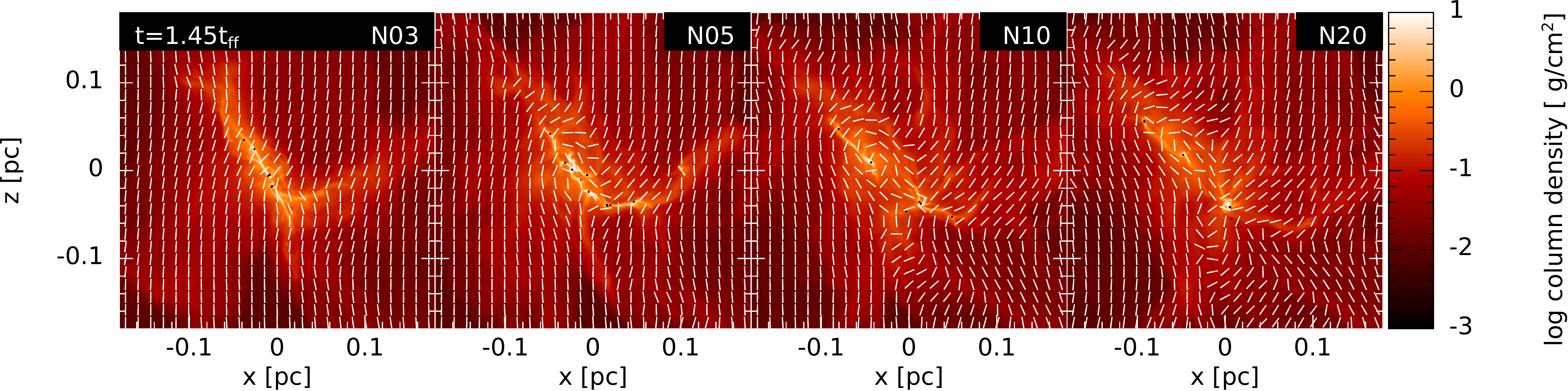}  
\includegraphics[width=1.0\textwidth]{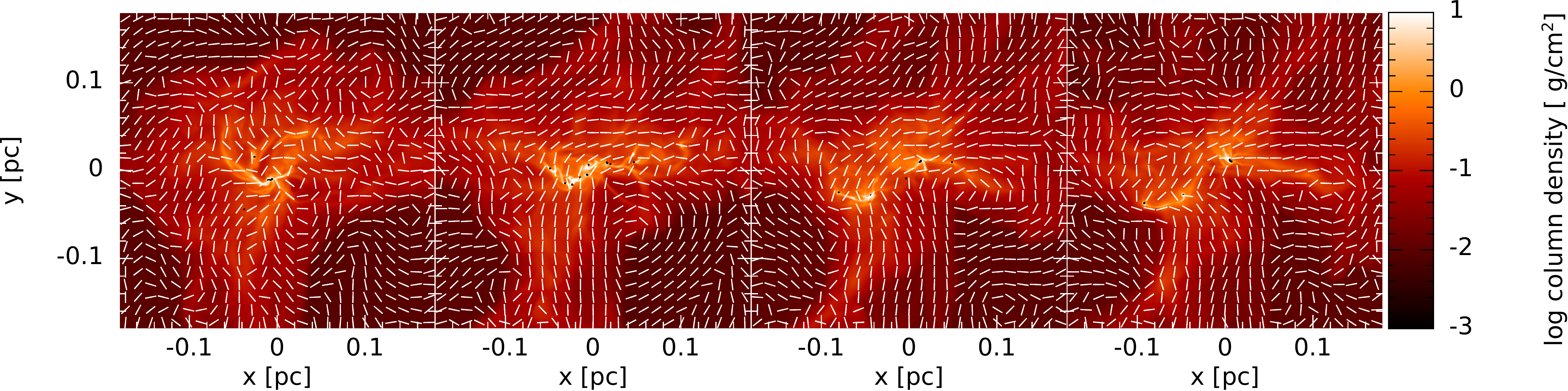}  
\caption{Gas column density of the non-ideal MHD models at \ttnow{} in the $xz$- (top) and $xy$- (bottom) planes.  The white lines represent the direction of the magnetic field.  While remaining mostly vertical on the large scale in the $xz$-plane, the magnetic field becomes more twisted near the filament in the weaker magnetic field models.  The magnetic field lines are mostly vertical at scales larger than shown here.  In \emph{N03}, the magnetic field crosses the primary filament approximately perpendicularly, while components parallel to the filaments appear in the models with weaker initial magnetic field strengths.  Given the lack of initial preferential magnetic field direction in the $xy$-plane, the strong magnetic field lines better flow perpendicularly through the dense filaments, and the weak magnetic field lines flow parallel to the low-density filaments.  On this scale, there is no significant difference between the ideal and non-ideal MHD models of the same initial magnetic field strength.}
\label{fig:Bvec:global}
\end{figure*}

Fig.~\ref{fig:Bvec:global} shows the gas column density over-plotted with the magnetic field direction in both the $xz$- and $xy$-planes.
Given the initial magnetic field of $\bm{B} = -B_0\hat{\bm{z}}$, the mostly vertical magnetic field in the $xz$-plane is to be expected (top row).   For stronger initial magnetic field strengths, the field remains closer to the initial conditions as the cluster evolves, especially on scales larger than shown in the figure.  In \emph{N03}, the magnetic field crosses the primary filament approximately perpendicularly, in agreement with Planck observations \citepeg{Planck2016or,Planck2016role} and optical polarisation measurements in Taurus molecular cloud \citep{Goldsmith+2008,Chapman+2011} and the Pipe Nebula \citep{AlvesFrancoGirart2008}, indicating that such magnetically controlled filament formation is likely to occur in the ISM; we note that the Planck observations \citep{Planck2016role} typically resolve down to 0.4~pc, whereas we can resolve the gas density in the filaments down to $\rho \sim 5\times10^{-3}$~pc.  In the models with weaker magnetic fields, the magnetic field also crosses the filaments perpendicularly, but there is also a considerable parallel component, especially in the regions surrounding the filament.  This parallel component is more prominent in the less dense filament and in the weaker magnetic field models.

From observations, it is expected that the magnetic field is perpendicular to the dense filaments and parallel to the low-density filaments, and our results broadly agree with this, although our agreement is more tenuous in the initially weakly magnetised models.  This tenuous agreement may also be a result of the initial preferential direction of the magnetic field.  When we observe the cluster from the $xy$-plane (bottom row of Fig.~\ref{fig:Bvec:global}), the magnetic field better follows the observationally expected paths, since in this plane, there is no preferential magnetic field direction.  In \emph{N20}, the filament is only dense enough in a few regions such that the magnetic field crosses it perpendicularly.

\subsubsection{B vs $\rho$}

\begin{figure}
\centering
\includegraphics[width=1.0\columnwidth]{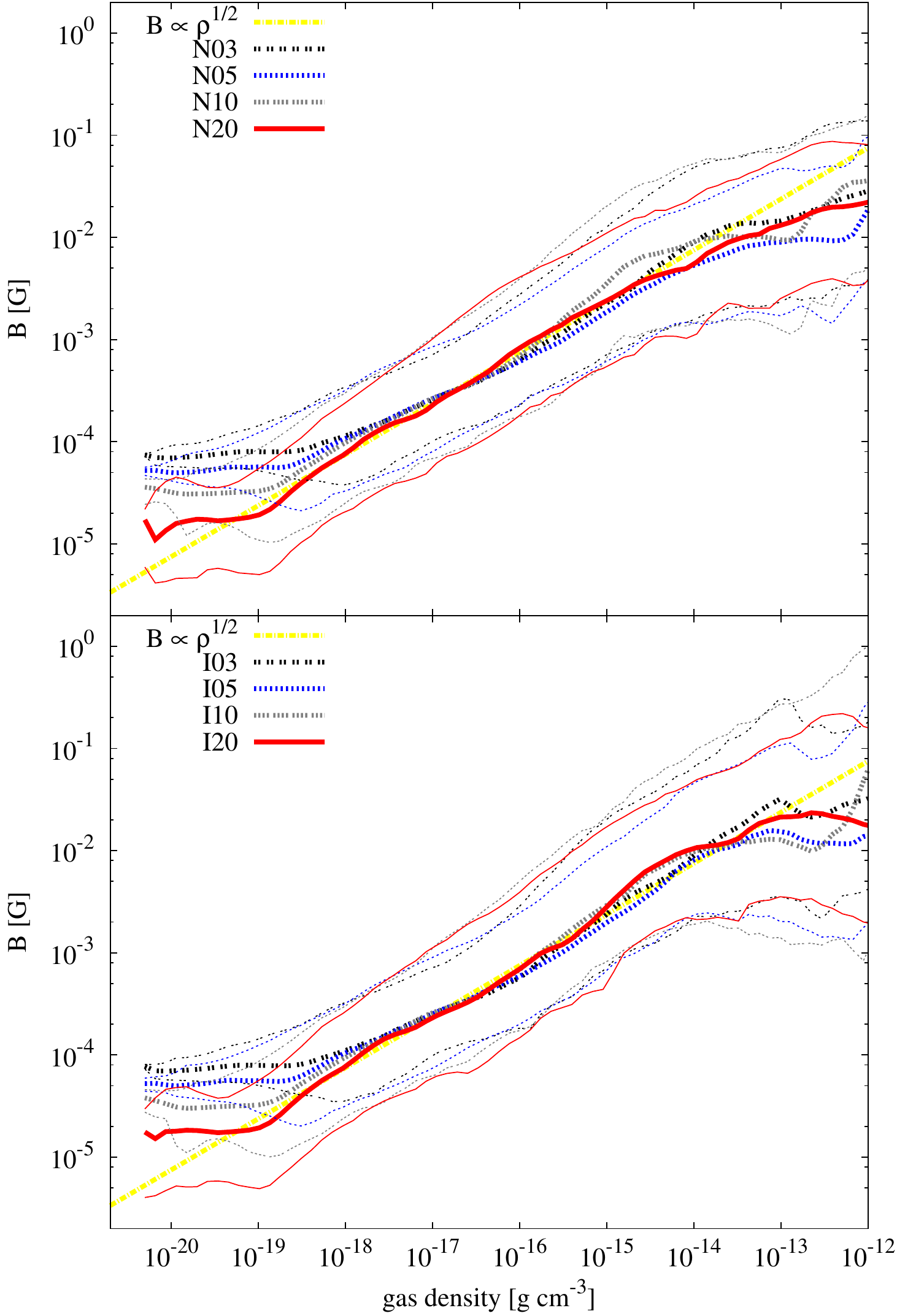}  
\caption{Magnetic field strength as a function of gas density for the non-ideal (top) and ideal (bottom) MHD models.  Thick lines give mean values averaged over each snapshot just prior to the formation of each star before \tnow{}, and the thin lines bracket 95 per cent of all magnetic field strengths at a given density.  For \rhogs{-17}, the magnetic field strength becomes independent of the initial conditions. }
\label{fig:RhoVB}
\end{figure}

\fig{fig:RhoVB} shows the mean magnetic field strength,
\begin{equation}
\label{eq:Bave}
B_\text{mean} =  10^{\left(\Sigma_i^N \log B_i\right)/N},
\end{equation}
along with the range containing 95 per cent of the magnetic field strengths, in the density range from $\rho = 5\times10^{-21}$ to $10^{-12}$~\gpercc{}; recall that the initial cloud density is $\rho_0 = 1.22\times10^{-19}$~\gpercc{}.  These curves are the average profiles of each snapshot just prior to each sink formation event before \tnow{}.  Although the calculations follow protostellar collapse to $\rho = 10^{-5}$~\gpercc{} before sink particle insertion, the numerical resolution is much lower than in our isolated studies of star formation and the artificial resistivity is likely to weaken the field within the first hydrostatic cores ($\gtrsim 10^{-12}$~\gpercc{}), thus we do not comment on the magnetic field at higher densities. 
 
At low densities (\rhols{-19}), the magnetic field strength remains close to the initial field strength; there is larger scatter around this value for the high-$\mu_0$ models, and little scatter for low-$\mu_0$ models.  At intermediate  densities ($10^{-17} \lesssim \rho/($\gpercc$) \lesssim 10^{-13}$), the relationship converges for the eight models, with the average magnetic field strength increases approximately as $B \propto \rho^{1/2}$;  in this range,  the average magnetic field strengths only differ by a factor of a few between the models, which is less than the initial spread.  All models also have approximately the same dispersion of field strengths in this range, which is approximately an order of magnitude.  

The overall scaling of $B \propto \rho^{1/2}$ in this range, and the relative independence of the average magnetic strength on the initial field strength is consistent with the results of recent moving mesh simulations of magnetised, turbulent molecular clouds of \cite{Mocz+2017}.  They find that $B \propto \rho^{2/3}$ for weakly magnetised clouds,  transitioning to $B \propto \rho^{1/2}$ for more strongly magnetised clouds, with the transition occurring at around a Alfv\'enic Mach number of unity.  Our initial conditions (magnetic field strength and density) are similar to their more highly magnetised cases.  It is interesting to note that they also find that the typical magnetic field strength at  densities \rhogs{-18} is independent of their initial magnetic field strength (see their fig.~4). They perform simulations in which the initial magnetic field strength is varied by two orders of magnitude (compared to the range of a factor of $\approx 7$ in our simulations).  They do not obtain as great a dispersion in magnetic field strengths at \rhogs{-15}, but this may simply be because they stop their calculations when the first region collapses.

The wide range of magnetic field strengths in all the models prior to the hydrostatic core phase (i.e. in the range $10^{-17} \lesssim \rho/($\gpercc$) \lesssim 10^{-13}$) suggest a wide range of magnetic initial conditions for star formation.  The maximum magnetic field strengths here are slightly higher than those presented in the isolated star formation study of \citet{BateTriccoPrice2014}, who used similar mass-to-flux ratios (but slightly stronger absolute magnetic field strengths).  Given the demonstration of the magnetic braking catastrophe \citep{AllenLiShu2003} in that study, this suggests that disc formation may be hindered in the current simulations, or at least in the regions permeated by strong magnetic fields as given by the upper limits shown in \fig{fig:RhoVB}).  Discs and disc formation will be further discussed in \secref{sec:ppd}.

There is no significant difference in results between the ideal and non-ideal MHD models (compare top to bottom panels of Fig.~\ref{fig:RhoVB}.)  In agreement with our previous studies \citepeg{WursterPriceBate2016,WursterBatePrice2018sd,WursterBatePrice2018hd}, this indicates that non-ideal MHD processes only affect high-density (\rhogs{-12}) gas.

\subsubsection{Non-ideal vs ideal MHD}
\label{sec:lss:ini}
In the outer regions of the cloud where \rhols{-18}, the ionisation fraction is $n_\text{e}/n \gtrsim 10^{-6}$.  This relatively high ionisation rate is a result of cosmic rays easily being able to ionise the low-density gas; in this gas, the temperature is too cold, $T \sim \mathcal{O}\left(10\right)$~K, and the dust grain population is too small to contribute to the ionisation fraction.  The drift velocity (i.e. $\bm{v}_\text{drift} = \bm{v}_\text{neutral} - \bm{v}_\text{ion}$) in the bulk of the cloud is \sm$10^{-4}-10^{-1}$~\kms, suggesting that the ions are well-coupled to the neutral particles.  Given the high ionisation fractions, the non-ideal processes are too weak to contribute significantly to the evolution of the gas.   Therefore, the evolution of the large-scale structure is essentially independent of whether ideal or non-ideal MHD is assumed.

Comparison of the bottom two rows of Fig.~\ref{fig:columndensity:globalevol} confirms this. The quantity of gas with \rhogs{-15} (c.f. \fig{fig:gasmass}; ideal models not shown) is also not strongly sensitive to non-ideal MHD effects.  At any given time, the mass of gas with \rhogs{-15} in an ideal model is slightly lower than its non-ideal counterpart, which is reasonable since magnetic fields resist gravitational collapse, and the non-ideal processes weaken the local magnetic field.  This decrease of mass in dense gas, however, occurs mainly on small scales, as discussed in Section~\ref{sec:sss} below.

The ionisation fraction decreases for increasing density within the cloud, reaching fractions of $n_\text{e}/n \sim 10^{-9}$ in filaments of \rhoapprox{-16}, and fractions of $n_\text{e}/n \sim 10^{-14}$ in protostellar discs. Hence we expect non-ideal MHD to act mainly on small scales.

\subsection{Small scale structure}
\label{sec:sss}
\begin{figure*}
\centering
\includegraphics[width=1.0\textwidth]{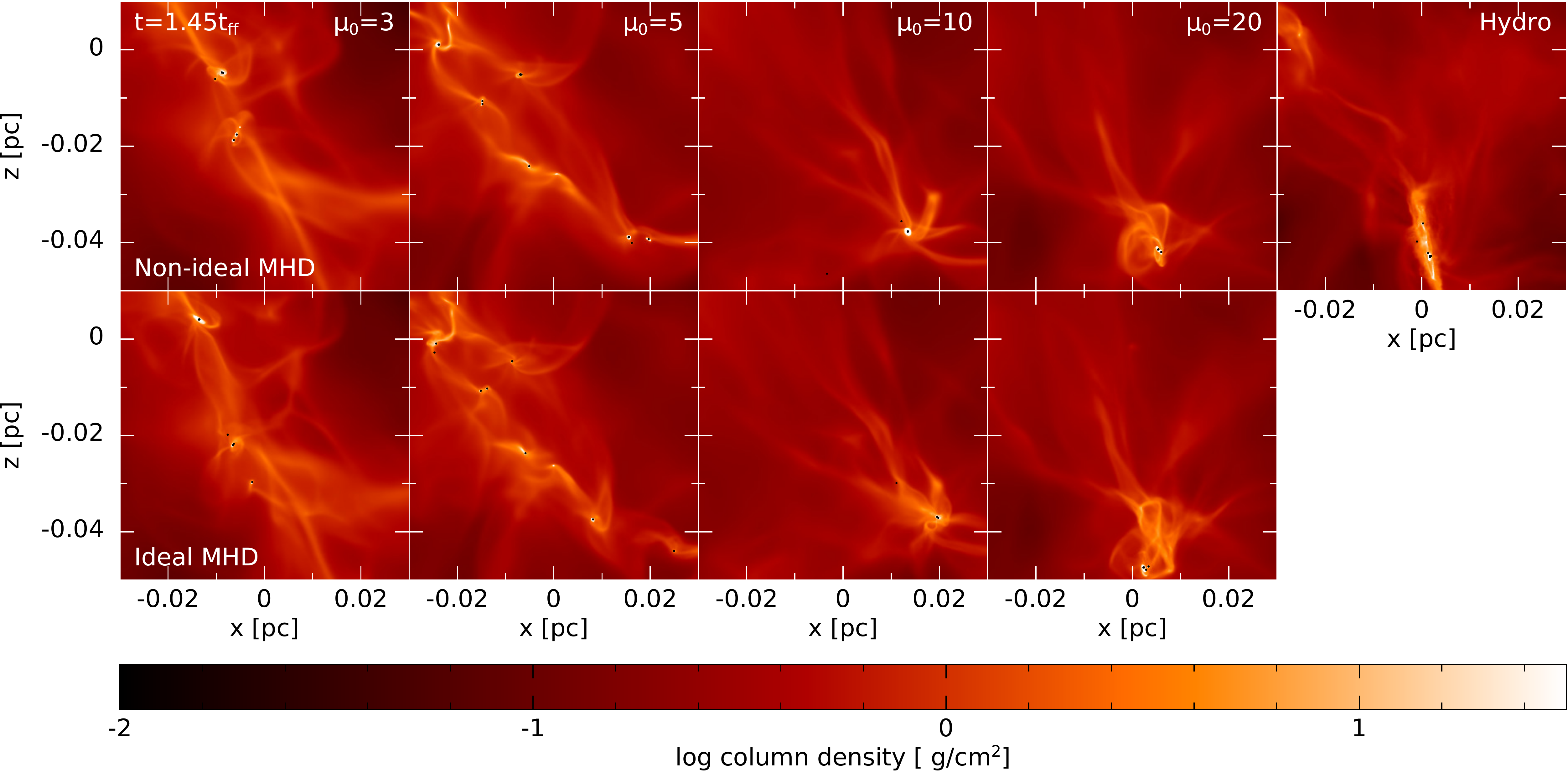}  
\caption{Gas column density of a small region of the cluster at \ttnow \ for each model.  Each black dot represent the location of a sink particle (not to scale).  On the small scales, decreasing the initial magnetic field yields more clumpy and less filamentary objects (i.e. increasing $\mu_0$; left to right).  Removing the non-ideal effects results in less dense structures and filaments, indicating that the non-ideal processes are important on small scales.}
\label{fig:columndensity:local}
\end{figure*}

Fig.~\ref{fig:columndensity:local} shows a $0.06 \times 0.06$ pc$^2$ region of each model to highlight differences caused by changing the initial magnetic field strength, or by using ideal instead of non-ideal MHD.  Each panel shows the same spatial region for comparison.  Since each model was initialised with the same velocity field, the differences are a direct result of the initial magnetic field strength and non-ideal MHD processes.
\begin{figure*}
\centering
\includegraphics[width=1.0\textwidth]{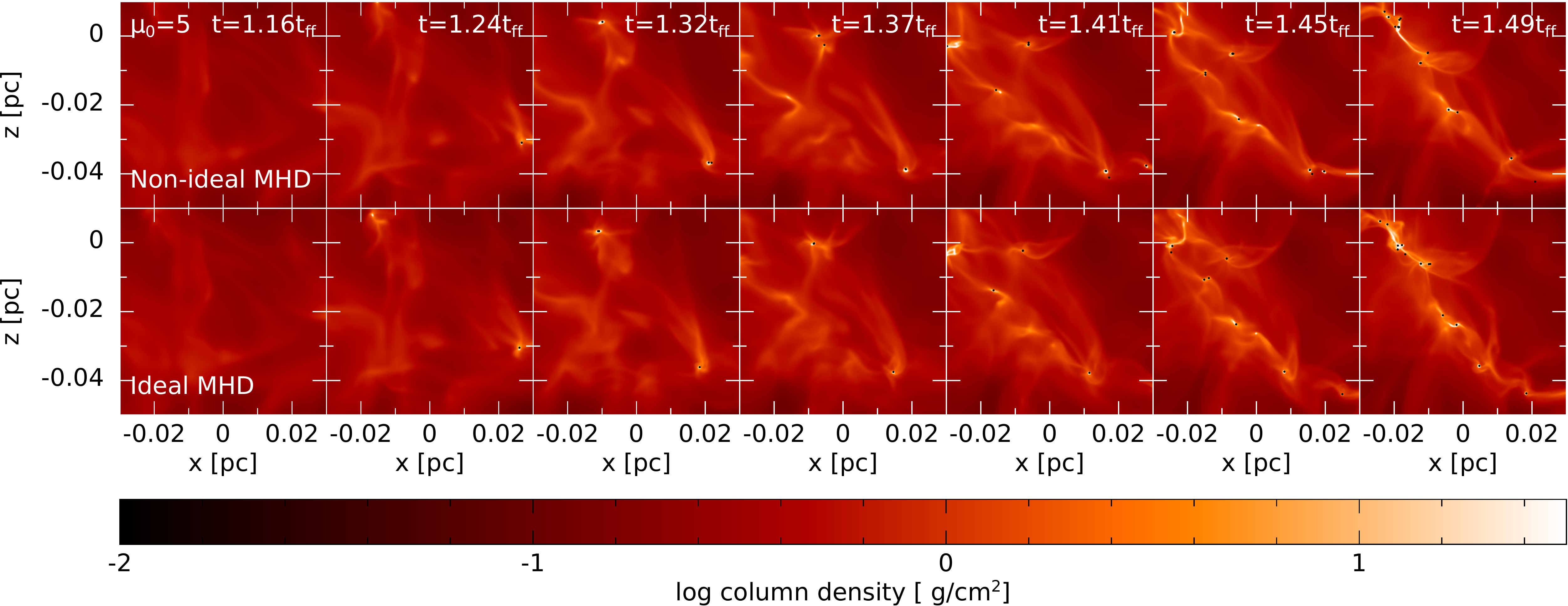}  
\caption{Evolution of the gas column density of a small region of the cluster for the models with \mueq{5}, comparing non-ideal MHD (top) to ideal MHD (top).  Each black dot represent the location of a sink particle (not to scale).  The initial evolution of the general gas structure is similar between the two models, although slightly denser clumps form in \emph{N05}.  As the simulations evolve, the dense gas structures begin to vary, as can be seen by $t = 1.49$~\tff \ (final column).   }
\label{fig:columndensity:evol5}
\end{figure*} 

In the hydrodynamic and weak field ($\mu_0=10$ and $20$) models, Fig.~\ref{fig:columndensity:local} shows a single dense region in each panel that does not strongly connect to any other large scale structure through dense filaments. Although the clump in the hydro model (right panel) appears filamentary in the figure, it is in fact pancake-shaped.  The structure in these models forms from the collapse of a short filament, and rotates around the stars as the entire substructure moves throughout the cloud.  Each of these dense regions hosts or has hosted several stars, and there are several filaments emanating from these dense regions.  

The small scale structure of the stronger field models ($\mu_0 = 5$ and $3$, left two columns) are more filamentary, with several small clumps embedded within the filament; each individual clump is smaller than in the weakly magnetised models. There are fewer filaments in these models, and the ones that exist tend to be denser and narrower than in their weakly magnetised counterparts.  

On this scale, there appears to be a trend from filamentary to broken filaments to clumpy as the initial magnetic field strength decreases.   As suggested by the large-scale structure, the magnetic field is approximately perpendicular to the dense filaments.  

Discs are visible around many of the stars, and these will be discussed in Section~\ref{sec:ppd} below. 

\begin{figure}
\centering
\includegraphics[width=1.0\columnwidth]{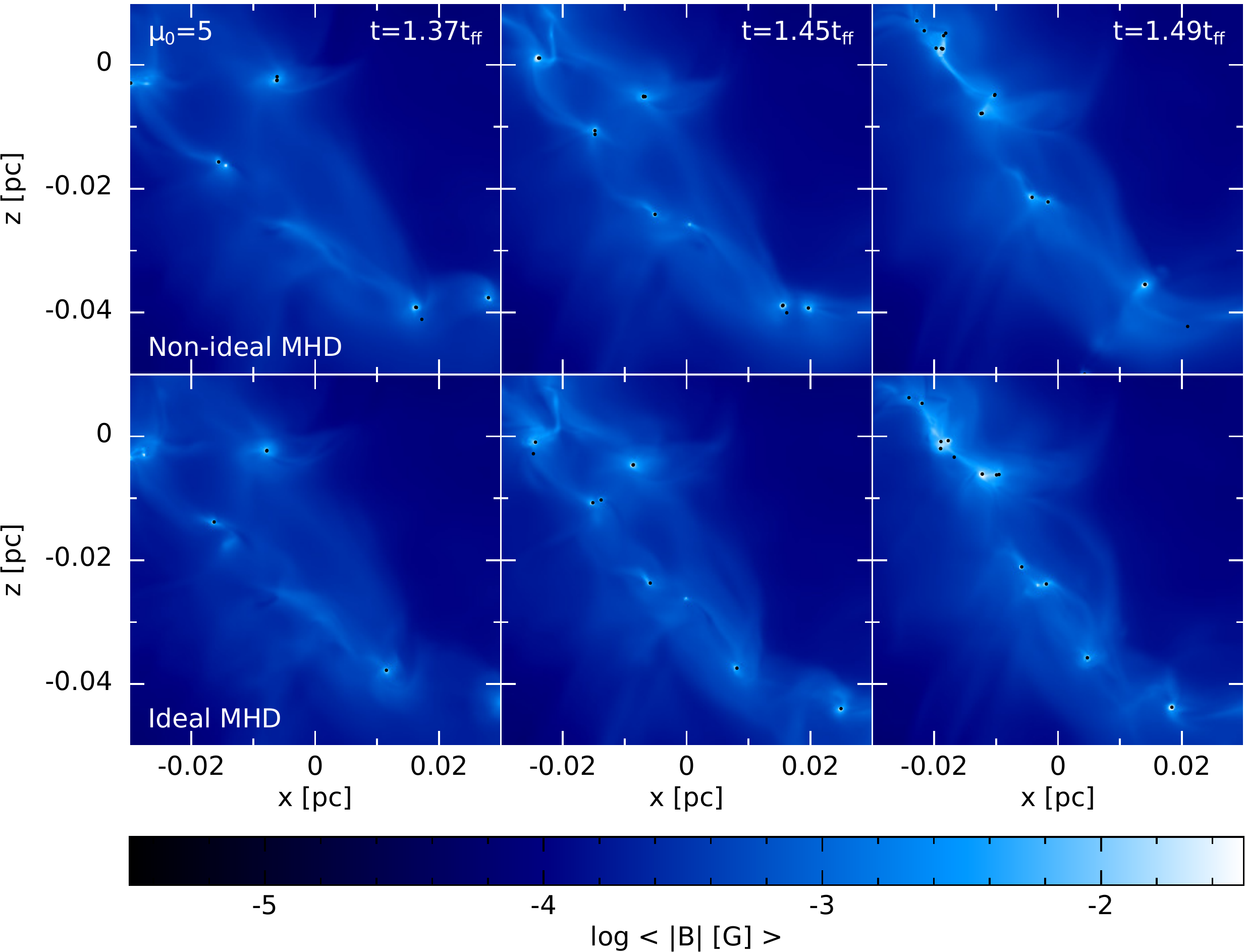}  
\caption{Evolution of the magnetic field strength of a small region of the cluster for the models with \mueq{5}, comparing non-ideal MHD (top) to ideal MHD (bottom).  These images correspond to the final three columns in \fig{fig:columndensity:evol5}.  In general, the field is stronger near the stars than in the background.  The clumps in \emph{I05} are denser and hence more strongly magnetised at $t = 1.49$~\tff \ (final column) than in \emph{N05}.}
\label{fig:B:evol5}
\end{figure} 

\begin{figure*}
\centering
\includegraphics[width=1.0\textwidth]{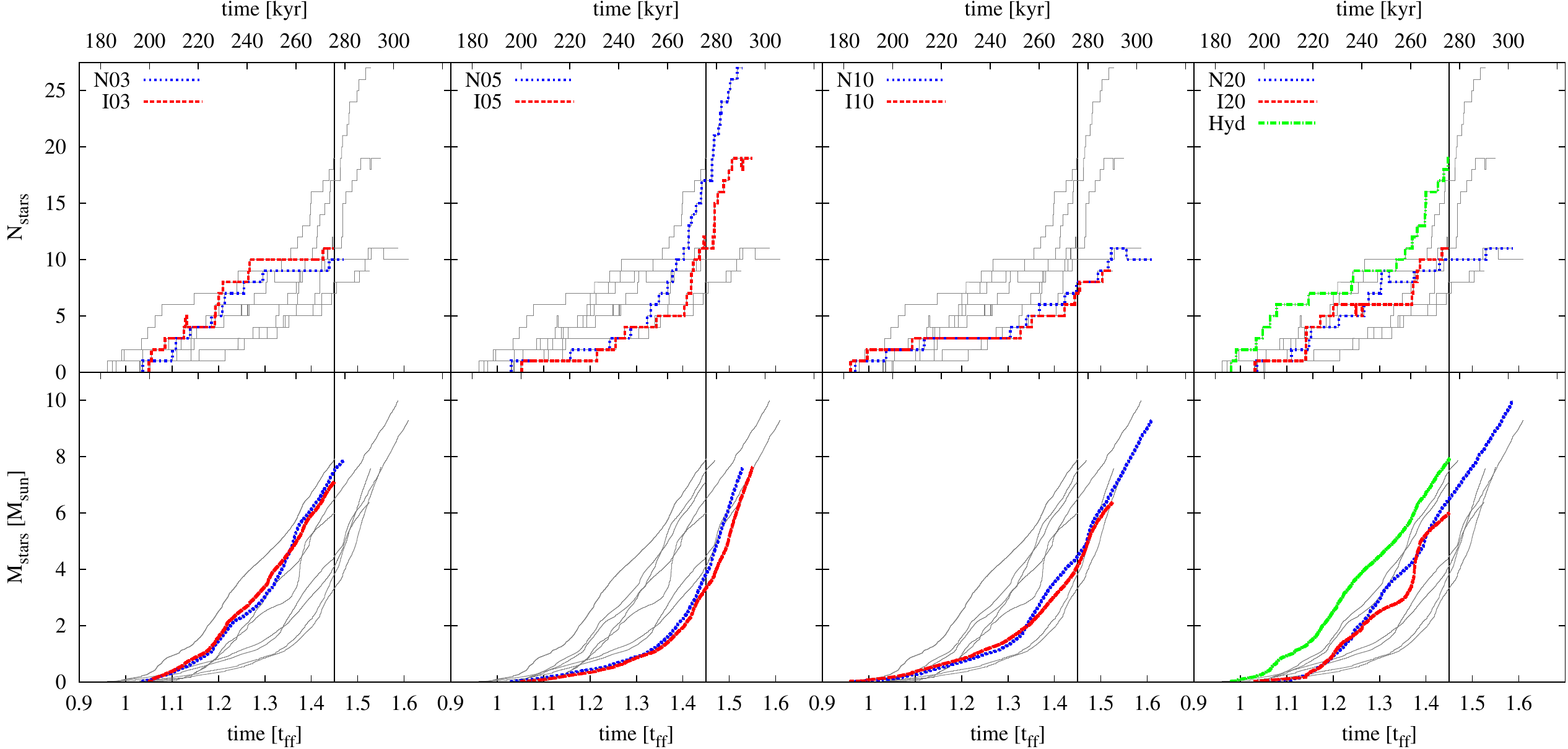}  
\caption{The total number of stars (top) and the total mass in stars (bottom).   For a better comparison, all nine models are plotted in each panel (thin grey lines), and two or three models are highlighted.  The vertical line is at \ttnow, which is the final time of our slowest models.  The hydro model typically forms stars earlier in the simulation, and more mass is accreted onto the stars than in the magnetised models.  At each magnetic field strength, both the ideal and non-ideal MHD models form stars at similar rates, and similar masses are in the stars. }
\label{fig:stars}
\end{figure*} 

\subsubsection{Non-ideal vs ideal MHD}
\label{sec:sss:ini}
The differences caused by the inclusion of non-ideal effects are primarily found on small scales.  As an example, Figs.~\ref{fig:columndensity:evol5} and \ref{fig:B:evol5} show the evolution of a small region of the gas column density and the density-weighted line-of-sight averaged magnetic field strength (i.e. $ \left<|B|\right> =  \int |B| \rho \text{d}y' / \int \rho \text{d}y'$), comparing \emph{N05} (top rows) to \emph{I05} (bottom rows).   In this region, the gas structure is indistinguishable between the two models at $t = 1.16$~\tff.   At 1.41~\tff, the primary filament is slightly more clumpy in \emph{I05}, and by 1.49~\tff, the structure of the primary filament has noticeable differences, as indicated by where the stars form in the two models.  This again confirms that non-ideal MHD affects the density on small scales, and that there is a greater divergence between an ideal/non-ideal pair as the simulations continue to evolve.   Throughout the evolution, there is slightly more dense gas in the non-ideal models compared to their ideal MHD counterparts (c.f. \fig{fig:gasmass} and associated discussion), although this is not necessarily clear from the image.

\fig{fig:B:evol5} shows that the magnetic field is stronger in the dense regions around the stars (i.e. the discs) compared to the background. Except at 1.49~\tff, there is no noticeable difference in strength between the ideal and non-ideal MHD models; however, this stronger magnetic field at 1.49~\tff{} corresponds to a region of slightly higher density in \emph{I05} (final column of \fig{fig:columndensity:evol5}).  This is consistent with \fig{fig:RhoVB} which suggest that all models have similar mean magnetic field strengths at a given gas density and an order of magnitude scatter.

\subsection{Stellar populations}
\label{sec:sp}

Sink particles with a radius of 0.5~au are inserted when \rhoxeq{-5} is reached.  Given our sink radius and insertion criteria, each sink particle represents one star or brown dwarf, thus we use `star' and `sink particle' interchangeably. 

\subsubsection{Star formation rate}
\fig{fig:stars} shows the total number of stars and the total mass in stars as a function of time.  Table~\ref{table:stars} lists information about the stellar and disc populations at \ttnow \ and at the end of the respective simulations, \tend.

\begin{table*}
\begin{center}
\begin{tabular}{c c c c c c c c c c c c c c}
\hline
Name              & $M^\text{c}_\text{stars}$ & $N^\text{c}_\text{stars}$  & $N^\text{c}_\text{mergers}$  & $N^\text{c}_\text{systems}$  & $N^\text{c}_\text{discs}$  & 
$t_\text{final}$ & $M^\text{f}_\text{stars}$ & $N^\text{f}_\text{stars}$    & $N^\text{f}_\text{mergers}$  &  $N^\text{f}_\text{systems}$  & $N^\text{f}_\text{discs}$  \\
 &  [\Msun] & & &  1, 2, 3, 4 & 1, 2, 3, 4 & [\tff]  & [\Msun] & & &  1, 2, 3, 4 & 1, 2, 3, 4 &\\
\hline 
\emph{N03}    & 7.51 & 10 &  0 & 3, 0, 1, 1 &  2 (5), 0 (3), 1, 1   & 1.46 & 7.72 & 10 & 0 & 3, 0, 1, 1 &   2 (7), 0 (3), 1, 1   \\ 
\emph{N05}    & 3.83 & 17 &  0 & 4, 1, 1, 2 &  4 (13), 1 (4), 1, 2 & 1.53 & 7.57 & 27 & 0 & 8, 5, 3, 0 &   3 (17), 5 (8), 3, 0 \\ 
\emph{N10}    & 4.48 &   8 &  0 & 2, 1, 0, 1 &  2 (7), 1 (2), 0, 1   & 1.61 & 9.29 & 10 & 1 & 3, 0, 1, 1 &   2 (9), 0 (2), 1, 1   \\ 
\emph{N20}    & 6.51 & 10 & 1 & 2, 1, 2, 0 &   0 (8), 1 (3), 2, 0   & 1.59 & 9.99 & 11 & 1 & 2, 3, 1, 0 &   0 (7), 3 (4), 1, 0   \\ 
\emph{Hyd}    & 7.93 & 19 & 0 & 6, 0, 3, 1 &  4 (14), 0 (4), 3, 1                                                                                           \\ 
\emph{I03}     & 7.10 &  11 & 1 & 2, 1, 1, 1&    1 (3), 0 (3), 1, 1                                                                                            \\ 
\emph{I05}     & 3.34 & 11 & 1 & 5, 3, 0, 0  &4 (10), 3 (3), 0, 0  & 1.55 & 7.64 & 19 & 2 & 10, 1, 1, 1 & 4 (11), 1 (3), 1, 1  \\ 
\emph{I10}     & 4.14 &   7 & 0 & 3, 0, 0, 1  & 3 (6), 0 (1), 0, 1  & 1.52 & 6.37 &   9  & 0 &   3, 1, 0, 1 &  3 (5), 1 (2), 0, 1   \\ 
\emph{I20}     & 6.00 & 11 & 2 & 4, 0, 1, 1  & 1 (6), 0 (2), 1, 1                                                                                             \\ 
\hline
\end{tabular}
\caption{Summary of the stellar population properties at the common time of \ttnow \ (denoted by superscript c), and at the final time of each simulation (the final times are listed in the $t_\text{final}$ column, and denoted by superscript f). Entries are not duplicated for the models with $t_\text{final} =$ \tnow.  $M_\text{stars}$ is the total mass in stars, $N_\text{stars}$ is the total number of stars, $N_\text{mergers}$ is the number of mergers prior to the common or final time, $N_\text{systems}$ is the number of stellar systems with 1, 2, 3 or 4 stars, respectively, and $N_\text{discs}$ is the number of discs around single, binary, triple, and quadruple stars; the number in brackets includes the total number of discs of each classification that are also part of higher order stellar systems.  }
\label{table:stars} 
\end{center}
\end{table*}

By \ttnow, there are 8--19 stars, depending on the simulation.  The hydro model has the highest number of stars, which is to be expected since it lacks magnetic support against gravitational collapse.  However, for the magnetised models, the number of stars does not appear to correlate with the initial magnetic field strength or whether or not we include non-ideal MHD.  The magnetised model with the highest number of stars is \emph{N05}, which has a relatively strong initial magnetic field. Caution is required when comparing stellar populations at a given time, since, for example, \emph{N05} and \emph{I05} are undergoing an epoch of rapid star formation over the $\approx 0.1$~\tff\ before they are stopped.

The integrated star formation rate (i.e., the total mass in stars as a function of time; bottom row of Fig.~\ref{fig:stars}) depends on the initial magnetic field strength in a (nearly) systematic way.  The hydro model always has the most mass in stars, followed by the models with $\mu_0 = 20$, $10$ and $5$, respectively.  Thus, amongst these models, we find that stronger magnetic fields slow the star formation rate, which is in agreement with \citet{PriceBate2008,PriceBate2009} and the analysis of the dense gas presented above.  We further find that the star formation rate does not depend on whether ideal or non-ideal MHD is assumed.

The models with the strongest initial magnetic field strengths (\emph{N03} and \emph{I03}) do not follow the above trend.  At any given time $t \gtrsim 1.1$~\tff, the integrated star formation rate is greater in these models than the other magnetised models and only slightly lower than \emph{Hyd}.  The strong magnetic field qualitatively changes the collapse, with a quicker collapse along the magnetic field lines giving a larger quantity of dense gas (recall \fig{fig:gasmass}).  Despite the strong magnetic fields, this larger quantity of dense gas fuels accretion onto the stars.  This deviation from the systematic trend is in contradiction with \citet{PriceBate2008,PriceBate2009}, however, it may depend on the initial velocity field.

Many stars remain near the clump in which they were born, however, several stars are ejected into the low-density medium due to interactions with other stellar systems; these stars are typically ejected with low ($v \lesssim 5$~\kms) velocity, and we observe only four high-velocity ($v \gtrsim 5$~\kms)  stars by \tend.  Not all stars accrete at similar rates, and by \tend, several stars in each model have stopped accreting gas.  Most of these non-accreting stars have been ejected into the low-density medium and are amongst the lowest mass stars in our suite.  Despite the inclusion of magnetic fields and our low number of stars, we observe ejected and non-accreting stars, as seen previously in the literature \citepeg{BateBonnellBromm2003,BateBonnell2005}.

\subsubsection{Stellar mass distributions}

\begin{figure*}
\centering
\includegraphics[width=0.9\textwidth]{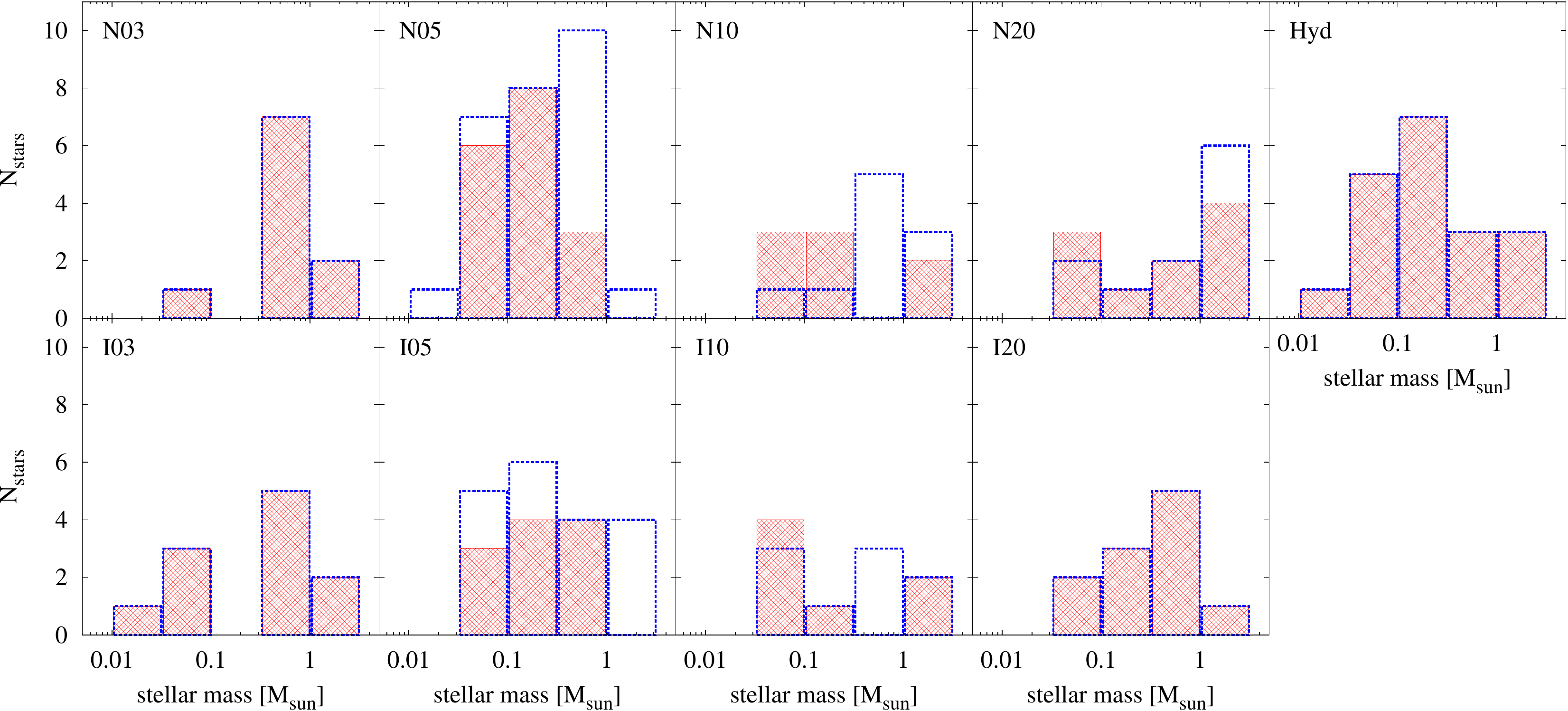}  
\caption{The stellar mass distribution for each model.  The red hatched bins represent the mass distribution at the common time of \ttnow, and the bins outlined in blue represent the mass distribution at the end of the simulation, \ttend.  Each bin has a width of 0.5 dex.  There are no obvious trends amongst the models, and identifying trends is unreliable given the low number of stars. }
\label{fig:massdistribution}
\end{figure*}

\fig{fig:massdistribution} shows the number of stars in each mass bin at both \ttnow{} (red hatched) and at \tend{} (blue outline).  Given the small number of stars and that the majority of them are still accreting, it is not practical to calculate an IMF, or explicitly comment on stellar population.
Considerably fewer stars are produced in our simulations compared to the magnetised barotropic calculations of \citet{PriceBate2008}, which produced 17--23 and were evolved to  1.27--1.53~\tff.  This is because introducing radiative feedback reduces small-scale fragmentation \citep{Bate2009rfb}.  As expected, the numbers are more similar to those produced by the magnetised radiation hydrodynamical calculations of \citet{PriceBate2009}, which produced 3--10 objects and were evolved to 1.36--1.54~\tff.  These earlier magnetised star formation calculations used Euler potentials to model ideal MHD, and the earlier radiative transfer scheme of \cite{WhitehouseBate2006} which did not include the diffuse interstellar medium treatment and separate gas and dust temperatures.  Despite these differences, the results are qualitatively similar.  

\subsubsection{Multiplicity}
\label{sec:stars:binary}
The majority 
of the stars are born as single stars and then become part of multiple systems when they become gravitationally bound during a close interaction (similar to, e.g., \citealp{Bate2012, Bate2018}; \citealt{Seifried+2013}).  Many of these stars are born $\sim\mathcal{O}\left(10^2 - 10^3\right)$~au and $\sim\mathcal{O}\left(0.1 - 10\right)$~kyr apart, implying that most of our binary systems are not primordial.   

We locate systems hierarchically up to systems containing four mutually bound stars; this is performed simultaneously with determining disc properties, as described in \secref{sec:ppd}.  First, for each star+circumstellar disc, we determine if it is bound to its closest star+circumstellar disc; if so, then this is a binary pair.  Once all the binaries are found,  we determine if each binary is bound to its closest star or binary system.  For the systems with three stars, we search to determine if the nearest star is single and also bound, for a system of four stars.  In our clusters, higher order systems typically include one or two binaries.  The number of stellar system of each population is listed in the $N_\text{systems}$ column in \tabref{table:stars}.  Although there are more systems with only one star than higher order systems, these single-star systems represent $\lesssim 45$ per cent of the total number of stars in any given model.

Excluding mergers and one lone exception, once a binary is formed, it persists for the duration of the simulation.  The higher order systems, however, are created and destroyed with time, which is particularly noticeable in \emph{N05} and \emph{I05}, which both form many new stars after \tnow.  

This high number of multiplicity is consistent with previous hydrodynamic studies \citepeg{BateBonnellBromm2003,Bate2009rfb,Bate2012,Bate2019}, suggesting that magnetic fields do not significantly hinder or promote the formation of binaries or higher order systems.  

\subsubsection{Mergers}
\label{sec:stars:merger}
When two stars come within 27~\Rsun, we considered them to have merged (we increased this from the default of 6~\Rsun \ for computational efficiency since following inspiralling stars requires a small computational timestep).  During a merger, we replace the two sink particles with a single sink with the centre of mass properties of its progenitors. The $N_\text{mergers}$ columns of Table~\ref{table:stars} lists the number of mergers prior to \tnow \ and \tend.

There are two classes of mergers in our simulations: stars that `collide' during fly-by and merge, and those that form a binary system with a decaying major axis and ultimately merge.  In both \emph{I03} and \emph{I20}, the stars form a binary system before merging, whereas the remainder of the mergers occur during a fly-by (where a fly-by is indistinguishable from a binary with a large major axis and high eccentricity).  

In \emph{I03}, the stars were born with a separation of \sm260~au, and quickly became gravitationally bound.  Their orbit quickly decayed, and within 14~kyr of the birth of the younger star, the periastron separation decreased to \sm27~\Rsun \ and the stars merged.  At the time of merger, their apastron was \sm110~\Rsun.

In \emph{I20}, the binary system initially has a stable average separation of \sm5~au before a close encounter causes the orbit become more eccentric, but maintaining a similar apastron.  However, a second encounter causes the orbit to decay while becoming more eccentric until the stars ultimately merge at periastron.  This is a merger of two stars of similar masses.  Then, \sm2.5~kyr later, a low mass star `collides' with this star in a fly-by and merges.  Thus, between the two mergers, the primary star rapidly increases its mass, and at \tnow \ is the second most massive star in our suite (the most massive star is the star that underwent a merger in \emph{I03}).  

In reality, due to the large merger radius, many of these systems may have formed spectroscopic binaries rather than merger products.

\subsection{Protostellar disc properties}
\label{sec:ppd}

We track three separate classifications of discs: circumstellar discs (i.e. a disc around a single star), circumbinary discs (i.e. a common disc around a binary star system), and circumsystem discs (i.e. discs around a system of either three or four stars).  Thus, it is possible for several discs to be associated with each star.  The number of each of these classifications of discs is listed in the $N_\text{discs}$ column of \tabref{table:stars}, where the total number of circumstellar and circumbinary discs is listed in parentheses and the number not in parentheses lists the number of discs associated with the highest order systems from the $N_\text{systems}$ column.  

To calculate the mass of the discs, we follow the prescription of \citet{Bate2018}.  For each star, we first sort all the particles by distance.  We check the closest particle to determine if it is bound to the sink, has an eccentricity of $e < 0.3$ and has density\footnote{Unlike \citet{Bate2018}, we include the density threshold to prevent our discs from including any low-density filamentary material.} \rhoge{-14}; if so, it is added to the sink+circumstellar disc system.  Each consecutive particle is checked and added to the system if the criteria are met.  We do this for all the particles within 2000~au of the star or until another star is encountered.  To determine higher order discs, we determine if stellar systems (star+circumstellar discs) are mutually bound, as described in \secref{sec:stars:binary}.  If so, then the above process is repeated, but using the bound pair rather than star+circumstellar disc.  We record the circumbinary disc independently from the circumstellar discs for later analysis.  This process is repeated with each new system up to a maximum of four stars per system.  The mass of each disc is the total mass of the gas that has been added to the disc.  The radius of each disc, $R_\text{disc}$, is the radius that includes 63.2 per cent of the disc mass.  

\subsubsection{Formation history}
Stars tend to form in isolation (recall \secref{sec:stars:binary}), thus the initial discs are small, circumstellar discs, assuming a primordial disc forms at all.  However, the bulk flow of the gas promotes interactions between the stellar systems.  As such, the systems are frequently promoted to or demoted from higher order systems due to these interactions.  As stars approach and become bound, their discs usually merge forming a larger circumsystem disc while often retaining small circumstellar discs.  Although disc growth is a common consequence, interactions with unbound stars (i.e. fly-by interactions) or bound stars on large orbits can cause a reduction in the disc size through tidal stripping, or totally disrupt the disc.  Thus, the masses and radii of our discs are rapidly changing, as previously seen in the hydrodynamical simulations by \citet{Bate2018}.  We thus focus our analysis on the discs instantaneously existing at \tnow, similar to observing discs in a particular star forming region at a particular time.

\begin{figure*}
\centering
\includegraphics[width=0.45\textwidth]{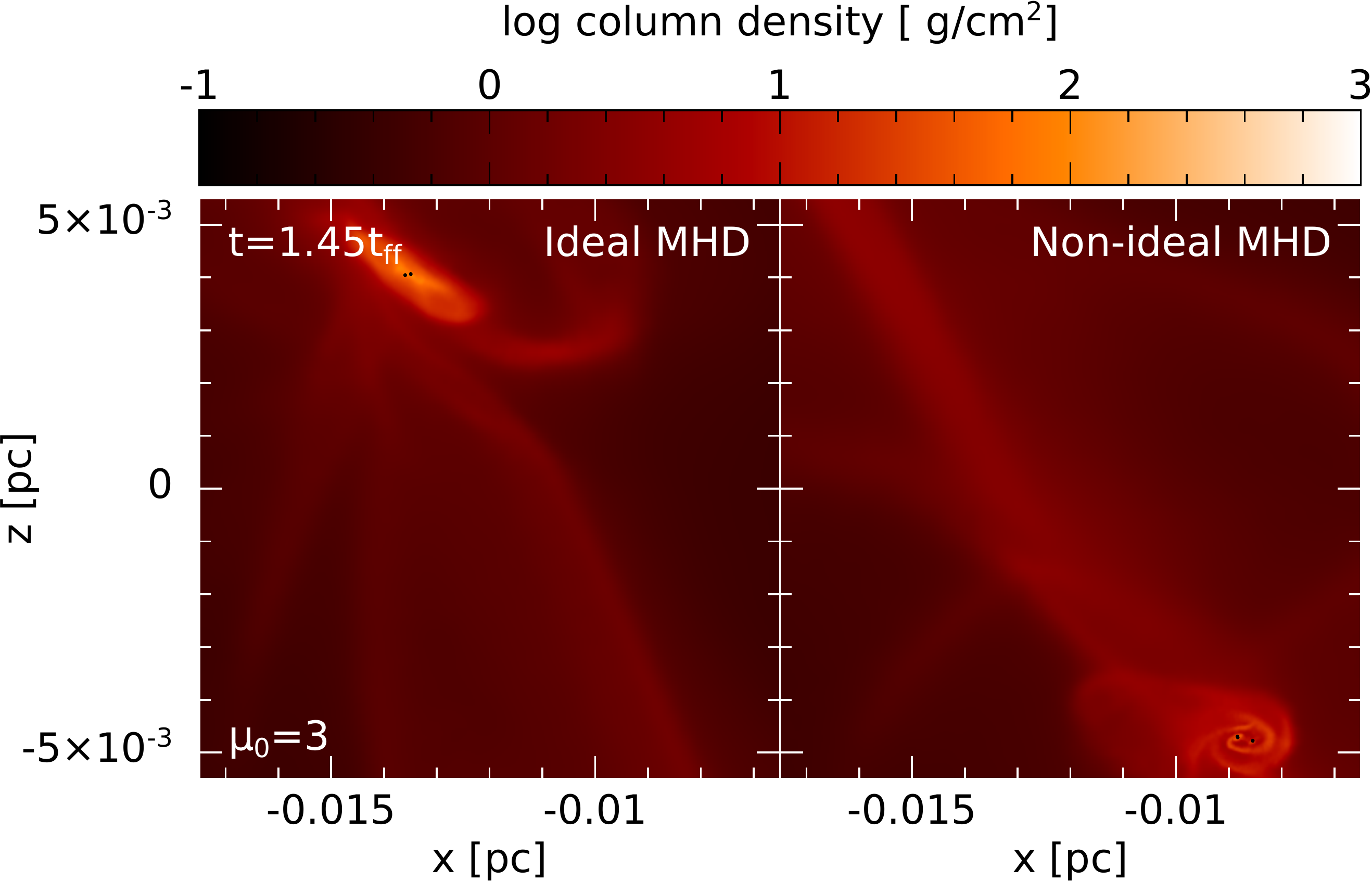}  
\includegraphics[width=0.45\textwidth]{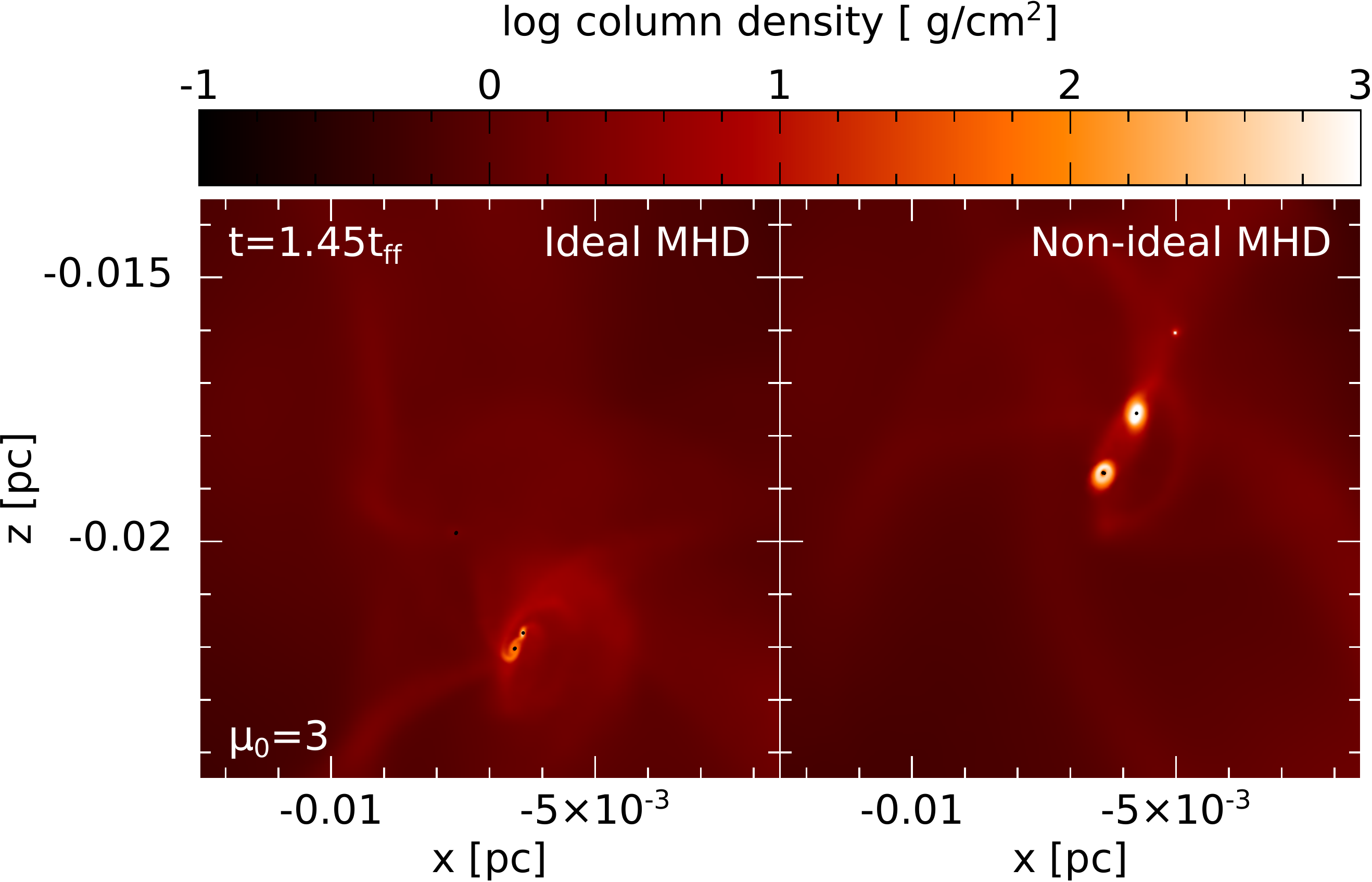}  
\includegraphics[width=0.45\textwidth]{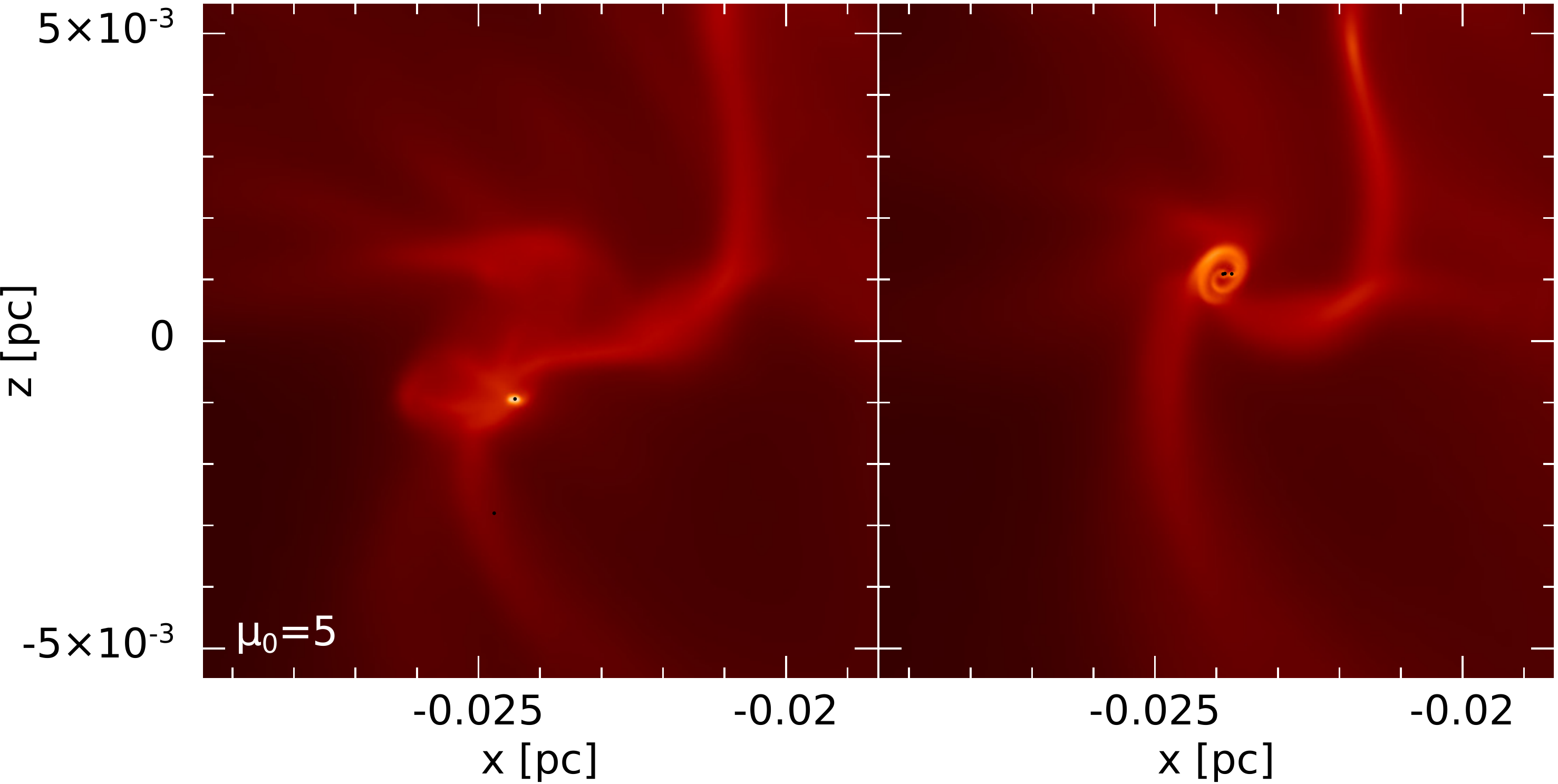}  
\includegraphics[width=0.45\textwidth]{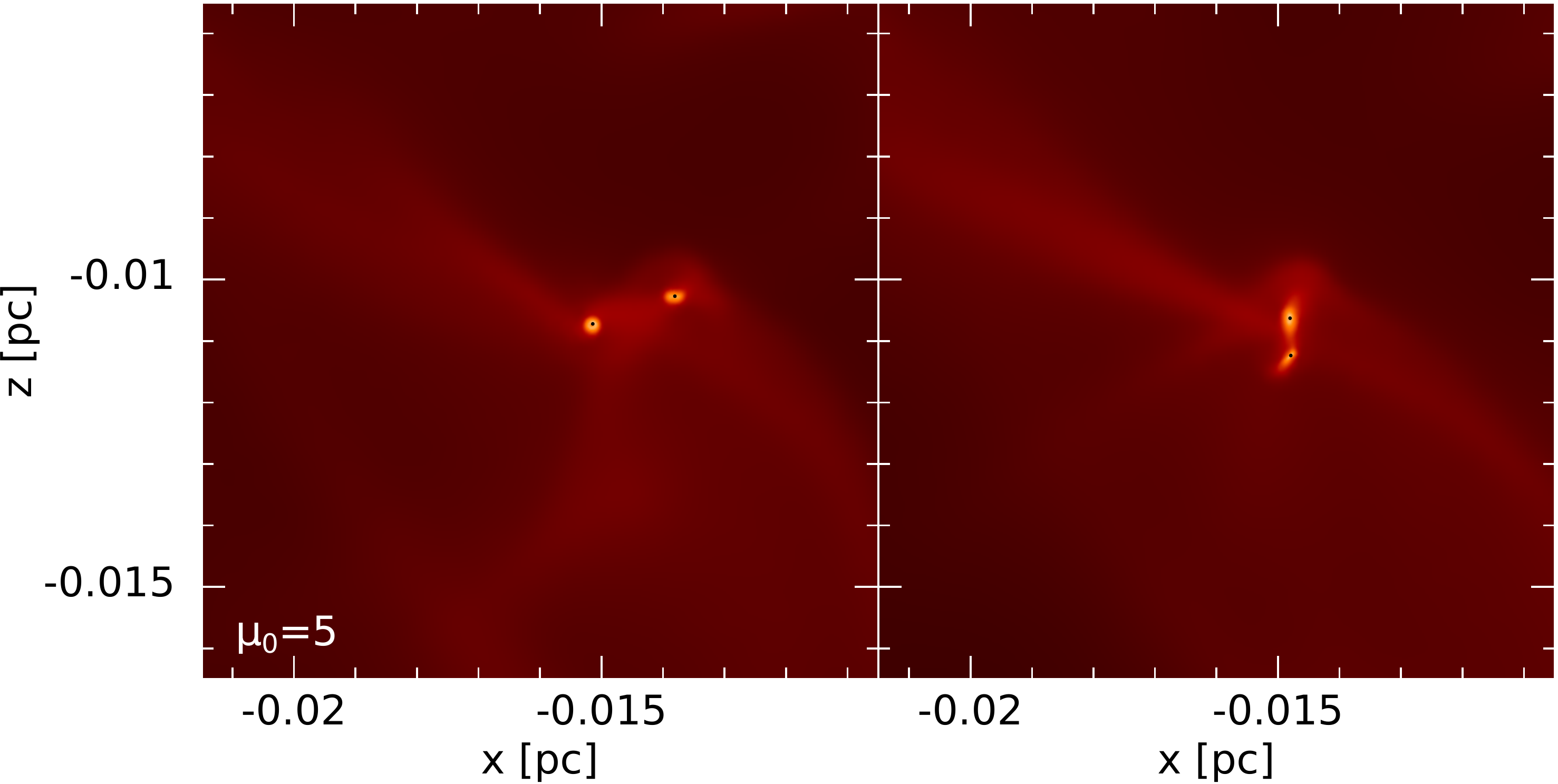}  
\includegraphics[width=0.45\textwidth]{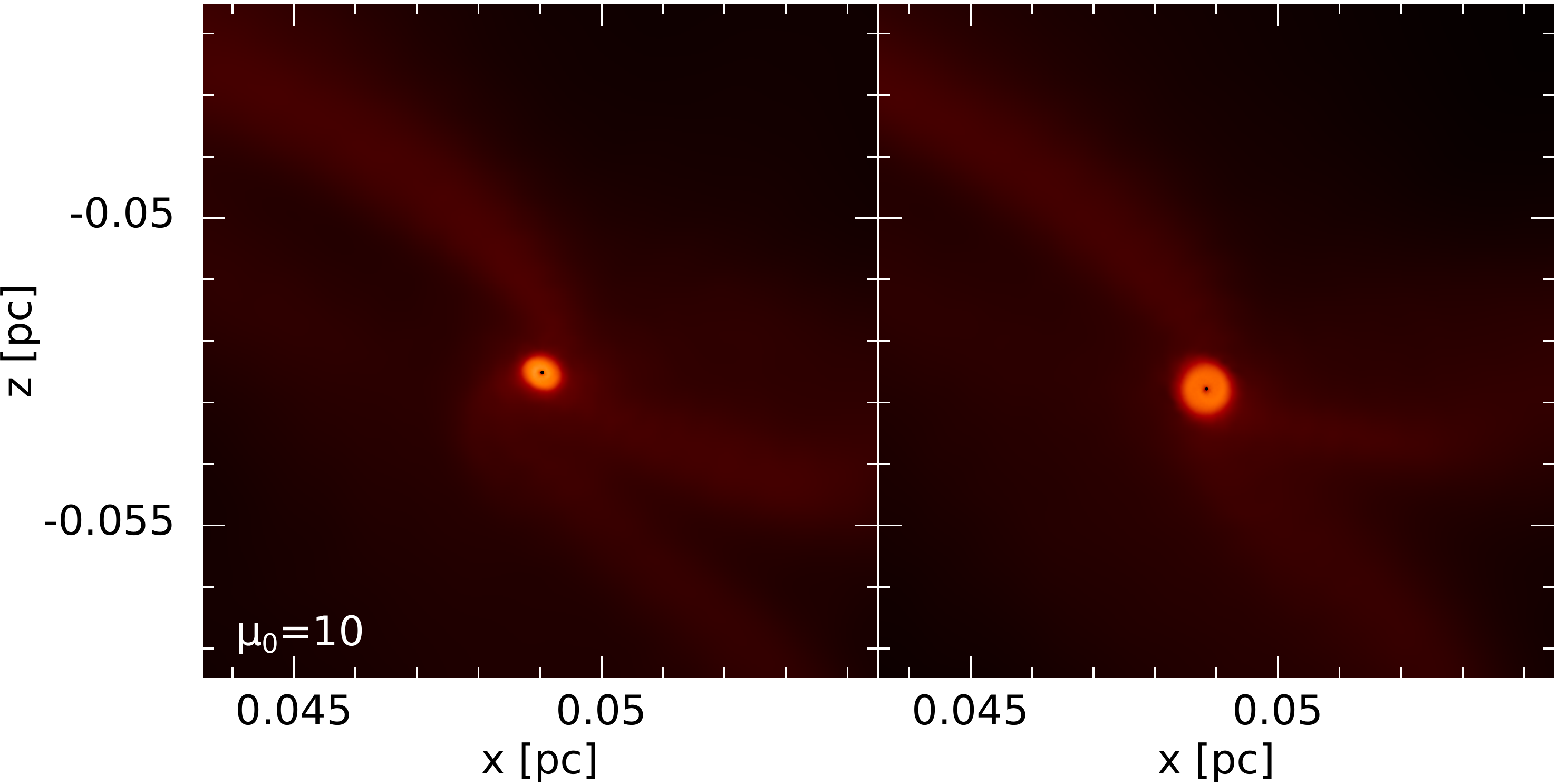}  
\includegraphics[width=0.45\textwidth]{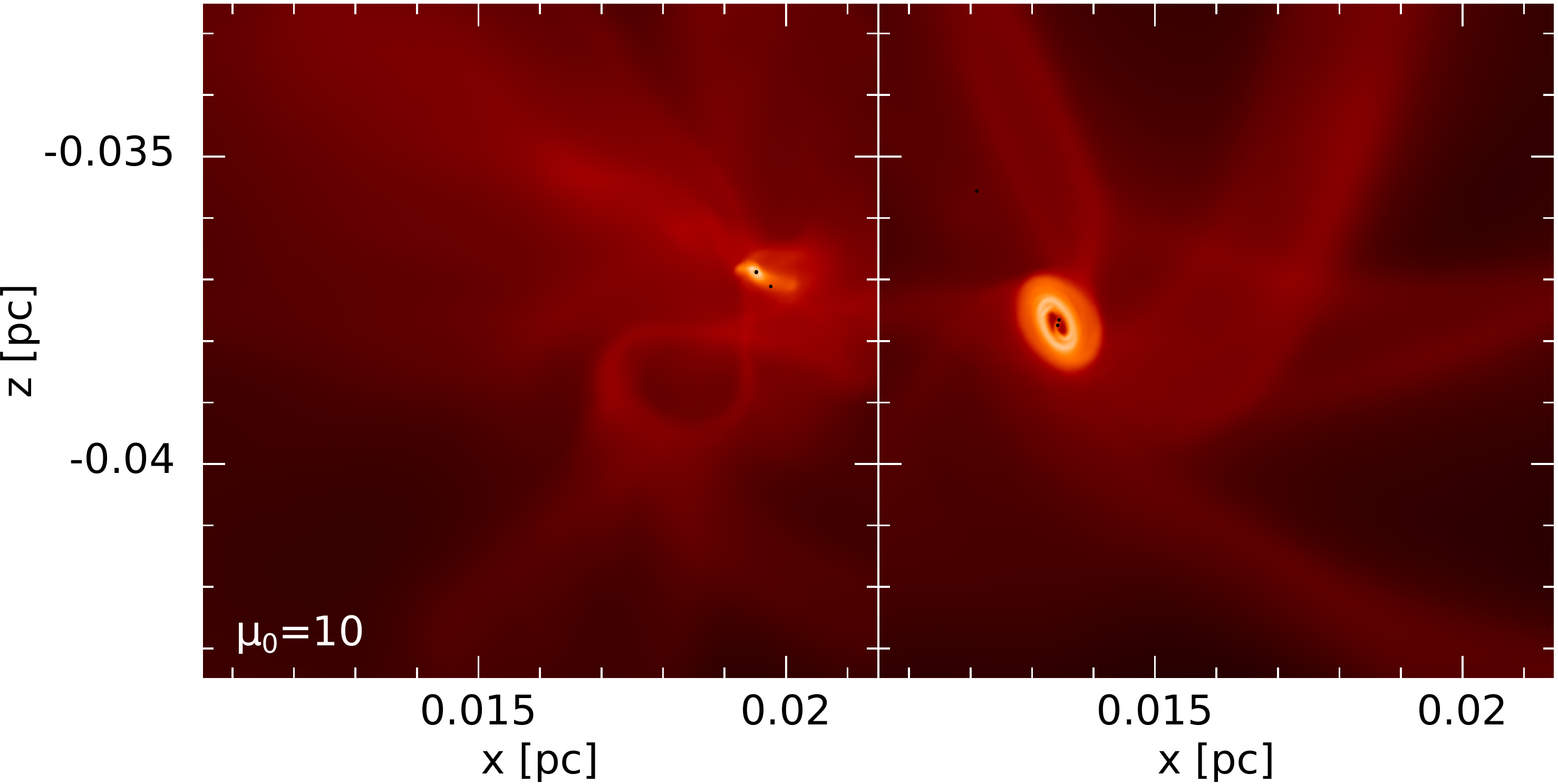}  
\includegraphics[width=0.45\textwidth]{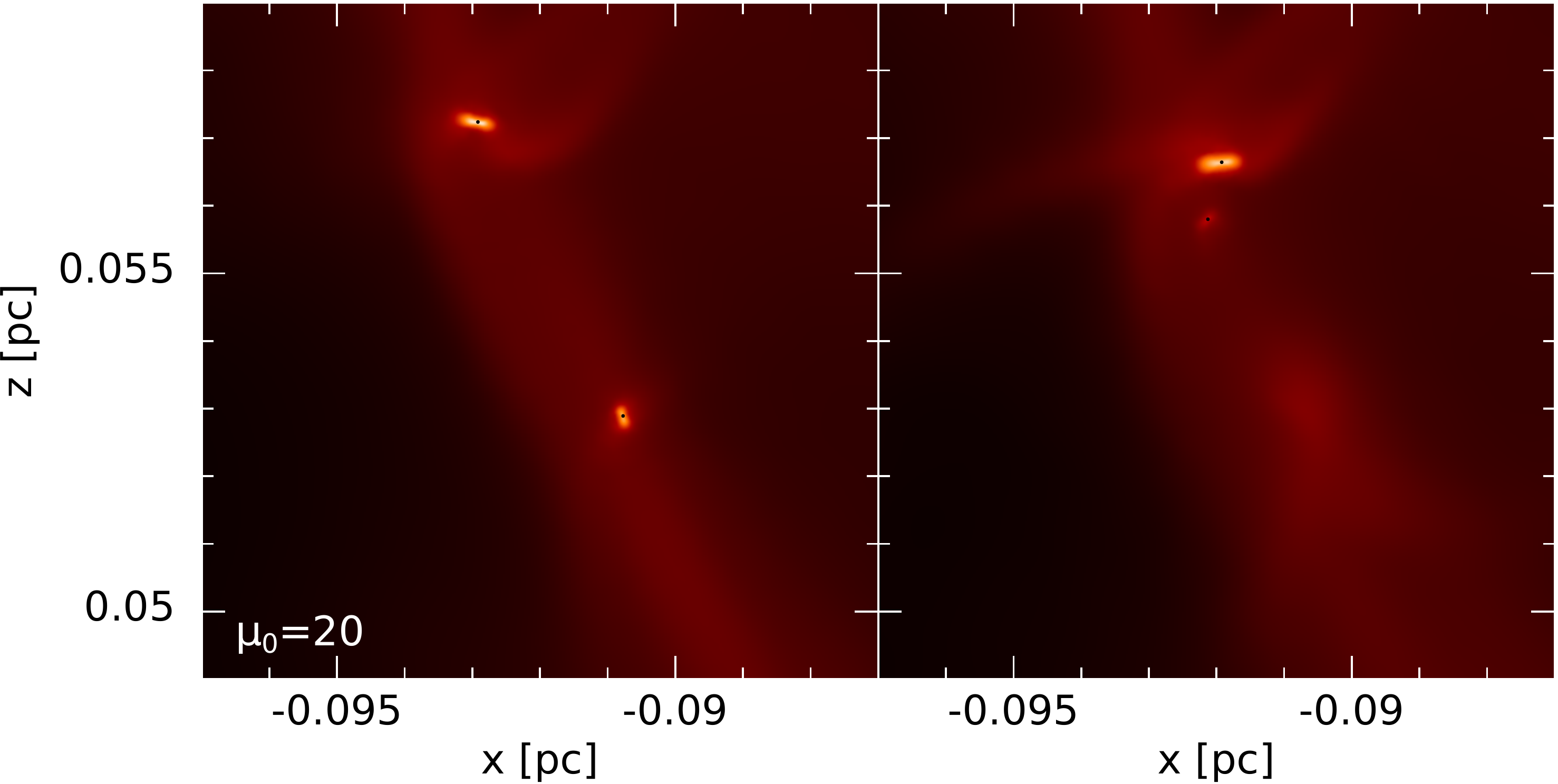}  
\includegraphics[width=0.45\textwidth]{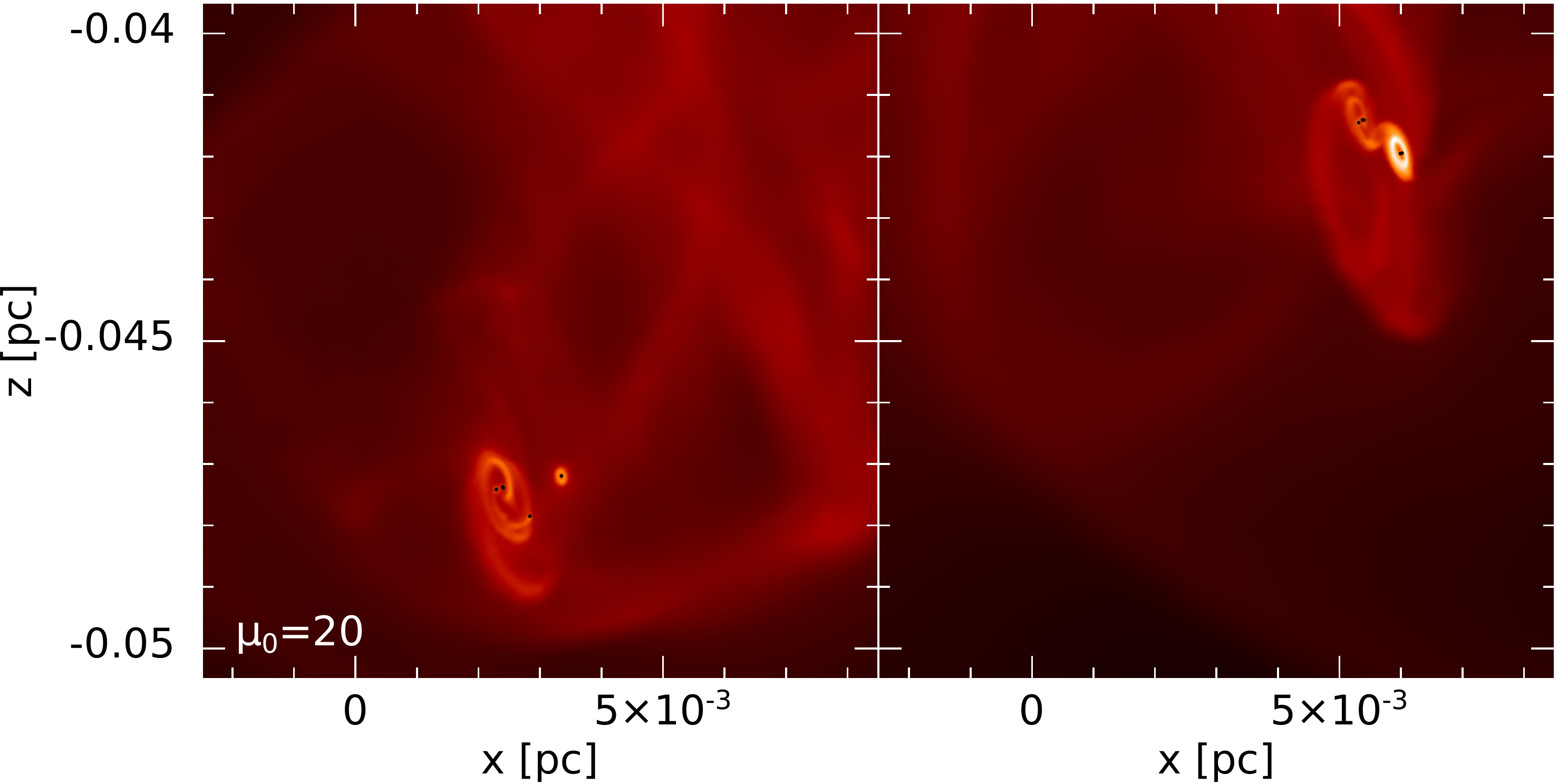}  
\includegraphics[width=0.255\textwidth]{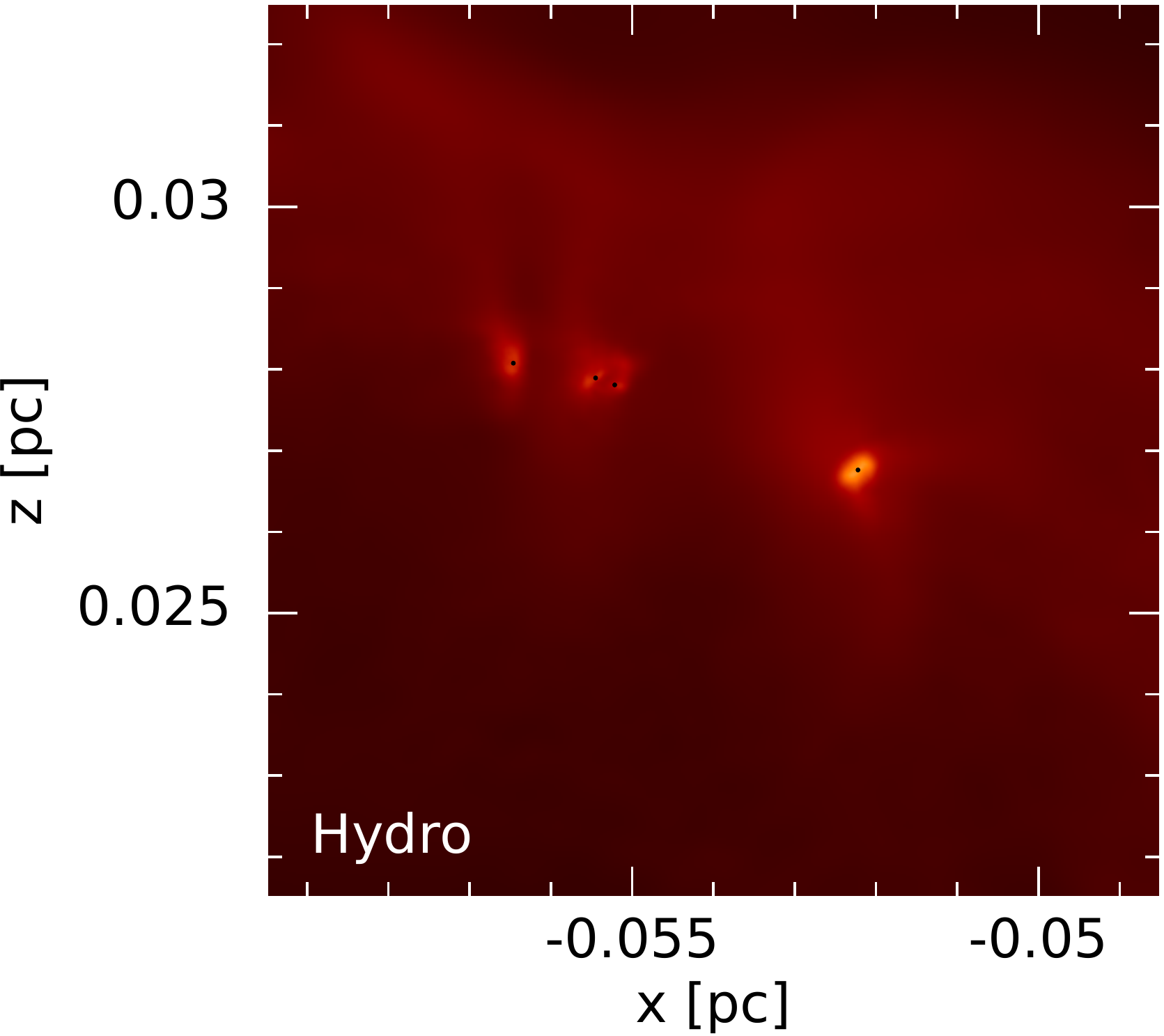}  
\includegraphics[width=0.255\textwidth]{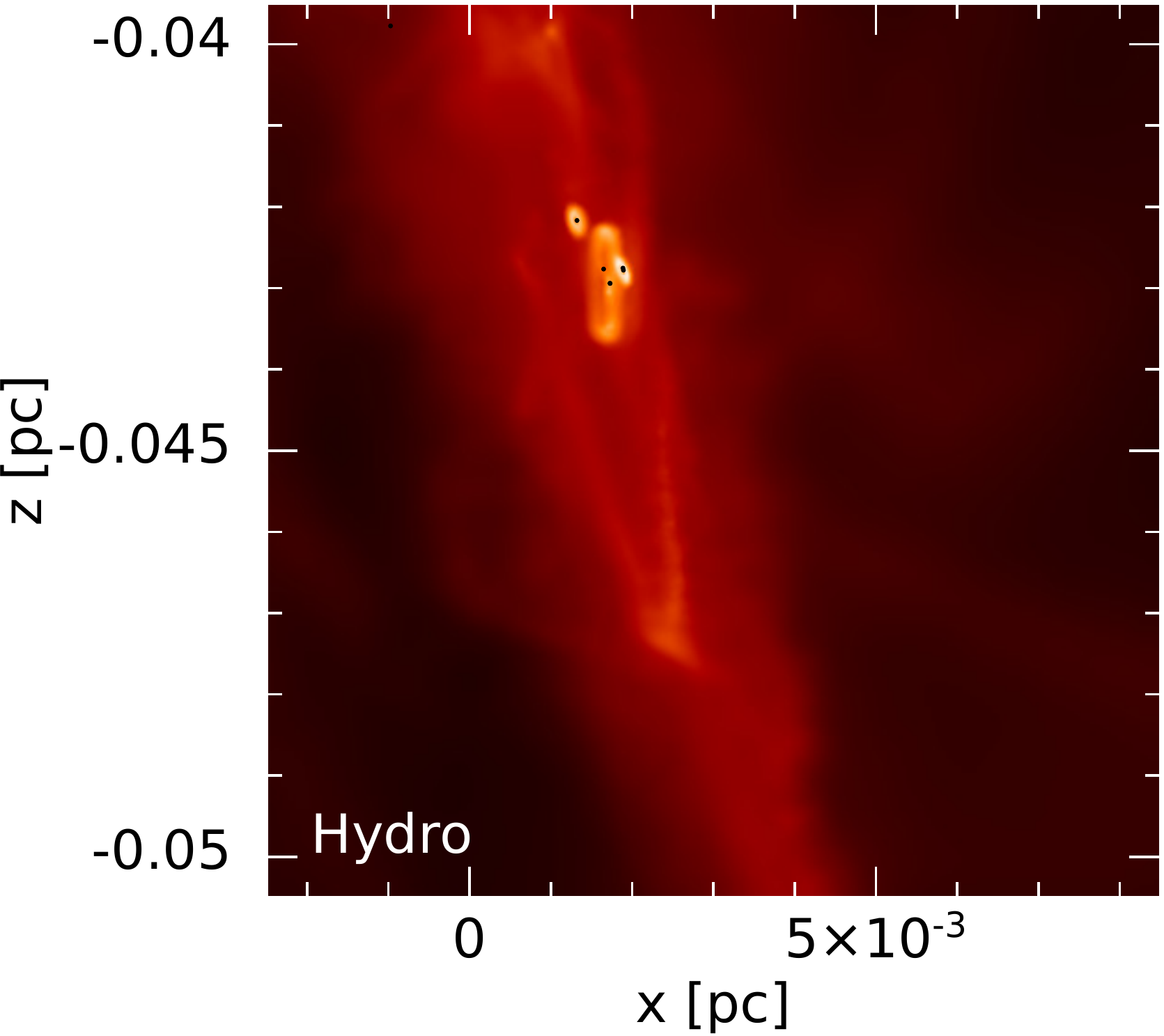}  
\caption{Gas column density of two disc-containing regions for each model.  Each panel has the same spatial dimensions and density range; the separation between ticks is 207~au.  Each ideal/non-ideal pair is of the same spatial region for a direct comparison of the effect of the non-ideal processes.  The right-most panels of the \mueq{20} models and the hydro model are of the same spatial region.  Each black dot represent the location of a sink particle, and in several of the panels, the stars have formed a close multiple system.  This is a representative snapshot, since discs are continually growing and being tidally disrupted due to interactions with other stars and/or discs.}
\label{fig:columndensity:discs}
\end{figure*}

\fig{fig:columndensity:discs} shows the column density of two disc-containing regions of each model; each panel has the same spatial scale, and we show the same region for each each ideal/non-ideal pair (left and right panels in each column, respectively).  From a visual comparison, we observe the strong influence of the non-ideal MHD processes on the small-scale evolution.  

Many of the panels in \fig{fig:columndensity:discs} show isolated circumstellar/circumsystem discs (e.g. the left-hand panels of the \mueq{10} and $20$ models and \emph{Hyd}) or bound systems (e.g. the mutually bound discs in the right-hand panels of the \mueq{3} and $5$ models).  At this time, these discs are relatively smooth, and a reasonable analysis of the disc properties can be performed.    

The discs in the left-hand panels of \emph{N03} and \emph{N05} are undergoing a violent change as a result of their recent interactions.  More dramatically, disc interaction is occurring at this time in the right-hand panels of the \mueq{20} models and \emph{Hyd}.  A smooth disc in \emph{I20} became disrupted at $t \approx 1.39$~\tff{} yielding the disrupted system in the figure, while gas has been stripped from one disc and accreted onto the other in \emph{N20}.  Thus, as discussed by \cite{Bate2018}, the formation history of discs in a cluster is violent.

\begin{figure*}
\includegraphics[width=\textwidth]{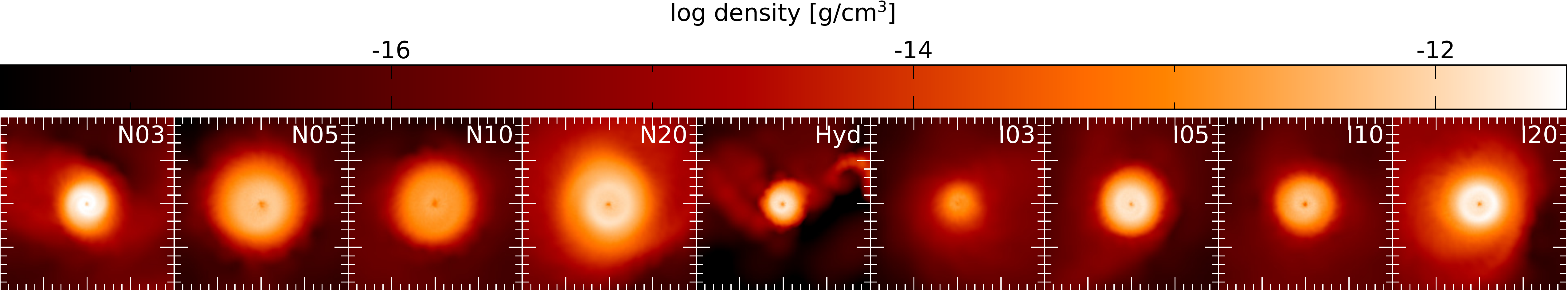}  
\includegraphics[width=\textwidth]{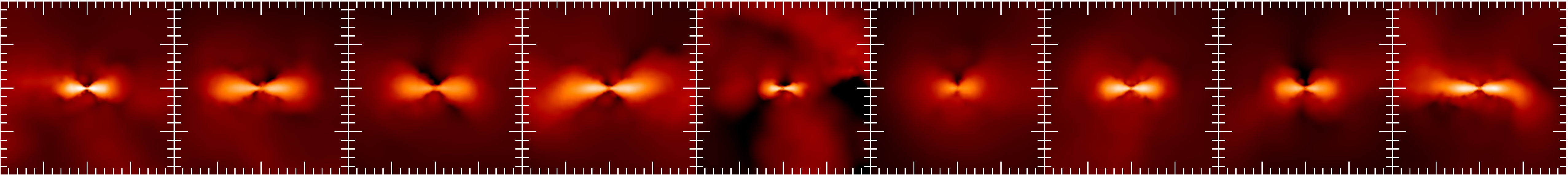}  
\includegraphics[width=\textwidth]{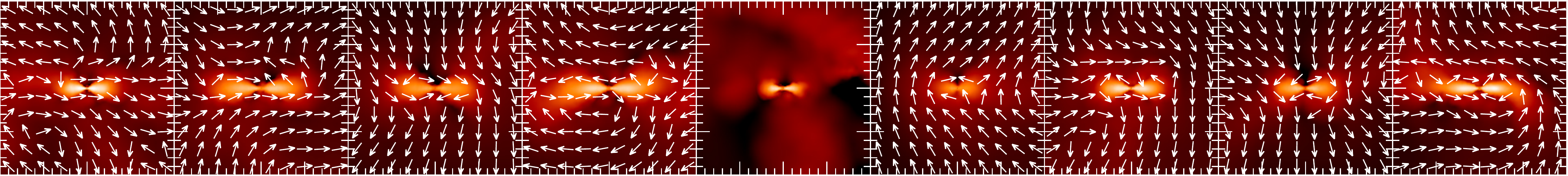}  
\includegraphics[width=\textwidth]{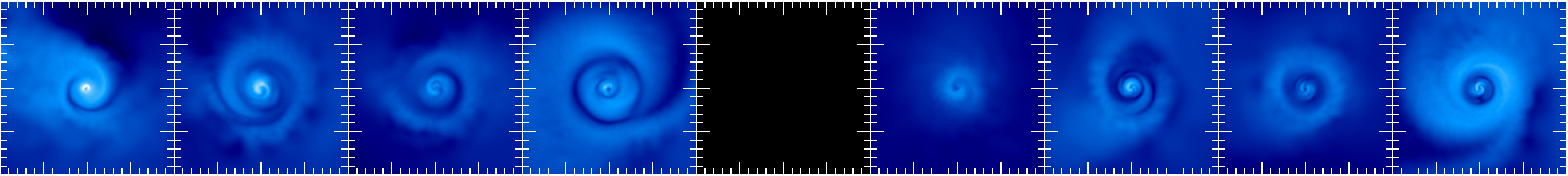}  
\includegraphics[width=\textwidth]{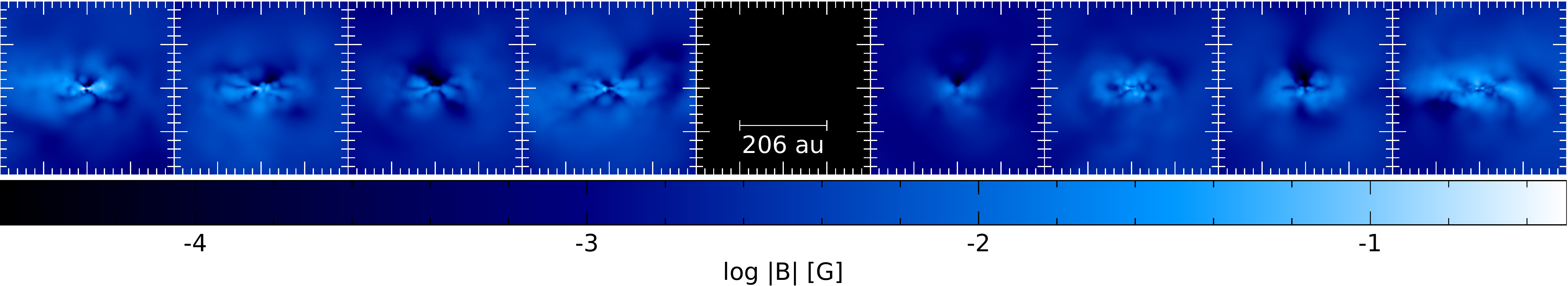}  
\caption{A cross-section of the largest, most well-defined circumstellar disc from each model at \ttnow.  From top to bottom: Face-on gas density, edge-on gas density, edge-on gas density over-plotted with magnetic field vectors representing direction only, face-on magnetic field strength, and edge-on magnetic field strength. All panels are slices through the centre of the host star.  Sink particles are not plotted, but the star is at the centre of each panel.  Major ticks represent a spatial scale of $5\times10^{-4}$~pc = 103~au.  There is no trend amongst the discs, indicating the the local environment is more important than the initial magnetic field strength of the cluster.  The large disc in each model implies that the magnetic braking catastrophe is a numerical problem that only arises from idealised initial conditions when modelling isolated stars.}
\label{fig:disc:One}
\end{figure*}

\fig{fig:disc:One} shows the gas density and magnetic field strength and direction in a slice through the centre of the most well-defined circumstellar disc in each model.  
These representative discs indicate that there is no trend amongst the models.  It is worth explicitly noting that large discs exist in each model, even those clusters with initially strong magnetic field strengths.  This suggests that the angular momentum required for disc formation originates from the turbulent velocity of the gas, and that any hindrance of disc formation by the magnetic field (i.e., magnetic braking) is relatively weak; this will be discussed further in \secref{sec:ppd:magbrak}.  The smaller discs are a result of their environment and the interaction with other stellar systems, rather than a dependence on the magnetic field; for example, the discs in \emph{Hyd} and \emph{N03} are orbiting a circumbinary disc which regulates the sizes of both discs.  The larger discs have not recently undergone any interaction with another stellar system, which allows their discs to grow.  This further suggests that discs formation and evolution is more strongly dependent on the local velocity field than the local magnetic field.

The majority of the discs are not as well-defined as in \fig{fig:disc:One}.  Several single stars do not have circumstellar discs; many of these are low mass stars that have been ejected from their birth clump, thus were likely stripped of their primordial disc (if they even had one) and are not in an environment that is gas-rich enough for the disc to reform.  Many of the circumsystem discs are being influenced and disrupted by the orbits of their host stars, and several are undergoing interactions where the discs is being rapid augmented or destroyed.  For further discussion, see \citet{Bate2018}; for the complete disc population in our study, see  Appendix~\ref{app:discs}.

The second row of \fig{fig:disc:One} does not reveal any jets or outflows, which is true for all the discs at \tnow.  This is likely a consequence of the constant changes in disc orientation due to close encounters, and our limited numerical resolution preventing disc wind formation.

\subsubsection{Disc sizes}
\begin{figure}
\centering
\includegraphics[width=0.98\columnwidth]{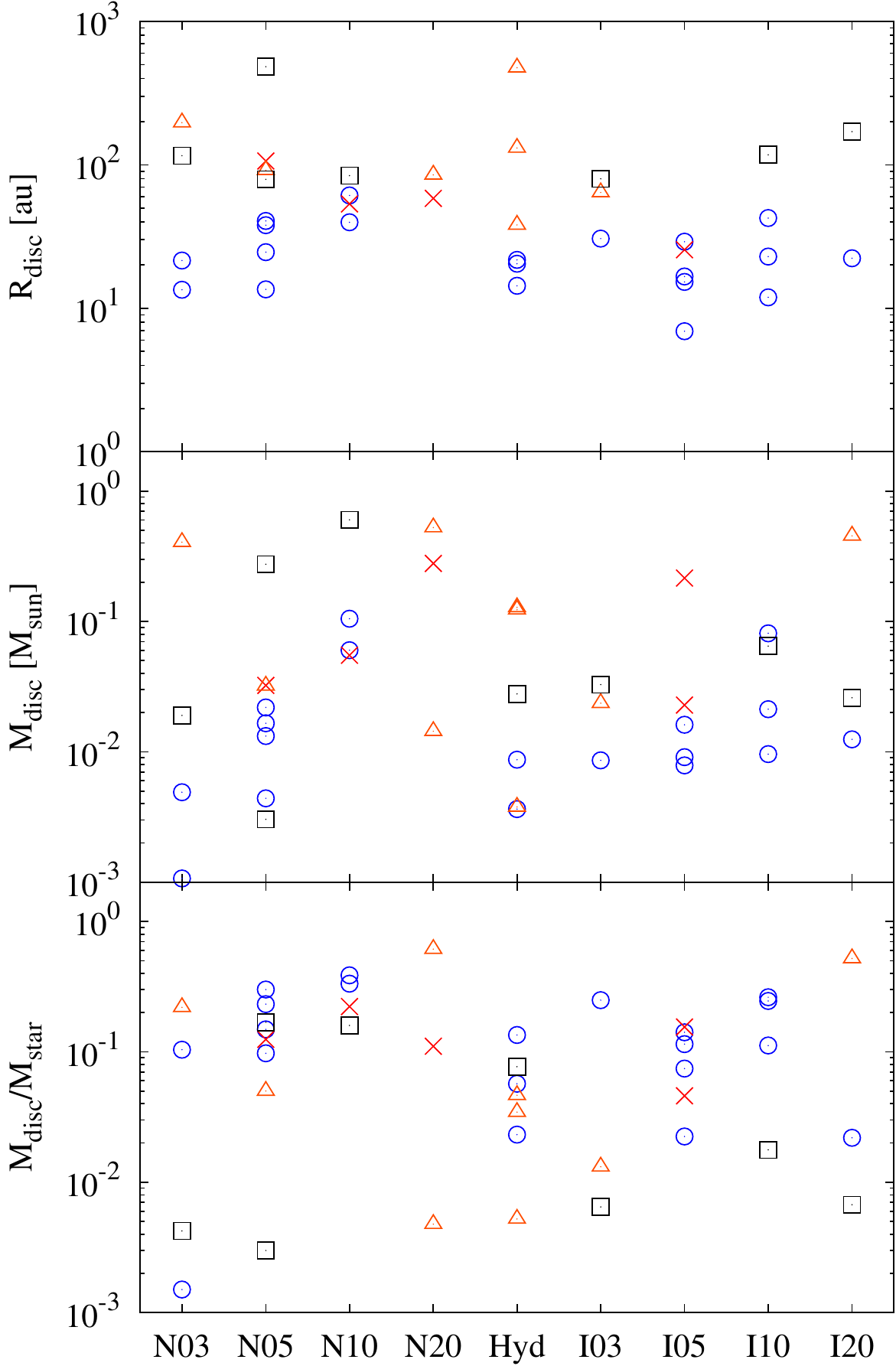}  
\caption{The disc radius (top), mass (middle) and disc-to-stellar mass ratio (bottom) of the highest-order discs at \tnow.  The circles represent circumstellar discs, the x's represent circumbinary discs, the triangles represent circumsystem discs about three stars, and squares represent circumsystem discs about four stars.  Trends exist amongst the hierarchy of discs, but not across the various models.}
\label{fig:disc:RM}
\end{figure}

\fig{fig:disc:RM} shows the disc radius, mass and disc-to-stellar mass ratios of the highest-order discs at \tnow.  The radius is for the highest order disc in each system (i.e. the disc that surrounds all the stars in the bound stellar system; this corresponds to the unbracketed numbers in \tabref{table:stars}), while the mass is a sum of the masses of all the discs in the system; thus, there may not necessarily be a corresponding radius to each mass if the stellar system does not include a circumsystem disc (e.g. \emph{I20}). 

Circumstellar discs of $10 \lesssim r/\text{au} \lesssim 80$ and circumsystem discs of $30 \lesssim r/\text{au} \lesssim 500$ form in our models with no obvious dependence on our simulation parameters.   Several $10 \lesssim r/\text{au} \lesssim 30$ circumstellar discs form even in our strongest magnetic field models (\emph{N03} and \emph{I03}), confirming that the examples in \fig{fig:disc:One} are a good representation.

We must be cautious about resolution effects in our disc analysis.  The mass of an SPH particle is $m_\text{p} = 10^{-5}$~\Msun, thus our discs typically contain $10^2 - 10^4$ particles.  Despite the small numbers, they are resolved as per the Jeans mass \citep{BateBurkert1997} and the Toomre-Mass \citep{Nelson2006} criteria.  However, \citet{Nelson2006} suggests that the scale height must be resolved by at least four smoothing lengths in the midplane of the disc, in order to prevent numerical fragmentation.  Our discs do not meet this criteria, but we also do not observe any disc fragmentation.

\subsubsection{Disc masses}
As with the radius, disc masses show no obvious dependence on simulation parameters. At \tnow, there are 10 massive discs ($M > 0.1$~\Msun), only one of which is a circumstellar disc.  Many of these massive discs survive to the end of the simulation, however, the survival of massive discs is not guaranteed.  Throughout the evolution of the clusters, there are several discs with $M > 0.1$~\Msun, which are then partially or totally disrupted by close encounters.  This is to be expected since these massive discs typically form in high-density regions into which new stars are gravitationally attracted, and are typically extended discs whose outer regions are easy to strip off.

Slightly more than half of the discs have masses of $M_\text{disc} > 0.1 M_\text{star}$, where $M_\text{star}$ is the total stellar mass in the system, with the higher order discs typically having smaller ratios.  We also find that most binaries are comprised of nearly equal mass stars.

\subsubsection{Magnetic field strength and geometry}
\begin{figure}
\centering
\includegraphics[width=0.98\columnwidth]{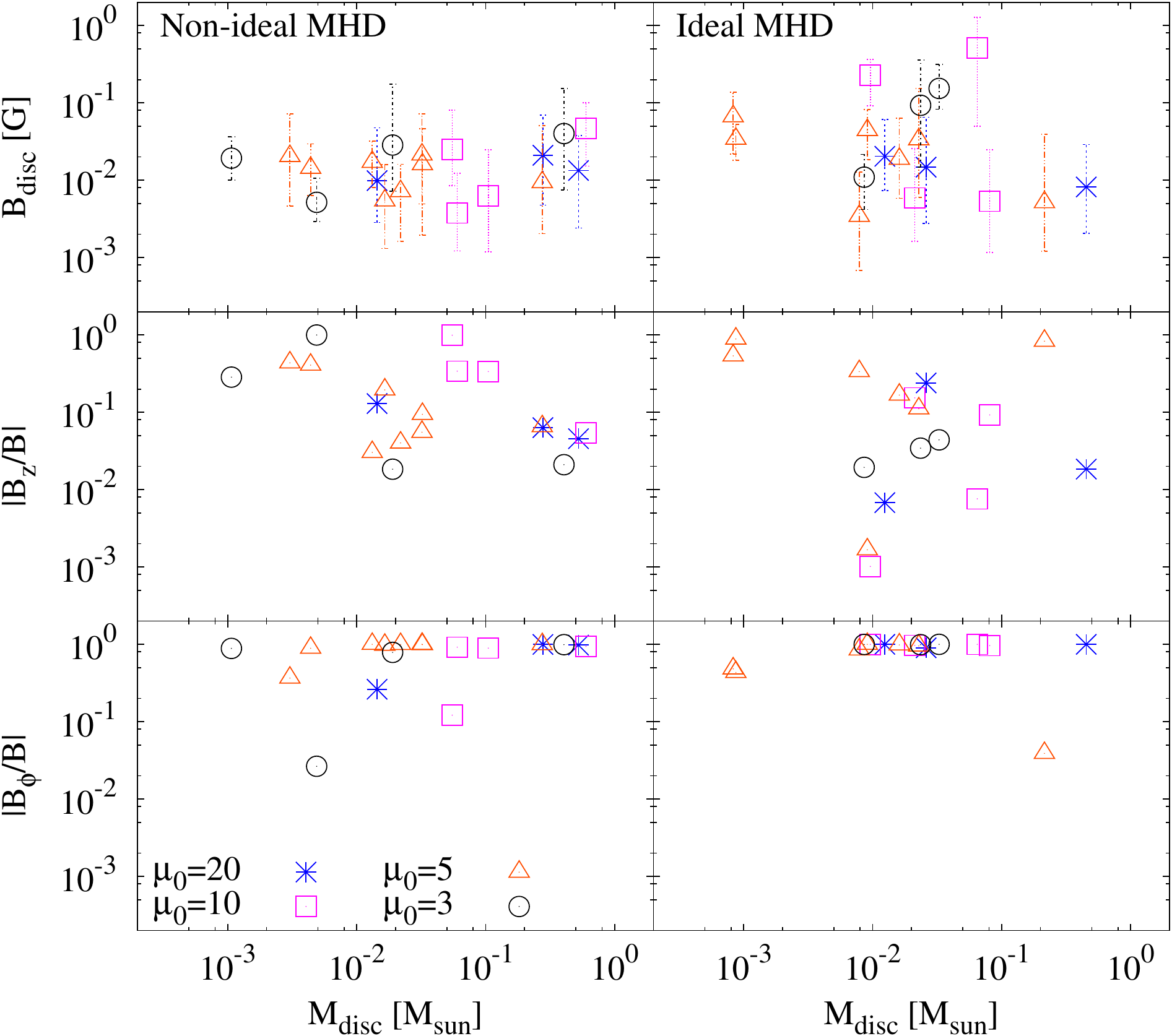}  
\caption{Top: the average magnetic field strength in the discs, with the vertical bars covering 95 per cent of the range of strengths in each disc.  Middle and bottom: The fraction of the magnetic field component perpendicular to the plane of the disc ($B_\text{z}$) and the azimuthal field in the disc ($B_\phi$), respectively.  There is no trend amongst the models in the magnetic field strength (top row), but the values in the non-ideal suite tend to span a smaller range. In most of the discs, the azimuthal (toroidal) field is dominant over the vertical component of the field.}
\label{fig:disc:B}
\end{figure}

The structure of the magnetic field of the representative circumstellar discs is shown in the third and fourth row of \fig{fig:disc:One}, and the arrows in the fifth row indicate the direction of the magnetic field.  Despite the reasonably smooth density profiles (top row), spiral structures exist in the magnetic field \citep[see discussion in][]{WursterBatePrice2018ff}, resulting in a range of magnetic field strengths in each disc.  The average magnetic field strength of each disc and the range containing 95 per cent of the field strengths is show in the top row of \fig{fig:disc:B}.
The average field strength is $0.005 \lesssim B/\text{G} \lesssim 0.5$, with the range in any given disc spanning up to \sm1.5~dex.  The disc magnetic field strength is independent of the initial field strength, and the non-ideal processes moderate the magnetic field strengths in the discs, resulting in a narrower range of field strengths in the non-ideal suite than the ideal suite.

The middle and bottom rows of \fig{fig:disc:B} show the fraction of the magnetic field component in the discs that is parallel to the disc axis (middle row) and in the azimuthal direction (bottom row).  The average magnetic field components are calculated by
\begin{equation}
\label{eq:Bave}
B^j_\text{mean} =  \frac{ N10^{\left[\sum_{(B^j_i > 0)}^N \log (B^j_i)\right]/N }  - n10^{\left[\sum_{(B^j_i < 0)}^n \log (-B^j_i)\right]/n}} {N + n}
\end{equation}
where $j \in \{\phi, z\}$, and we individually sum over the $N$ positive and $n$ negative values before performing a linear weighted average of the two terms.  For the circumbinary and circumsystem discs, we include the contribution from the lower order discs.  Unsurprisingly, there is a strong azimuthal component in most of the discs.  

Although the initial magnetic field is $-B_0\hat{\bm{z}}$, there is no preference for aligned or anti-aligned magnetic fields, thus the Hall effect is expected to be important in approximately half of the discs \citep{Tsukamoto+2015hall}; however, other effects may be dominant such that the contribution from the Hall effect is minimal \citep{WursterBate2019}.  

When comparing the poloidal to toroidal magnetic field in the region around the disc, we find that the poloidal component is typically dominant, which is consistent with studies of isolated star formation where the initial magnetic field is (anti-)aligned with the rotation axis \citepeg{BateTriccoPrice2014,WursterBatePrice2018sd}.   The larger-scale magnetic field (out to \sm800~au) typically reflects the small-scale magnetic field (out to \sm200~au), although in a few cases, a horizontal field twists near the disc to become vertical.  

\subsubsection{Is there a magnetic braking catastrophe?}
\label{sec:ppd:magbrak}

The presence of large ($r \gtrsim 20$~au), massive ($M \gtrsim 0.01$ M$_\odot$) discs in \emph{all} of our calculations suggests that their formation and structure primarily depends on the local environment. Disc formation occurs regardless of the initial magnetic field strength of the progenitor cloud and whether or not non-ideal MHD is employed.   As shown in \fig{fig:RhoVB}, the magnetic field strength in the dense gas ($\rho \gtrsim 10^{-17}${\gpercc}) is similar for all models, but varies by \sm1~dex at each density.  Thus, the local magnetic field strength in star forming clumps does not simply reflect that of the initial strength in the cluster.   As a result, large rotationally supported discs form even in ideal MHD models with initially strong magnetic fields.

This suggests that the magnetic braking catastrophe is not a problem in realistic environments due to the turbulent and dynamic nature of the environment and that turbulence and interactions are more important than magnetic fields for disc formation and evolution; this is in agreement with \citet{Seifried+2013}.   Our results also suggest a universal initial condition in local star forming regions (or at least a limited set of initial conditions) that may even be independent of the large scale magnetic field of its host environment.  Further investigation of this is beyond the scope of this study, but this possibility should be explored in the future.

Several single stars in our suite do not have circumstellar discs, however, investigating their history suggests that this is a result of their dynamical ejection into a gas-poor environment, rather than as a result of the magnetic fields and magnetic braking catastrophe.  Thus, it is likely that nearly all stars host discs, even if only briefly.

Although we have learned much from investigating the magnetic braking catastrophe in idealised simulations of the formation of isolated stars, it appears the main problem is the artificial initial conditions that were employed.

\section{Discussion}
\label{sec:discussion}

\subsection{Previous cluster scale calculations}

Magnetised hydrodynamical simulations of star formation in stellar groups and clusters have lagged up to a decade behind purely hydrodynamical simulations.  The first such calculations that attempted to resolve individual stellar systems were the ideal MHD simulations of \cite{PriceBate2008} using a barotropic equation of state, and \cite{PriceBate2009} that included radiative transfer and a realistic gas equation of state.  These SPH simulations employed Euler potentials to model the magnetic fields, limiting the magnetic field geometries that could be modelled.  However, they successfully demonstrated the two main effects of magnetic fields on star cluster formation --- the formation of different structures in magnetised molecular clouds compared to those modelled purely with hydrodynamics (e.g. anisotropic turbulent motions, low-density striations, and magnetised voids), and the star formation rate is reduced with increasing magnetic field strength due to magnetic support on large scales.  Subsequent grid-based calculations have confirmed the reduction of the star formation rate with magnetic fields \citep[e.g.][]{PadoanNordlund2011,PadoanHaugbolleNordlund2012,FederrathKlessen2012,Myers+2014}, but the magnitude of the effect is only at the level of factors of two or three times slower with strong fields compared to the star formation rate obtained using pure hydrodynamics.

Regarding the effects on stellar masses, the situation is less clear.  The calculations of \cite{PriceBate2008,PriceBate2009} produced only small numbers of stars and brown dwarfs and did not find any strong evidence for an effect of magnetic fields on their characteristic mass.  The calculations of \cite{Myers+2014} produced $\approx 90$ objects each, and they found that strong fields may increase the characteristic mass by a factor of two or three.  However, the more recent calculations of \cite{Cunningham+2018} which also include protostellar outflows and produce $\approx 20$--$70$ objects each find no evidence for a shift in the characteristic mass for calculations performed with driven turbulence, and a weak trend for a characteristic mass that decreases with increasing magnetic field strength for calculations performed with decaying turbulence. Thus, with the limited statistics currently available, there is no clear dependence of stellar masses on magnetic field strength.  Our results do not shed any further light on this question --- the numbers of objects formed are too small to detect any weak dependence of the stellar mass function on magnetic fields, should one exist.  It seems certain, however, that magnetic fields do not play the dominant role in determining the IMF.  Instead, the characteristic stellar mass seems to be set primarily by radiative feedback from protostars \citep{Bate2009rfb,Bate2012,Bate2014,Bate2019,Krumholz2011,KrumholzKleinMcKee2012} and by thermodynamic processes \citep{LeeHennebelle2018b}.

The calculations we present here extend the investigation of magnetised star formation in two main ways beyond the studies mentioned above.  First, they are the first calculations of the formation of stellar groups that treat the three main non-ideal MHD effects (ambipolar diffusion, the Hall effect, and Ohmic diffusion). We find that the non-ideal processes have little effect on the large-scale structures ($\gtrsim 0.05$~pc).  However, including non-ideal MHD does alter structures on smaller scales (i.e., the filaments and cores) and, thus, changes the details of the cloud fragmentation and distribution of protostars.  There is no evidence from our current calculations that including non-ideal MHD leads to any changes in the statistical properties of stellar systems, but calculations that produce at least an order of magnitude more objects will be required to investigate this question further.  Second, unlike in most of the above MHD calculations, protostellar discs are resolved.   The calculations of \cite{PriceBate2008,PriceBate2009} resolved discs as small as $\sim 10$~au, but they used Euler potentials to model the magnetic fields which did not allow magnetic fields to be wound up.   Our calculations employ sink particles with accretion radii of only 0.5~au, meaning that binary and multiple protostellar systems are resolved, along with protostellar discs with radii as small as $\approx 10$ au.

\subsection{Discs and binaries}

The effects of both ideal and non-ideal magnetic fields have been thoroughly investigated in studies of isolated star formation \citepeg{AllenLiShu2003,PriceBate2007,HennebelleFromang2008,MellonLi2008,Tomida+2010rmhd,Tomida+2010llc,Commercon+2010,Seifried+2011,Tomida+2013,BateTriccoPrice2014}; see \citet{WursterLi2018} for a recent review.  Early studies found that large, protostellar discs did not form in the presence of strong, ideal magnetic fields --- the so-called `magnetic braking catastrophe' \citep{AllenLiShu2003,Galli+2006,PriceBate2007,HennebelleFromang2008}.  These models assumed simple initial conditions, with ordered initial magnetic fields aligned parallel to the rotation axis.  However, with misaligned magnetic fields \citepeg{JoosHennebelleCiardi2012,LewisBatePrice2015,LewisBate2017}, turbulent velocity fields \citepeg{MachidaMatsumoto2011,Seifried+2012,Seifried+2013,Santoslima+2013,Joos+2013,LewisBate2018} or non-ideal MHD \citepeg{Tsukamoto+2015oa,Tsukamoto+2015hall,WursterPriceBate2016,Tsukamoto+2017,Vaytet+2018,WursterBatePrice2018sd,WursterBatePrice2018hd,WursterBatePrice2018ff}, substantial protostellar discs can be formed.  The key differences between these simulations and the earlier more idealised models are that the discs are subject to weaker magnetic braking and angular momentum transport due to complex field/rotation geometries and/or weaker magnetic fields (e.g. due to non-ideal effects and/or turbulent or grid-scale magnetic reconnection).

Similarly, past studies of the collapse of isolated magnetised molecular cloud cores have shown that magnetic fields can inhibit fragmentation and multiple star formation \citep{HoskingWhitworth2004frag,PriceBate2007,HennebelleTeyssier2008,Commercon+2010,CommerconHennebelleHenning2011}.  However, \cite{WursterPriceBate2017} and \citet{WursterBate2019} showed that whether or not a dense core fragments depends on its initial density (sub-)structure, rotation, and field geometry rather than just on the strength of the magnetic field.  Thus, in a turbulent molecular cloud in which dense cores are likely to have considerable substructure and rotation and fields that are misaligned with the net rotation, fragmentation is quite possible even with strong fields.  

\cite{Seifried+2013} performed calculations of ideally magnetised turbulent molecular cloud cores with masses up to 1000~M$_\odot$, that produce multiple protostellar objects and resolved discs.  However, their initial conditions were chosen to be strongly centrally-condensed ($\rho \propto r^{3/2}$) which favours massive star formation, rather than to study the formation of low-mass stellar groups.  The overwhelming majority of their stars formed via gravitational collapse of distinct overdense regions, then multiple systems formed via capture rather than disc fragmentation; their discs tended to have masses $0.05 \lesssim M/$\Msun$ \lesssim 0.1$ and radii of  $50 \lesssim r/\text{au} \lesssim 150$.  In general, similar to the work we present here, their study concluded that discs form as a result of the turbulence within star-forming regions, independent of the the presence of strong magnetic fields, which quickly become disordered.

In our calculations, protostars are surrounded by discs with a variety of sizes, ranging from $\approx 10-500$~au.  The multiple systems form from the fragmentation of filaments, rather than disc fragmentation.  Similar results are obtained from turbulent radiation hydrodynamical simulations without magnetic fields \citep{Bate2012,Bate2018,Bate2019}.  The statistical properties of the multiple systems and discs do not display any obvious trends with initial magnetic field strength, or with whether or not non-ideal MHD processes are included.  Thus, it seems that it is the density and velocity structure and dynamical interactions between protostars that dominate the properties of multiple systems and discs, rather than whether or not magnetic fields are present.

\section{Summary and conclusion}
\label{sec:conclusion}
We have presented a suite of radiation non-ideal magnetohydrodynamic simulations modelling star formation in a 50~\Msun \ molecular cloud.  Each model was initialised with the same turbulent velocity field, and was threaded by a uniform magnetic field; we tested four magnetic field strengths plus a purely hydrodynamic model.  For each magnetised model, we performed both an ideal and non-ideal MHD version, where the latter included Ohmic resistivity, ambipolar diffusion and the Hall effect.   Sink particles with radii of 0.5~au were inserted at the opacity limit for fragmentation such that each particle represented one star.  

Our key results are as follows:
\begin{enumerate}

\item \emph{Magnetic fields influence the large scale structure}.  The initial mass-to-flux ratio was found to mainly affect the large scale evolution, with models with stronger magnetic fields slower to form high-density clumps due to increased magnetic support.  However, our models with the strongest magnetic field defied this trend, with gas rapidly collapsing along the magnetic field lines.  With strong magnetic fields, the field lines tend to be perpendicular to dense filaments, while with weak magnetic fields there is a non-trivial parallel component near the filament.  Magnetic fields tend to be parallel to low-density filaments.  

\item \emph{Initial conditions are quickly erased}.  Magnetic field strengths in the high-density gas (\rhogs{-17}) were found to be independent of the initial mass-to-flux ratio and of the non-ideal MHD processes.   At a given high density, the magnetic field strength and range was similar for each model and spanned approximately one order of magnitude.  Thus, individual star forming clumps contain a wide range of magnetic field strengths. 

\item \emph{Non-ideal MHD acts mainly on small scales}. Small scale structures tend to be filamentary for models with strong initial magnetic fields and more clumpy for models with weak initial magnetic fields.  Non-ideal MHD, while not affecting the large scale evolution, was found to play a more significant role on $\lesssim 0.05$ pc scales. We found the main effect was to create tighter filaments and to reduce the magnetic field strength in gas in and around the protostellar discs.

\item \emph{Star formation is violent and chaotic}. 
Binary and hierarchical triple/quadruple systems form in our models. Binary stars form primarily by gravitational capture, and nearly all of the binaries survived to the end of the simulation, with their orbits and ellipticities strongly influenced by interactions with other stars.  The number of stars does not depend on the initial magnetic field strength, however, there is a general trend of decreasing total stellar mass as the initial magnetic field strength is increased; our strongest field models defied this trend.

\item \emph{There is no magnetic braking catastrophe}.  
Protostellar discs with radii of \sm$10 - 80$~au form in \emph{all} of our models, and circumsystem discs have radii up to \sm500~au.  The magnetic field strength in the protostellar discs is independent of the initial magnetic field strength of the molecular cloud, and the non-ideal processes yield a narrower range of disc field strengths than when assuming ideal MHD.  Thus, non-ideal MHD may not be required for disc formation, but it does regulate the magnetic field in discs.  The presence of large, massive discs in every model suggests that the magnetic braking catastrophe only arises when modelling the formation of isolated stars from idealised initial conditions, and that in the presence of strong global magnetic fields, turbulence promotes disc formation.

\end{enumerate}

In all of our models, we form stars, stellar systems, and protostellar discs, even in clusters with initially strong magnetic field strengths.  Given the weak or non-existent trends in the properties that we investigated, this suggests that observed objects (and possibly trends) can be reproduced, even when including realistic strong magnetic fields.  Thus, magnetic fields are not a hinderance to numerical star and protostellar disc formation.

\section*{Acknowledgements}

We would like to thank the referee, Dr Daniel Seifried, for useful comments that greatly improved the quality of this manuscript.
JW and MRB acknowledge support from the European Research Council under the European Community's Seventh Framework Programme (FP7/2007- 2013 grant agreement no. 339248). DJP received funding via Australian Research Council grants FT130100034, DP130102078 and DP180104235.   Calculations and analyses for this paper were performed on the University of Exeter Supercomputer, Isca, which is part of the University of Exeter High-Performance Computing (HPC) facility, and on the DiRAC Data Intensive service at Leicester, operated by the University of Leicester IT Services, which forms part of the STFC DiRAC HPC Facility (www.dirac.ac.uk). The equipment was funded by BEIS capital funding via STFC capital grants ST/K000373/1 and ST/R002363/1 and STFC DiRAC Operations grant ST/R001014/1. DiRAC is part of the National e-Infrastructure.
The research data supporting this publication will be openly available from the University of Exeter's institutional repository.
Several figures were made using \textsc{splash} \citep{Price2007}.  

\bibliography{clusters}

\begin{thebibliography}{}
\makeatletter
\relax
\def\mn@urlcharsother{\let\do\@makeother \do\$\do\&\do\#\do\^\do\_\do\%\do\~}
\def\mn@doi{\begingroup\mn@urlcharsother \@ifnextchar [ {\mn@doi@}
  {\mn@doi@[]}}
\def\mn@doi@[#1]#2{\def\@tempa{#1}\ifx\@tempa\@empty \href
  {http://dx.doi.org/#2} {doi:#2}\else \href {http://dx.doi.org/#2} {#1}\fi
  \endgroup}
\def\mn@eprint#1#2{\mn@eprint@#1:#2::\@nil}
\def\mn@eprint@arXiv#1{\href {http://arxiv.org/abs/#1} {{\tt arXiv:#1}}}
\def\mn@eprint@dblp#1{\href {http://dblp.uni-trier.de/rec/bibtex/#1.xml}
  {dblp:#1}}
\def\mn@eprint@#1:#2:#3:#4\@nil{\def\@tempa {#1}\def\@tempb {#2}\def\@tempc
  {#3}\ifx \@tempc \@empty \let \@tempc \@tempb \let \@tempb \@tempa \fi \ifx
  \@tempb \@empty \def\@tempb {arXiv}\fi \@ifundefined
  {mn@eprint@\@tempb}{\@tempb:\@tempc}{\expandafter \expandafter \csname
  mn@eprint@\@tempb\endcsname \expandafter{\@tempc}}}

\bibitem[\protect\citeauthoryear{{Allen}, {Li}  \& {Shu}}{{Allen}
  et~al.}{2003}]{AllenLiShu2003}
{Allen} A.,  {Li} Z.-Y.,   {Shu} F.~H.,  2003, \mn@doi [\apj] {10.1086/379243},
  \href {http://adsabs.harvard.edu/abs/2003ApJ...599..363A} {599, 363}

\bibitem[\protect\citeauthoryear{{Alves}, {Franco}  \& {Girart}}{{Alves}
  et~al.}{2008}]{AlvesFrancoGirart2008}
{Alves} F.~O.,  {Franco} G.~A.~P.,   {Girart} J.~M.,  2008, \mn@doi [\aap]
  {10.1051/0004-6361:200810091}, \href
  {http://adsabs.harvard.edu/abs/2008A%26A...486L..13A} {486, L13}

\bibitem[\protect\citeauthoryear{{Asplund}, {Grevesse}, {Sauval}  \&
  {Scott}}{{Asplund} et~al.}{2009}]{Asplund+2009}
{Asplund} M.,  {Grevesse} N.,  {Sauval} A.~J.,   {Scott} P.,  2009, \mn@doi
  [\araa] {10.1146/annurev.astro.46.060407.145222}, \href
  {http://adsabs.harvard.edu/abs/2009ARA%26A..47..481A} {47, 481}

\bibitem[\protect\citeauthoryear{{Bate}}{{Bate}}{2009a}]{Bate2009sp}
{Bate} M.~R.,  2009a, \mn@doi [\mnras] {10.1111/j.1365-2966.2008.14106.x},
  \href {http://adsabs.harvard.edu/abs/2009MNRAS.392..590B} {392, 590}

\bibitem[\protect\citeauthoryear{{Bate}}{{Bate}}{2009b}]{Bate2009rfb}
{Bate} M.~R.,  2009b, \mn@doi [\mnras] {10.1111/j.1365-2966.2008.14165.x},
  \href {http://adsabs.harvard.edu/abs/2009MNRAS.392.1363B} {392, 1363}

\bibitem[\protect\citeauthoryear{{Bate}}{{Bate}}{2009c}]{Bate2009ics}
{Bate} M.~R.,  2009c, \mn@doi [\mnras] {10.1111/j.1365-2966.2009.14970.x},
  \href {http://adsabs.harvard.edu/abs/2009MNRAS.397..232B} {397, 232}

\bibitem[\protect\citeauthoryear{{Bate}}{{Bate}}{2012}]{Bate2012}
{Bate} M.~R.,  2012, \mn@doi [\mnras] {10.1111/j.1365-2966.2011.19955.x}, \href
  {http://adsabs.harvard.edu/abs/2012MNRAS.419.3115B} {419, 3115}

\bibitem[\protect\citeauthoryear{{Bate}}{{Bate}}{2014}]{Bate2014}
{Bate} M.~R.,  2014, \mn@doi [\mnras] {10.1093/mnras/stu795}, \href
  {http://adsabs.harvard.edu/abs/2014MNRAS.442..285B} {442, 285}

\bibitem[\protect\citeauthoryear{{Bate}}{{Bate}}{2018}]{Bate2018}
{Bate} M.~R.,  2018, \mn@doi [\mnras] {10.1093/mnras/sty169}, \href
  {http://adsabs.harvard.edu/abs/2018MNRAS.475.5618B} {475, 5618}

\bibitem[\protect\citeauthoryear{{Bate}}{{Bate}}{2019}]{Bate2019}
{Bate} M.~R.,  2019, \mn@doi [\mnras] {10.1093/mnras/stz103}, \href
  {https://ui.adsabs.harvard.edu/abs/2019MNRAS.484.2341B} {484, 2341}

\bibitem[\protect\citeauthoryear{{Bate} \& {Bonnell}}{{Bate} \&
  {Bonnell}}{2005}]{BateBonnell2005}
{Bate} M.~R.,  {Bonnell} I.~A.,  2005, \mn@doi [\mnras]
  {10.1111/j.1365-2966.2004.08593.x}, \href
  {http://adsabs.harvard.edu/abs/2005MNRAS.356.1201B} {356, 1201}

\bibitem[\protect\citeauthoryear{{Bate} \& {Burkert}}{{Bate} \&
  {Burkert}}{1997}]{BateBurkert1997}
{Bate} M.~R.,  {Burkert} A.,  1997, \mn@doi [\mnras]
  {10.1093/mnras/288.4.1060}, \href
  {http://adsabs.harvard.edu/abs/1997MNRAS.288.1060B} {288, 1060}

\bibitem[\protect\citeauthoryear{{Bate} \& {Keto}}{{Bate} \&
  {Keto}}{2015}]{BateKeto2015}
{Bate} M.~R.,  {Keto} E.~R.,  2015, \mn@doi [\mnras] {10.1093/mnras/stv451},
  \href {http://adsabs.harvard.edu/abs/2015MNRAS.449.2643B} {449, 2643}

\bibitem[\protect\citeauthoryear{{Bate}, {Bonnell}  \& {Price}}{{Bate}
  et~al.}{1995}]{BateBonnellPrice1995}
{Bate} M.~R.,  {Bonnell} I.~A.,   {Price} N.~M.,  1995, \mn@doi [\mnras]
  {10.1093/mnras/277.2.362}, \href
  {http://adsabs.harvard.edu/abs/1995MNRAS.277..362B} {277, 362}

\bibitem[\protect\citeauthoryear{{Bate}, {Bonnell}  \& {Bromm}}{{Bate}
  et~al.}{2003}]{BateBonnellBromm2003}
{Bate} M.~R.,  {Bonnell} I.~A.,   {Bromm} V.,  2003, \mn@doi [\mnras]
  {10.1046/j.1365-8711.2003.06210.x}, \href
  {http://adsabs.harvard.edu/abs/2003MNRAS.339..577B} {339, 577}

\bibitem[\protect\citeauthoryear{{Bate}, {Tricco}  \& {Price}}{{Bate}
  et~al.}{2014}]{BateTriccoPrice2014}
{Bate} M.~R.,  {Tricco} T.~S.,   {Price} D.~J.,  2014, \mn@doi [\mnras]
  {10.1093/mnras/stt1865}, \href
  {http://adsabs.harvard.edu/abs/2014MNRAS.437...77B} {437, 77}

\bibitem[\protect\citeauthoryear{{Benz}}{{Benz}}{1990}]{Benz1990}
{Benz} W.,  1990, in {Buchler} J.~R.,  ed., Numerical Modelling of Nonlinear
  Stellar Pulsations Problems and Prospects. p.~269

\bibitem[\protect\citeauthoryear{{Bergin} \& {Tafalla}}{{Bergin} \&
  {Tafalla}}{2007}]{BerginTafalla2007}
{Bergin} E.~A.,  {Tafalla} M.,  2007, \mn@doi [\araa]
  {10.1146/annurev.astro.45.071206.100404}, \href
  {http://adsabs.harvard.edu/abs/2007ARA%26A..45..339B} {45, 339}

\bibitem[\protect\citeauthoryear{{Boley}, {Hartquist}, {Durisen}  \&
  {Michael}}{{Boley} et~al.}{2007}]{Boley+2007}
{Boley} A.~C.,  {Hartquist} T.~W.,  {Durisen} R.~H.,   {Michael} S.,  2007,
  \mn@doi [\apjl] {10.1086/512235}, \href
  {http://adsabs.harvard.edu/abs/2007ApJ...656L..89B} {656, L89}

\bibitem[\protect\citeauthoryear{{B{\o}rve}, {Omang}  \& {Trulsen}}{{B{\o}rve}
  et~al.}{2001}]{BorveOmangTrulsen2001}
{B{\o}rve} S.,  {Omang} M.,   {Trulsen} J.,  2001, \mn@doi [\apj]
  {10.1086/323228}, \href {http://adsabs.harvard.edu/abs/2001ApJ...561...82B}
  {561, 82}

\bibitem[\protect\citeauthoryear{{Burkert} \& {Hartmann}}{{Burkert} \&
  {Hartmann}}{2004}]{BurkertHartmann2004}
{Burkert} A.,  {Hartmann} L.,  2004, \mn@doi [\apj] {10.1086/424895}, \href
  {http://adsabs.harvard.edu/abs/2004ApJ...616..288B} {616, 288}

\bibitem[\protect\citeauthoryear{{Chapman}, {Goldsmith}, {Pineda}, {Clemens},
  {Li}  \& {Kr{\v c}o}}{{Chapman} et~al.}{2011}]{Chapman+2011}
{Chapman} N.~L.,  {Goldsmith} P.~F.,  {Pineda} J.~L.,  {Clemens} D.~P.,  {Li}
  D.,   {Kr{\v c}o} M.,  2011, \mn@doi [\apj] {10.1088/0004-637X/741/1/21},
  \href {https://ui.adsabs.harvard.edu/abs/2011ApJ...741...21C} {741, 21}

\bibitem[\protect\citeauthoryear{{Commer{\c c}on}, {Hennebelle}, {Audit},
  {Chabrier}  \& {Teyssier}}{{Commer{\c c}on} et~al.}{2010}]{Commercon+2010}
{Commer{\c c}on} B.,  {Hennebelle} P.,  {Audit} E.,  {Chabrier} G.,
  {Teyssier} R.,  2010, \mn@doi [\aap] {10.1051/0004-6361/200913597}, \href
  {http://adsabs.harvard.edu/abs/2010A%26A...510L...3C} {510, L3}

\bibitem[\protect\citeauthoryear{{Commer{\c{c}}on}, {Hennebelle}  \&
  {Henning}}{{Commer{\c{c}}on} et~al.}{2011}]{CommerconHennebelleHenning2011}
{Commer{\c{c}}on} B.,  {Hennebelle} P.,   {Henning} T.,  2011, \mn@doi [\apj]
  {10.1088/2041-8205/742/1/L9}, \href
  {https://ui.adsabs.harvard.edu/abs/2011ApJ...742L...9C} {742, L9}

\bibitem[\protect\citeauthoryear{{Cox}, {Harris}, {Looney}, {Li}, {Yang},
  {Tobin}  \& {Stephens}}{{Cox} et~al.}{2018}]{Cox+2018}
{Cox} E.~G.,  {Harris} R.~J.,  {Looney} L.~W.,  {Li} Z.-Y.,  {Yang} H.,
  {Tobin} J.~J.,   {Stephens} I.,  2018, \mn@doi [\apj]
  {10.3847/1538-4357/aaacd2}, \href
  {http://adsabs.harvard.edu/abs/2018ApJ...855...92C} {855, 92}

\bibitem[\protect\citeauthoryear{{Crutcher}}{{Crutcher}}{2012}]{Crutcher2012}
{Crutcher} R.~M.,  2012, \mn@doi [\araa] {10.1146/annurev-astro-081811-125514},
  \href {http://adsabs.harvard.edu/abs/2012ARA%26A..50...29C} {50, 29}

\bibitem[\protect\citeauthoryear{{Cunningham}, {Krumholz}, {McKee}  \&
  {Klein}}{{Cunningham} et~al.}{2018}]{Cunningham+2018}
{Cunningham} A.~J.,  {Krumholz} M.~R.,  {McKee} C.~F.,   {Klein} R.~I.,  2018,
  \mn@doi [\mnras] {10.1093/mnras/sty154}, \href
  {http://adsabs.harvard.edu/abs/2018MNRAS.476..771C} {476, 771}

\bibitem[\protect\citeauthoryear{{Dale}}{{Dale}}{2017}]{Dale2017}
{Dale} J.~E.,  2017, \mn@doi [\mnras] {10.1093/mnras/stx028}, \href
  {http://adsabs.harvard.edu/abs/2017MNRAS.467.1067D} {467, 1067}

\bibitem[\protect\citeauthoryear{{Dale}, {Ngoumou}, {Ercolano}  \&
  {Bonnell}}{{Dale} et~al.}{2014}]{DaleNgoumouErcolanoBonnell2014}
{Dale} J.~E.,  {Ngoumou} J.,  {Ercolano} B.,   {Bonnell} I.~A.,  2014, \mn@doi
  [\mnras] {10.1093/mnras/stu816}, \href
  {http://adsabs.harvard.edu/abs/2014MNRAS.442..694D} {442, 694}

\bibitem[\protect\citeauthoryear{{Dunham}, {Chen}, {Arce}, {Bourke}, {Schnee}
  \& {Enoch}}{{Dunham} et~al.}{2011}]{Dunham+2011}
{Dunham} M.~M.,  {Chen} X.,  {Arce} H.~G.,  {Bourke} T.~L.,  {Schnee} S.,
  {Enoch} M.~L.,  2011, \mn@doi [\apj] {10.1088/0004-637X/742/1/1}, \href
  {http://adsabs.harvard.edu/abs/2011ApJ...742....1D} {742, 1}

\bibitem[\protect\citeauthoryear{{Fall}, {Krumholz}  \& {Matzner}}{{Fall}
  et~al.}{2010}]{FallKrumholzMatzner2010}
{Fall} S.~M.,  {Krumholz} M.~R.,   {Matzner} C.~D.,  2010, \mn@doi [\apjl]
  {10.1088/2041-8205/710/2/L142}, \href
  {http://adsabs.harvard.edu/abs/2010ApJ...710L.142F} {710, L142}

\bibitem[\protect\citeauthoryear{{Federrath} \& {Klessen}}{{Federrath} \&
  {Klessen}}{2012}]{FederrathKlessen2012}
{Federrath} C.,  {Klessen} R.~S.,  2012, \mn@doi [\apj]
  {10.1088/0004-637X/761/2/156}, \href
  {https://ui.adsabs.harvard.edu/abs/2012ApJ...761..156F} {761, 156}

\bibitem[\protect\citeauthoryear{{Fehlberg}}{{Fehlberg}}{1969}]{Fehlberg1969}
{Fehlberg} E.,  1969, NASA Technical Report R-315

\bibitem[\protect\citeauthoryear{{Ferguson}, {Alexander}, {Allard}, {Barman},
  {Bodnarik}, {Hauschildt}, {Heffner-Wong}  \& {Tamanai}}{{Ferguson}
  et~al.}{2005}]{Ferguson+2005}
{Ferguson} J.~W.,  {Alexander} D.~R.,  {Allard} F.,  {Barman} T.,  {Bodnarik}
  J.~G.,  {Hauschildt} P.~H.,  {Heffner-Wong} A.,   {Tamanai} A.,  2005,
  \mn@doi [\apj] {10.1086/428642}, \href
  {http://adsabs.harvard.edu/abs/2005ApJ...623..585F} {623, 585}

\bibitem[\protect\citeauthoryear{{Franco}, {Alves}  \& {Girart}}{{Franco}
  et~al.}{2010}]{FrancoAlvesGirart2010}
{Franco} G.~A.~P.,  {Alves} F.~O.,   {Girart} J.~M.,  2010, \mn@doi [\apj]
  {10.1088/0004-637X/723/1/146}, \href
  {http://adsabs.harvard.edu/abs/2010ApJ...723..146F} {723, 146}

\bibitem[\protect\citeauthoryear{{Galli}, {Lizano}, {Shu}  \& {Allen}}{{Galli}
  et~al.}{2006}]{Galli+2006}
{Galli} D.,  {Lizano} S.,  {Shu} F.~H.,   {Allen} A.,  2006, \mn@doi [\apj]
  {10.1086/505257}, \href {http://adsabs.harvard.edu/abs/2006ApJ...647..374G}
  {647, 374}

\bibitem[\protect\citeauthoryear{{Geen}, {Hennebelle}, {Tremblin}  \&
  {Rosdahl}}{{Geen} et~al.}{2015}]{Geen+2015}
{Geen} S.,  {Hennebelle} P.,  {Tremblin} P.,   {Rosdahl} J.,  2015, \mn@doi
  [\mnras] {10.1093/mnras/stv2272}, \href
  {http://adsabs.harvard.edu/abs/2015MNRAS.454.4484G} {454, 4484}

\bibitem[\protect\citeauthoryear{{Geen}, {Watson}, {Rosdahl}, {Bieri},
  {Klessen}  \& {Hennebelle}}{{Geen} et~al.}{2018}]{Geen+2018}
{Geen} S.,  {Watson} S.~K.,  {Rosdahl} J.,  {Bieri} R.,  {Klessen} R.~S.,
  {Hennebelle} P.,  2018, \mn@doi [\mnras] {10.1093/mnras/sty2439}, \href
  {http://adsabs.harvard.edu/abs/2018MNRAS.481.2548G} {481, 2548}

\bibitem[\protect\citeauthoryear{{Gendelev} \& {Krumholz}}{{Gendelev} \&
  {Krumholz}}{2012}]{GendelevKrumholz2012}
{Gendelev} L.,  {Krumholz} M.~R.,  2012, \mn@doi [\apj]
  {10.1088/0004-637X/745/2/158}, \href
  {http://adsabs.harvard.edu/abs/2012ApJ...745..158G} {745, 158}

\bibitem[\protect\citeauthoryear{{Gerin} et~al.,}{{Gerin}
  et~al.}{2017}]{Gerin+2017}
{Gerin} M.,  et~al., 2017, \mn@doi [\aap] {10.1051/0004-6361/201630187}, \href
  {http://adsabs.harvard.edu/abs/2017A%26A...606A..35G} {606, A35}

\bibitem[\protect\citeauthoryear{{Girichidis} et~al.,}{{Girichidis}
  et~al.}{2016}]{Girichidis+2016}
{Girichidis} P.,  et~al., 2016, \mn@doi [\mnras] {10.1093/mnras/stv2742}, \href
  {http://adsabs.harvard.edu/abs/2016MNRAS.456.3432G} {456, 3432}

\bibitem[\protect\citeauthoryear{{Goldsmith}, {Heyer}, {Narayanan}, {Snell},
  {Li}  \& {Brunt}}{{Goldsmith} et~al.}{2008}]{Goldsmith+2008}
{Goldsmith} P.~F.,  {Heyer} M.,  {Narayanan} G.,  {Snell} R.,  {Li} D.,
  {Brunt} C.,  2008, \mn@doi [\apj] {10.1086/587166}, \href
  {http://adsabs.harvard.edu/abs/2008ApJ...680..428G} {680, 428}

\bibitem[\protect\citeauthoryear{{Hatchell} et~al.,}{{Hatchell}
  et~al.}{2013}]{Hatchell+2013}
{Hatchell} J.,  et~al., 2013, \mn@doi [\mnras] {10.1093/mnrasl/sls015}, \href
  {https://ui.adsabs.harvard.edu/abs/2013MNRAS.429L..10H} {429, L10}

\bibitem[\protect\citeauthoryear{{Heiles} \& {Crutcher}}{{Heiles} \&
  {Crutcher}}{2005}]{HeilesCrutcher2005}
{Heiles} C.,  {Crutcher} R.,  2005, in {Wielebinski} R.,  {Beck} R.,  eds,
  Lecture Notes in Physics, Berlin Springer Verlag Vol. 664, Cosmic Magnetic
  Fields. p.~137 (\mn@eprint {} {astro-ph/0501550}),
  \mn@doi{10.1007/11369875_7}

\bibitem[\protect\citeauthoryear{{Hennebelle} \& {Fromang}}{{Hennebelle} \&
  {Fromang}}{2008}]{HennebelleFromang2008}
{Hennebelle} P.,  {Fromang} S.,  2008, \mn@doi [\aap]
  {10.1051/0004-6361:20078309}, \href
  {http://adsabs.harvard.edu/abs/2008A%26A...477....9H} {477, 9}

\bibitem[\protect\citeauthoryear{{Hennebelle} \& {Inutsuka}}{{Hennebelle} \&
  {Inutsuka}}{2019}]{HennebelleInutsuka2019}
{Hennebelle} P.,  {Inutsuka} S.-i.,  2019, arXiv e-prints, \href
  {http://adsabs.harvard.edu/abs/2019arXiv190200798H} {}

\bibitem[\protect\citeauthoryear{{Hennebelle} \& {Teyssier}}{{Hennebelle} \&
  {Teyssier}}{2008}]{HennebelleTeyssier2008}
{Hennebelle} P.,  {Teyssier} R.,  2008, \mn@doi [\aap]
  {10.1051/0004-6361:20078310}, \href
  {http://adsabs.harvard.edu/abs/2008A%26A...477...25H} {477, 25}

\bibitem[\protect\citeauthoryear{{Hennebelle}, {Commer{\c c}on}, {Joos},
  {Klessen}, {Krumholz}, {Tan}  \& {Teyssier}}{{Hennebelle}
  et~al.}{2011}]{Hennebelle+2011}
{Hennebelle} P.,  {Commer{\c c}on} B.,  {Joos} M.,  {Klessen} R.~S.,
  {Krumholz} M.,  {Tan} J.~C.,   {Teyssier} R.,  2011, \mn@doi [\aap]
  {10.1051/0004-6361/201016052}, \href
  {http://adsabs.harvard.edu/abs/2011A%26A...528A..72H} {528, A72}

\bibitem[\protect\citeauthoryear{{Heyer}, {Vrba}, {Snell}, {Schloerb}, {Strom},
  {Goldsmith}  \& {Strom}}{{Heyer} et~al.}{1987}]{Heyer+1987}
{Heyer} M.~H.,  {Vrba} F.~J.,  {Snell} R.~L.,  {Schloerb} F.~P.,  {Strom}
  S.~E.,  {Goldsmith} P.~F.,   {Strom} K.~M.,  1987, \mn@doi [\apj]
  {10.1086/165678}, \href {http://adsabs.harvard.edu/abs/1987ApJ...321..855H}
  {321, 855}

\bibitem[\protect\citeauthoryear{{Hosking} \& {Whitworth}}{{Hosking} \&
  {Whitworth}}{2004}]{HoskingWhitworth2004frag}
{Hosking} J.~G.,  {Whitworth} A.~P.,  2004, \mn@doi [\mnras]
  {10.1111/j.1365-2966.2004.07274.x}, \href
  {http://adsabs.harvard.edu/abs/2004MNRAS.347.1001H} {347, 1001}

\bibitem[\protect\citeauthoryear{{Jones} \& {Bate}}{{Jones} \&
  {Bate}}{2018a}]{JonesBate2018ics}
{Jones} M.~O.,  {Bate} M.~R.,  2018a, \mn@doi [\mnras] {10.1093/mnras/sty1250},
  \href {http://adsabs.harvard.edu/abs/2018MNRAS.478.2650J} {478, 2650}

\bibitem[\protect\citeauthoryear{{Jones} \& {Bate}}{{Jones} \&
  {Bate}}{2018b}]{JonesBate2018fb}
{Jones} M.~O.,  {Bate} M.~R.,  2018b, \mn@doi [\mnras] {10.1093/mnras/sty1969},
  \href {http://adsabs.harvard.edu/abs/2018MNRAS.480.2562J} {480, 2562}

\bibitem[\protect\citeauthoryear{{Joos}, {Hennebelle}  \& {Ciardi}}{{Joos}
  et~al.}{2012}]{JoosHennebelleCiardi2012}
{Joos} M.,  {Hennebelle} P.,   {Ciardi} A.,  2012, \mn@doi [\aap]
  {10.1051/0004-6361/201118730}, \href
  {http://adsabs.harvard.edu/abs/2012A%26A...543A.128J} {543, A128}

\bibitem[\protect\citeauthoryear{{Joos}, {Hennebelle}, {Ciardi}  \&
  {Fromang}}{{Joos} et~al.}{2013}]{Joos+2013}
{Joos} M.,  {Hennebelle} P.,  {Ciardi} A.,   {Fromang} S.,  2013, \mn@doi
  [\aap] {10.1051/0004-6361/201220649}, \href
  {http://adsabs.harvard.edu/abs/2013A%26A...554A..17J} {554, A17}

\bibitem[\protect\citeauthoryear{{Kim}, {Kim}  \& {Ostriker}}{{Kim}
  et~al.}{2016}]{KimKimOstriker2016}
{Kim} J.-G.,  {Kim} W.-T.,   {Ostriker} E.~C.,  2016, \mn@doi [\apj]
  {10.3847/0004-637X/819/2/137}, \href
  {http://adsabs.harvard.edu/abs/2016ApJ...819..137K} {819, 137}

\bibitem[\protect\citeauthoryear{{Krumholz}}{{Krumholz}}{2011}]{Krumholz2011}
{Krumholz} M.~R.,  2011, \mn@doi [\apj] {10.1088/0004-637X/743/2/110}, \href
  {https://ui.adsabs.harvard.edu/abs/2011ApJ...743..110K} {743, 110}

\bibitem[\protect\citeauthoryear{{Krumholz}, {Klein}  \& {McKee}}{{Krumholz}
  et~al.}{2012}]{KrumholzKleinMcKee2012}
{Krumholz} M.~R.,  {Klein} R.~I.,   {McKee} C.~F.,  2012, \mn@doi [\apj]
  {10.1088/0004-637X/754/1/71}, \href
  {https://ui.adsabs.harvard.edu/abs/2012ApJ...754...71K} {754, 71}

\bibitem[\protect\citeauthoryear{{Lee} \& {Hennebelle}}{{Lee} \&
  {Hennebelle}}{2018a}]{LeeHennebelle2018a}
{Lee} Y.-N.,  {Hennebelle} P.,  2018a, \mn@doi [\aap]
  {10.1051/0004-6361/201731522}, \href
  {http://adsabs.harvard.edu/abs/2018A%26A...611A..88L} {611, A88}

\bibitem[\protect\citeauthoryear{{Lee} \& {Hennebelle}}{{Lee} \&
  {Hennebelle}}{2018b}]{LeeHennebelle2018b}
{Lee} Y.-N.,  {Hennebelle} P.,  2018b, \mn@doi [\aap]
  {10.1051/0004-6361/201731523}, \href
  {http://adsabs.harvard.edu/abs/2018A%26A...611A..89L} {611, A89}

\bibitem[\protect\citeauthoryear{{Lee} \& {Hennebelle}}{{Lee} \&
  {Hennebelle}}{2019}]{LeeHennebelle2019}
{Lee} Y.-N.,  {Hennebelle} P.,  2019, \mn@doi [\aap]
  {10.1051/0004-6361/201834428}, \href
  {http://adsabs.harvard.edu/abs/2019A%26A...622A.125L} {622, A125}

\bibitem[\protect\citeauthoryear{{Lewis} \& {Bate}}{{Lewis} \&
  {Bate}}{2017}]{LewisBate2017}
{Lewis} B.~T.,  {Bate} M.~R.,  2017, \mn@doi [\mnras] {10.1093/mnras/stx271},
  \href {http://adsabs.harvard.edu/abs/2017MNRAS.467.3324L} {467, 3324}

\bibitem[\protect\citeauthoryear{{Lewis} \& {Bate}}{{Lewis} \&
  {Bate}}{2018}]{LewisBate2018}
{Lewis} B.~T.,  {Bate} M.~R.,  2018, \mn@doi [\mnras] {10.1093/mnras/sty829},
  \href {http://adsabs.harvard.edu/abs/2018MNRAS.477.4241L} {477, 4241}

\bibitem[\protect\citeauthoryear{{Lewis}, {Bate}  \& {Price}}{{Lewis}
  et~al.}{2015}]{LewisBatePrice2015}
{Lewis} B.~T.,  {Bate} M.~R.,   {Price} D.~J.,  2015, \mn@doi [\mnras]
  {10.1093/mnras/stv957}, \href
  {http://adsabs.harvard.edu/abs/2015MNRAS.451..288L} {451, 288}

\bibitem[\protect\citeauthoryear{{Lindberg} et~al.,}{{Lindberg}
  et~al.}{2014}]{Lindberg+2014}
{Lindberg} J.~E.,  et~al., 2014, \mn@doi [\aap] {10.1051/0004-6361/201322651},
  \href {http://adsabs.harvard.edu/abs/2014A%26A...566A..74L} {566, A74}

\bibitem[\protect\citeauthoryear{{Liptai}, {Price}, {Wurster}  \&
  {Bate}}{{Liptai} et~al.}{2017}]{Liptai+2017}
{Liptai} D.,  {Price} D.~J.,  {Wurster} J.,   {Bate} M.~R.,  2017, \mn@doi
  [\mnras] {10.1093/mnras/stw2770}, \href
  {http://adsabs.harvard.edu/abs/2017MNRAS.465..105L} {465, 105}

\bibitem[\protect\citeauthoryear{{Low} \& {Lynden-Bell}}{{Low} \&
  {Lynden-Bell}}{1976}]{LowLyndenbell1976}
{Low} C.,  {Lynden-Bell} D.,  1976, \mn@doi [\mnras] {10.1093/mnras/176.2.367},
  \href {http://adsabs.harvard.edu/abs/1976MNRAS.176..367L} {176, 367}

\bibitem[\protect\citeauthoryear{{Machida} \& {Matsumoto}}{{Machida} \&
  {Matsumoto}}{2011}]{MachidaMatsumoto2011}
{Machida} M.~N.,  {Matsumoto} T.,  2011, \mn@doi [\mnras]
  {10.1111/j.1365-2966.2011.18349.x}, \href
  {http://adsabs.harvard.edu/abs/2011MNRAS.413.2767M} {413, 2767}

\bibitem[\protect\citeauthoryear{{McKee} \& {Ostriker}}{{McKee} \&
  {Ostriker}}{2007}]{MckeeOstriker2007}
{McKee} C.~F.,  {Ostriker} E.~C.,  2007, \mn@doi [\araa]
  {10.1146/annurev.astro.45.051806.110602}, \href
  {http://adsabs.harvard.edu/abs/2007ARA%26A..45..565M} {45, 565}

\bibitem[\protect\citeauthoryear{{Mellon} \& {Li}}{{Mellon} \&
  {Li}}{2008}]{MellonLi2008}
{Mellon} R.~R.,  {Li} Z.-Y.,  2008, \mn@doi [\apj] {10.1086/587542}, \href
  {http://adsabs.harvard.edu/abs/2008ApJ...681.1356M} {681, 1356}

\bibitem[\protect\citeauthoryear{{Mestel} \& {Spitzer}}{{Mestel} \&
  {Spitzer}}{1956}]{MestelSpitzer1956}
{Mestel} L.,  {Spitzer} Jr. L.,  1956, \mn@doi [\mnras]
  {10.1093/mnras/116.5.503}, \href
  {http://adsabs.harvard.edu/abs/1956MNRAS.116..503M} {116, 503}

\bibitem[\protect\citeauthoryear{{Mocz}, {Burkhart}, {Hernquist}, {McKee}  \&
  {Springel}}{{Mocz} et~al.}{2017}]{Mocz+2017}
{Mocz} P.,  {Burkhart} B.,  {Hernquist} L.,  {McKee} C.~F.,   {Springel} V.,
  2017, \mn@doi [\apj] {10.3847/1538-4357/aa6475}, \href
  {https://ui.adsabs.harvard.edu/abs/2017ApJ...838...40M} {838, 40}

\bibitem[\protect\citeauthoryear{{Mouschovias} \& {Spitzer}}{{Mouschovias} \&
  {Spitzer}}{1976}]{MouschoviasSpitzer1976}
{Mouschovias} T.~C.,  {Spitzer} Jr. L.,  1976, \mn@doi [\apj] {10.1086/154835},
  \href {http://adsabs.harvard.edu/abs/1976ApJ...210..326M} {210, 326}

\bibitem[\protect\citeauthoryear{{Myers}, {McKee}, {Cunningham}, {Klein}  \&
  {Krumholz}}{{Myers} et~al.}{2013}]{Myers+2013}
{Myers} A.~T.,  {McKee} C.~F.,  {Cunningham} A.~J.,  {Klein} R.~I.,
  {Krumholz} M.~R.,  2013, \mn@doi [\apj] {10.1088/0004-637X/766/2/97}, \href
  {http://adsabs.harvard.edu/abs/2013ApJ...766...97M} {766, 97}

\bibitem[\protect\citeauthoryear{{Myers}, {Klein}, {Krumholz}  \&
  {McKee}}{{Myers} et~al.}{2014}]{Myers+2014}
{Myers} A.~T.,  {Klein} R.~I.,  {Krumholz} M.~R.,   {McKee} C.~F.,  2014,
  \mn@doi [\mnras] {10.1093/mnras/stu190}, \href
  {http://adsabs.harvard.edu/abs/2014MNRAS.439.3420M} {439, 3420}

\bibitem[\protect\citeauthoryear{{Nakano} \& {Umebayashi}}{{Nakano} \&
  {Umebayashi}}{1986}]{NakanoUmebayashi1986}
{Nakano} T.,  {Umebayashi} T.,  1986, \mn@doi [\mnras]
  {10.1093/mnras/218.4.663}, \href
  {http://adsabs.harvard.edu/abs/1986MNRAS.218..663N} {218, 663}

\bibitem[\protect\citeauthoryear{{Nelson}}{{Nelson}}{2006}]{Nelson2006}
{Nelson} A.~F.,  2006, \mn@doi [\mnras] {10.1111/j.1365-2966.2006.11119.x},
  \href {http://adsabs.harvard.edu/abs/2006MNRAS.373.1039N} {373, 1039}

\bibitem[\protect\citeauthoryear{{Offner}, {Klein}, {McKee}  \&
  {Krumholz}}{{Offner} et~al.}{2009}]{Offner+2009}
{Offner} S.~S.~R.,  {Klein} R.~I.,  {McKee} C.~F.,   {Krumholz} M.~R.,  2009,
  \mn@doi [\apj] {10.1088/0004-637X/703/1/131}, \href
  {https://ui.adsabs.harvard.edu/abs/2009ApJ...703..131O} {703, 131}

\bibitem[\protect\citeauthoryear{{Ostriker}, {Stone}  \& {Gammie}}{{Ostriker}
  et~al.}{2001}]{OstrikerStoneGammie2001}
{Ostriker} E.~C.,  {Stone} J.~M.,   {Gammie} C.~F.,  2001, \mn@doi [\apj]
  {10.1086/318290}, \href {http://adsabs.harvard.edu/abs/2001ApJ...546..980O}
  {546, 980}

\bibitem[\protect\citeauthoryear{{Padoan} \& {Nordlund}}{{Padoan} \&
  {Nordlund}}{2002}]{PadoanNordlund2002}
{Padoan} P.,  {Nordlund} {\AA}.,  2002, \mn@doi [\apj] {10.1086/341790}, \href
  {http://adsabs.harvard.edu/abs/2002ApJ...576..870P} {576, 870}

\bibitem[\protect\citeauthoryear{{Padoan} \& {Nordlund}}{{Padoan} \&
  {Nordlund}}{2011}]{PadoanNordlund2011}
{Padoan} P.,  {Nordlund} {\r{A}}.,  2011, \mn@doi [\apj]
  {10.1088/2041-8205/741/1/L22}, \href
  {https://ui.adsabs.harvard.edu/abs/2011ApJ...741L..22P} {741, L22}

\bibitem[\protect\citeauthoryear{{Padoan}, {Haugb{\o}lle}  \&
  {Nordlund}}{{Padoan} et~al.}{2012}]{PadoanHaugbolleNordlund2012}
{Padoan} P.,  {Haugb{\o}lle} T.,   {Nordlund} {\r{A}}.,  2012, \mn@doi [\apj]
  {10.1088/2041-8205/759/2/L27}, \href
  {https://ui.adsabs.harvard.edu/abs/2012ApJ...759L..27P} {759, L27}

\bibitem[\protect\citeauthoryear{{Planck Collaboration} et~al.,}{{Planck
  Collaboration} et~al.}{2016a}]{Planck2016or}
{Planck Collaboration} et~al., 2016a, \mn@doi [\aap]
  {10.1051/0004-6361/201425044}, \href
  {http://adsabs.harvard.edu/abs/2016A%26A...586A.135P} {586, A135}

\bibitem[\protect\citeauthoryear{{Planck Collaboration} et~al.,}{{Planck
  Collaboration} et~al.}{2016b}]{Planck2016role}
{Planck Collaboration} et~al., 2016b, \mn@doi [\aap]
  {10.1051/0004-6361/201525896}, \href
  {http://adsabs.harvard.edu/abs/2016A%26A...586A.138P} {586, A138}

\bibitem[\protect\citeauthoryear{{Pollack}, {McKay}  \&
  {Christofferson}}{{Pollack} et~al.}{1985}]{PollackMckayChristofferson1985}
{Pollack} J.~B.,  {McKay} C.~P.,   {Christofferson} B.~M.,  1985, \mn@doi
  [Icarus] {10.1016/0019-1035(85)90069-7}, \href
  {http://adsabs.harvard.edu/abs/1985Icar...64..471P} {64, 471}

\bibitem[\protect\citeauthoryear{{Pollack}, {Hollenbach}, {Beckwith},
  {Simonelli}, {Roush}  \& {Fong}}{{Pollack} et~al.}{1994}]{Pollack+1994}
{Pollack} J.~B.,  {Hollenbach} D.,  {Beckwith} S.,  {Simonelli} D.~P.,  {Roush}
  T.,   {Fong} W.,  1994, \mn@doi [\apj] {10.1086/173677}, \href
  {http://adsabs.harvard.edu/abs/1994ApJ...421..615P} {421, 615}

\bibitem[\protect\citeauthoryear{{Price}}{{Price}}{2007}]{Price2007}
{Price} D.~J.,  2007, \mn@doi [\pasa] {10.1071/AS07022}, \href
  {http://adsabs.harvard.edu/abs/2007PASA...24..159P} {24, 159}

\bibitem[\protect\citeauthoryear{{Price}}{{Price}}{2010}]{Price2010}
{Price} D.~J.,  2010, \mn@doi [\mnras] {10.1111/j.1365-2966.2009.15763.x},
  \href {http://adsabs.harvard.edu/abs/2010MNRAS.401.1475P} {401, 1475}

\bibitem[\protect\citeauthoryear{{Price}}{{Price}}{2012}]{Price2012}
{Price} D.~J.,  2012, \mn@doi [Journal of Computational Physics]
  {10.1016/j.jcp.2010.12.011}, \href
  {http://adsabs.harvard.edu/abs/2012JCoPh.231..759P} {231, 759}

\bibitem[\protect\citeauthoryear{{Price} \& {Bate}}{{Price} \&
  {Bate}}{2007}]{PriceBate2007}
{Price} D.~J.,  {Bate} M.~R.,  2007, \mn@doi [\mnras]
  {10.1111/j.1365-2966.2007.11621.x}, \href
  {http://adsabs.harvard.edu/abs/2007MNRAS.377...77P} {377, 77}

\bibitem[\protect\citeauthoryear{{Price} \& {Bate}}{{Price} \&
  {Bate}}{2008}]{PriceBate2008}
{Price} D.~J.,  {Bate} M.~R.,  2008, \mn@doi [\mnras]
  {10.1111/j.1365-2966.2008.12976.x}, \href
  {http://adsabs.harvard.edu/abs/2008MNRAS.385.1820P} {385, 1820}

\bibitem[\protect\citeauthoryear{{Price} \& {Bate}}{{Price} \&
  {Bate}}{2009}]{PriceBate2009}
{Price} D.~J.,  {Bate} M.~R.,  2009, \mn@doi [\mnras]
  {10.1111/j.1365-2966.2009.14969.x}, \href
  {http://adsabs.harvard.edu/abs/2009MNRAS.398...33P} {398, 33}

\bibitem[\protect\citeauthoryear{{Price} \& {Monaghan}}{{Price} \&
  {Monaghan}}{2004a}]{PriceMonaghan2004a}
{Price} D.~J.,  {Monaghan} J.~J.,  2004a, \mn@doi [\mnras]
  {10.1111/j.1365-2966.2004.07345.x}, \href
  {http://adsabs.harvard.edu/abs/2004MNRAS.348..123P} {348, 123}

\bibitem[\protect\citeauthoryear{{Price} \& {Monaghan}}{{Price} \&
  {Monaghan}}{2004b}]{PriceMonaghan2004b}
{Price} D.~J.,  {Monaghan} J.~J.,  2004b, \mn@doi [\mnras]
  {10.1111/j.1365-2966.2004.07346.x}, \href
  {http://adsabs.harvard.edu/abs/2004MNRAS.348..139P} {348, 139}

\bibitem[\protect\citeauthoryear{{Price} \& {Monaghan}}{{Price} \&
  {Monaghan}}{2005}]{PriceMonaghan2005}
{Price} D.~J.,  {Monaghan} J.~J.,  2005, \mn@doi [\mnras]
  {10.1111/j.1365-2966.2005.09576.x}, \href
  {http://adsabs.harvard.edu/abs/2005MNRAS.364..384P} {364, 384}

\bibitem[\protect\citeauthoryear{{Price} \& {Monaghan}}{{Price} \&
  {Monaghan}}{2007}]{PriceMonaghan2007}
{Price} D.~J.,  {Monaghan} J.~J.,  2007, \mn@doi [\mnras]
  {10.1111/j.1365-2966.2006.11241.x}, \href
  {http://adsabs.harvard.edu/abs/2007MNRAS.374.1347P} {374, 1347}

\bibitem[\protect\citeauthoryear{{Price} et~al.,}{{Price}
  et~al.}{2018}]{Phantom2018}
{Price} D.~J.,  et~al., 2018, \mn@doi [\pasa] {10.1017/pasa.2018.25}, \href
  {http://adsabs.harvard.edu/abs/2018PASA...35...31P} {35, e031}

\bibitem[\protect\citeauthoryear{{Rees}}{{Rees}}{1976}]{Rees1976}
{Rees} M.~J.,  1976, \mn@doi [\mnras] {10.1093/mnras/176.3.483}, \href
  {http://adsabs.harvard.edu/abs/1976MNRAS.176..483R} {176, 483}

\bibitem[\protect\citeauthoryear{{Santos-Lima}, {de Gouveia Dal Pino}  \&
  {Lazarian}}{{Santos-Lima} et~al.}{2013}]{Santoslima+2013}
{Santos-Lima} R.,  {de Gouveia Dal Pino} E.~M.,   {Lazarian} A.,  2013, \mn@doi
  [\mnras] {10.1093/mnras/sts597}, \href
  {http://adsabs.harvard.edu/abs/2013MNRAS.429.3371S} {429, 3371}

\bibitem[\protect\citeauthoryear{{Seifried}, {Banerjee}, {Klessen}, {Duffin}
  \& {Pudritz}}{{Seifried} et~al.}{2011}]{Seifried+2011}
{Seifried} D.,  {Banerjee} R.,  {Klessen} R.~S.,  {Duffin} D.,   {Pudritz}
  R.~E.,  2011, \mn@doi [\mnras] {10.1111/j.1365-2966.2011.19320.x}, \href
  {http://adsabs.harvard.edu/abs/2011MNRAS.417.1054S} {417, 1054}

\bibitem[\protect\citeauthoryear{{Seifried}, {Banerjee}, {Pudritz}  \&
  {Klessen}}{{Seifried} et~al.}{2012}]{Seifried+2012}
{Seifried} D.,  {Banerjee} R.,  {Pudritz} R.~E.,   {Klessen} R.~S.,  2012,
  \mn@doi [\mnras] {10.1111/j.1745-3933.2012.01253.x}, \href
  {http://adsabs.harvard.edu/abs/2012MNRAS.423L..40S} {423, L40}

\bibitem[\protect\citeauthoryear{{Seifried}, {Banerjee}, {Pudritz}  \&
  {Klessen}}{{Seifried} et~al.}{2013}]{Seifried+2013}
{Seifried} D.,  {Banerjee} R.,  {Pudritz} R.~E.,   {Klessen} R.~S.,  2013,
  \mn@doi [\mnras] {10.1093/mnras/stt682}, \href
  {http://adsabs.harvard.edu/abs/2013MNRAS.432.3320S} {432, 3320}

\bibitem[\protect\citeauthoryear{{Sicilia-Aguilar}, {Henning}, {Linz},
  {Andr{\'e}}, {Stutz}, {Eiroa}  \& {White}}{{Sicilia-Aguilar}
  et~al.}{2013}]{Siciliaaguilar+2013}
{Sicilia-Aguilar} A.,  {Henning} T.,  {Linz} H.,  {Andr{\'e}} P.,  {Stutz} A.,
  {Eiroa} C.,   {White} G.~J.,  2013, \mn@doi [\aap]
  {10.1051/0004-6361/201220170}, \href
  {https://ui.adsabs.harvard.edu/abs/2013A%26A...551A..34S} {551, A34}

\bibitem[\protect\citeauthoryear{{Skinner} \& {Ostriker}}{{Skinner} \&
  {Ostriker}}{2015}]{SkinnerOstriker2015}
{Skinner} M.~A.,  {Ostriker} E.~C.,  2015, \mn@doi [\apj]
  {10.1088/0004-637X/809/2/187}, \href
  {http://adsabs.harvard.edu/abs/2015ApJ...809..187S} {809, 187}

\bibitem[\protect\citeauthoryear{{Stephens} et~al.,}{{Stephens}
  et~al.}{2014}]{Stephens+2014}
{Stephens} I.~W.,  et~al., 2014, \mn@doi [\nat] {10.1038/nature13850}, \href
  {http://adsabs.harvard.edu/abs/2014Natur.514..597S} {514, 597}

\bibitem[\protect\citeauthoryear{{Stephens} et~al.,}{{Stephens}
  et~al.}{2017}]{Stephens+2017}
{Stephens} I.~W.,  et~al., 2017, \mn@doi [\apj] {10.3847/1538-4357/aa998b},
  \href {http://adsabs.harvard.edu/abs/2017ApJ...851...55S} {851, 55}

\bibitem[\protect\citeauthoryear{{Tachihara}, {Onishi}, {Mizuno}  \&
  {Fukui}}{{Tachihara} et~al.}{2002}]{Tachihara+2002}
{Tachihara} K.,  {Onishi} T.,  {Mizuno} A.,   {Fukui} Y.,  2002, \mn@doi [\aap]
  {10.1051/0004-6361:20020180}, \href
  {http://adsabs.harvard.edu/abs/2002A%26A...385..909T} {385, 909}

\bibitem[\protect\citeauthoryear{{Tobin} et~al.,}{{Tobin}
  et~al.}{2015}]{Tobin+2015}
{Tobin} J.~J.,  et~al., 2015, \mn@doi [\apj] {10.1088/0004-637X/805/2/125},
  \href {http://adsabs.harvard.edu/abs/2015ApJ...805..125T} {805, 125}

\bibitem[\protect\citeauthoryear{{Tomida}, {Tomisaka}, {Matsumoto}, {Ohsuga},
  {Machida}  \& {Saigo}}{{Tomida} et~al.}{2010a}]{Tomida+2010rmhd}
{Tomida} K.,  {Tomisaka} K.,  {Matsumoto} T.,  {Ohsuga} K.,  {Machida} M.~N.,
  {Saigo} K.,  2010a, \mn@doi [\apjl] {10.1088/2041-8205/714/1/L58}, \href
  {http://adsabs.harvard.edu/abs/2010ApJ...714L..58T} {714, L58}

\bibitem[\protect\citeauthoryear{{Tomida}, {Machida}, {Saigo}, {Tomisaka}  \&
  {Matsumoto}}{{Tomida} et~al.}{2010b}]{Tomida+2010llc}
{Tomida} K.,  {Machida} M.~N.,  {Saigo} K.,  {Tomisaka} K.,   {Matsumoto} T.,
  2010b, \mn@doi [\apjl] {10.1088/2041-8205/725/2/L239}, \href
  {http://adsabs.harvard.edu/abs/2010ApJ...725L.239T} {725, L239}

\bibitem[\protect\citeauthoryear{{Tomida}, {Tomisaka}, {Matsumoto}, {Hori},
  {Okuzumi}, {Machida}  \& {Saigo}}{{Tomida} et~al.}{2013}]{Tomida+2013}
{Tomida} K.,  {Tomisaka} K.,  {Matsumoto} T.,  {Hori} Y.,  {Okuzumi} S.,
  {Machida} M.~N.,   {Saigo} K.,  2013, \mn@doi [\apj]
  {10.1088/0004-637X/763/1/6}, \href
  {http://adsabs.harvard.edu/abs/2013ApJ...763....6T} {763, 6}

\bibitem[\protect\citeauthoryear{{Tricco} \& {Price}}{{Tricco} \&
  {Price}}{2013}]{TriccoPrice2013}
{Tricco} T.~S.,  {Price} D.~J.,  2013, \mn@doi [\mnras]
  {10.1093/mnras/stt1776}, \href
  {http://adsabs.harvard.edu/abs/2013MNRAS.436.2810T} {436, 2810}

\bibitem[\protect\citeauthoryear{{Tricco}, {Price}  \& {Bate}}{{Tricco}
  et~al.}{2016}]{TriccoPriceBate2016}
{Tricco} T.~S.,  {Price} D.~J.,   {Bate} M.~R.,  2016, \mn@doi [Journal of
  Computational Physics] {10.1016/j.jcp.2016.06.053}, \href
  {http://adsabs.harvard.edu/abs/2016JCoPh.322..326T} {322, 326}

\bibitem[\protect\citeauthoryear{{Tsukamoto}, {Iwasaki}, {Okuzumi}, {Machida}
  \& {Inutsuka}}{{Tsukamoto} et~al.}{2015a}]{Tsukamoto+2015oa}
{Tsukamoto} Y.,  {Iwasaki} K.,  {Okuzumi} S.,  {Machida} M.~N.,   {Inutsuka}
  S.,  2015a, \mn@doi [\mnras] {10.1093/mnras/stv1290}, \href
  {http://adsabs.harvard.edu/abs/2015MNRAS.452..278T} {452, 278}

\bibitem[\protect\citeauthoryear{{Tsukamoto}, {Iwasaki}, {Okuzumi}, {Machida}
  \& {Inutsuka}}{{Tsukamoto} et~al.}{2015b}]{Tsukamoto+2015hall}
{Tsukamoto} Y.,  {Iwasaki} K.,  {Okuzumi} S.,  {Machida} M.~N.,   {Inutsuka}
  S.,  2015b, \mn@doi [\apjl] {10.1088/2041-8205/810/2/L26}, \href
  {http://adsabs.harvard.edu/abs/2015ApJ...810L..26T} {810, L26}

\bibitem[\protect\citeauthoryear{{Tsukamoto}, {Okuzumi}, {Iwasaki}, {Machida}
  \& {Inutsuka}}{{Tsukamoto} et~al.}{2017}]{Tsukamoto+2017}
{Tsukamoto} Y.,  {Okuzumi} S.,  {Iwasaki} K.,  {Machida} M.~N.,   {Inutsuka}
  S.-i.,  2017, \mn@doi [\pasj] {10.1093/pasj/psx113}, \href
  {http://adsabs.harvard.edu/abs/2017PASJ...69...95T} {69, 95}

\bibitem[\protect\citeauthoryear{{Umebayashi} \& {Nakano}}{{Umebayashi} \&
  {Nakano}}{1990}]{UmebayashiNakano1990}
{Umebayashi} T.,  {Nakano} T.,  1990, \mn@doi [\mnras]
  {10.1093/mnras/243.1.103}, \href
  {http://adsabs.harvard.edu/abs/1990MNRAS.243..103U} {243, 103}

\bibitem[\protect\citeauthoryear{{Vaytet}, {Commer{\c c}on}, {Masson},
  {Gonz{\'a}lez}  \& {Chabrier}}{{Vaytet} et~al.}{2018}]{Vaytet+2018}
{Vaytet} N.,  {Commer{\c c}on} B.,  {Masson} J.,  {Gonz{\'a}lez} M.,
  {Chabrier} G.,  2018, \mn@doi [\aap] {10.1051/0004-6361/201732075}, \href
  {http://adsabs.harvard.edu/abs/2018A%26A...615A...5V} {615, A5}

\bibitem[\protect\citeauthoryear{{Walch}, {Whitworth}, {Bisbas}, {W{\"u}nsch}
  \& {Hubber}}{{Walch} et~al.}{2012}]{Walch+2012}
{Walch} S.~K.,  {Whitworth} A.~P.,  {Bisbas} T.,  {W{\"u}nsch} R.,   {Hubber}
  D.,  2012, \mn@doi [\mnras] {10.1111/j.1365-2966.2012.21767.x}, \href
  {http://adsabs.harvard.edu/abs/2012MNRAS.427..625W} {427, 625}

\bibitem[\protect\citeauthoryear{{Whitehouse} \& {Bate}}{{Whitehouse} \&
  {Bate}}{2006}]{WhitehouseBate2006}
{Whitehouse} S.~C.,  {Bate} M.~R.,  2006, \mn@doi [\mnras]
  {10.1111/j.1365-2966.2005.09950.x}, \href
  {http://cdsads.u-strasbg.fr/abs/2006MNRAS.367...32W} {367, 32}

\bibitem[\protect\citeauthoryear{{Whitehouse}, {Bate}  \&
  {Monaghan}}{{Whitehouse} et~al.}{2005}]{WhitehouseBateMonaghan2005}
{Whitehouse} S.~C.,  {Bate} M.~R.,   {Monaghan} J.~J.,  2005, \mn@doi [\mnras]
  {10.1111/j.1365-2966.2005.09683.x}, \href
  {http://adsabs.harvard.edu/abs/2005MNRAS.364.1367W} {364, 1367}

\bibitem[\protect\citeauthoryear{{Wurster}}{{Wurster}}{2016}]{Wurster2016}
{Wurster} J.,  2016, \mn@doi [\pasa] {10.1017/pasa.2016.34}, \href
  {http://adsabs.harvard.edu/abs/2016PASA...33...41W} {33, e041}

\bibitem[\protect\citeauthoryear{{Wurster} \& {Bate}}{{Wurster} \&
  {Bate}}{2019}]{WursterBate2019}
{Wurster} J.,  {Bate} M.~R.,  2019, \mn@doi [\mnras] {10.1093/mnras/stz1023},
  \href {http://adsabs.harvard.edu/abs/2019MNRAS.486.2587W} {486, 2587}

\bibitem[\protect\citeauthoryear{{Wurster} \& {Li}}{{Wurster} \&
  {Li}}{2018}]{WursterLi2018}
{Wurster} J.,  {Li} Z.-Y.,  2018, \mn@doi [Frontiers in Astronomy and Space
  Sciences] {10.3389/fspas.2018.00039}, \href
  {http://adsabs.harvard.edu/abs/2018FrASS...5...39W} {5, 39}

\bibitem[\protect\citeauthoryear{{Wurster}, {Price}  \& {Ayliffe}}{{Wurster}
  et~al.}{2014}]{WursterPriceAyliffe2014}
{Wurster} J.,  {Price} D.~J.,   {Ayliffe} B.,  2014, \mn@doi [\mnras]
  {10.1093/mnras/stu1524}, \href
  {http://adsabs.harvard.edu/abs/2014MNRAS.444.1104W} {444, 1104}

\bibitem[\protect\citeauthoryear{{Wurster}, {Price}  \& {Bate}}{{Wurster}
  et~al.}{2016}]{WursterPriceBate2016}
{Wurster} J.,  {Price} D.~J.,   {Bate} M.~R.,  2016, \mn@doi [\mnras]
  {10.1093/mnras/stw013}, \href
  {http://adsabs.harvard.edu/abs/2016MNRAS.457.1037W} {457, 1037}

\bibitem[\protect\citeauthoryear{{Wurster}, {Bate}, {Price}  \&
  {Tricco}}{{Wurster} et~al.}{2017a}]{Wurster+2017}
{Wurster} J.,  {Bate} M.~R.,  {Price} D.~J.,   {Tricco} T.~S.,  2017a,
  preprint, \href {http://adsabs.harvard.edu/abs/2017arXiv170607721W} {}
  (\mn@eprint {arXiv} {1706.07721})

\bibitem[\protect\citeauthoryear{{Wurster}, {Price}  \& {Bate}}{{Wurster}
  et~al.}{2017b}]{WursterPriceBate2017}
{Wurster} J.,  {Price} D.~J.,   {Bate} M.~R.,  2017b, \mn@doi [\mnras]
  {10.1093/mnras/stw3181}, \href
  {http://adsabs.harvard.edu/abs/2017MNRAS.466.1788W} {466, 1788}

\bibitem[\protect\citeauthoryear{{Wurster}, {Bate}  \& {Price}}{{Wurster}
  et~al.}{2018a}]{WursterBatePrice2018sd}
{Wurster} J.,  {Bate} M.~R.,   {Price} D.~J.,  2018a, \mn@doi [\mnras]
  {10.1093/mnras/stx3339}, \href
  {http://adsabs.harvard.edu/abs/2018MNRAS.475.1859W} {475, 1859}

\bibitem[\protect\citeauthoryear{{Wurster}, {Bate}  \& {Price}}{{Wurster}
  et~al.}{2018b}]{WursterBatePrice2018hd}
{Wurster} J.,  {Bate} M.~R.,   {Price} D.~J.,  2018b, \mn@doi [\mnras]
  {10.1093/mnras/sty2212}, \href
  {http://adsabs.harvard.edu/abs/2018MNRAS.480.4434W} {480, 4434}

\bibitem[\protect\citeauthoryear{{Wurster}, {Bate}  \& {Price}}{{Wurster}
  et~al.}{2018c}]{WursterBatePrice2018ff}
{Wurster} J.,  {Bate} M.~R.,   {Price} D.~J.,  2018c, \mn@doi [\mnras]
  {10.1093/mnras/sty2438}, \href
  {http://adsabs.harvard.edu/abs/2018MNRAS.481.2450W} {481, 2450}

\makeatother
\end{thebibliography}

\appendix
\section{Resolution study}
\label{app:resolution}

\begin{figure*}
\includegraphics[width=0.95\textwidth]{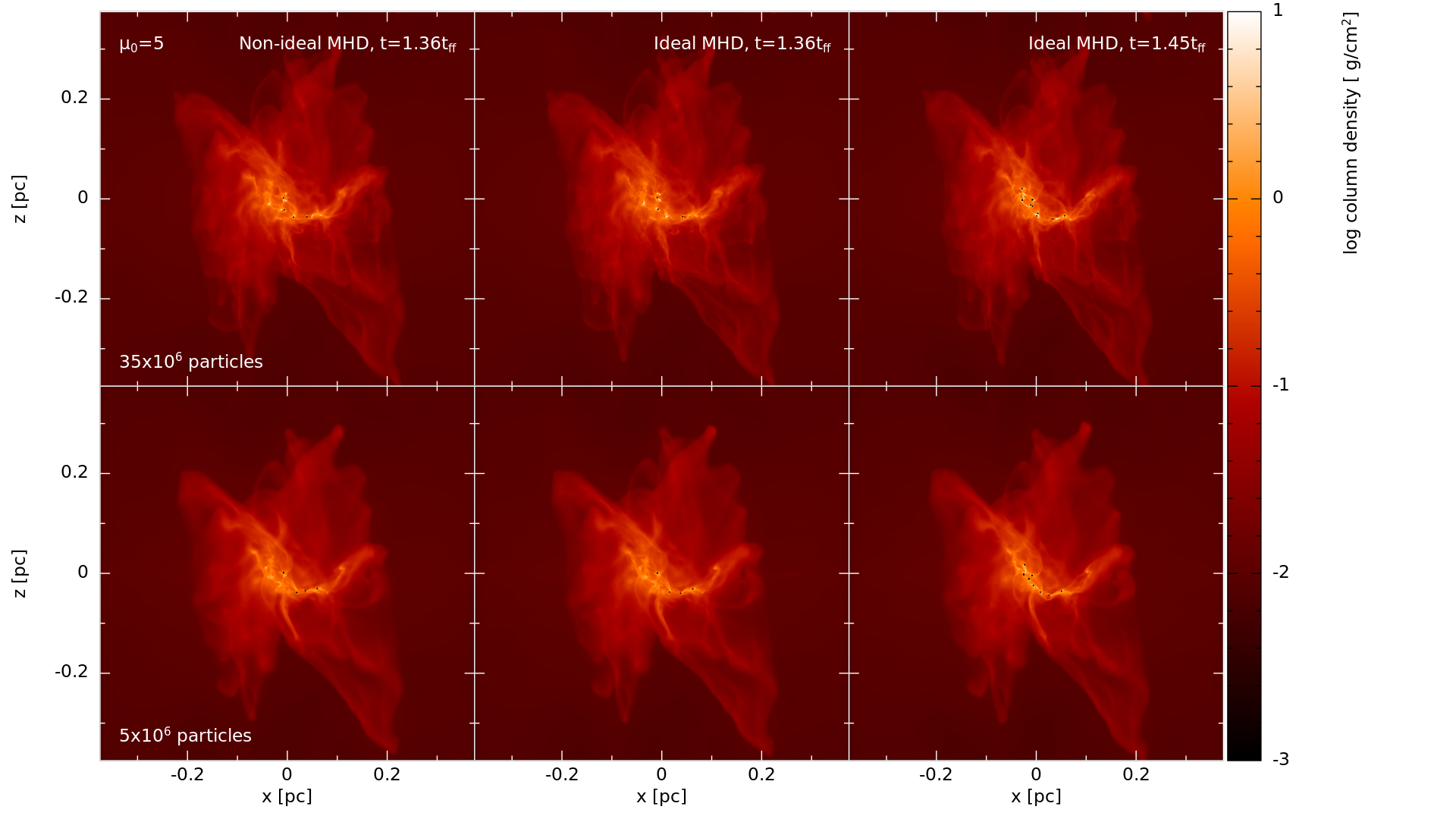}  
\caption{Global structure of the gas column density for models with \mueq{5} at \tendNh \ (first two columns) and at \tendIh \ (final column).  The models in the top row have $35\times 10^6$ particles in the initial sphere and the models in the bottom row have $5\times10^6$ particles.  Each black dot represent the location of a sink particle (not to scale).  The qualitative structure is similar between the resolutions, although the high resolution models have more filaments and clumps, and better well-defined structures.  
To capture the detail, we plot this image at a higher resolution than the other images, thus a direct comparison between this and other figures is unreliable. }
\label{fig:res:columndensity:global}
\end{figure*} 
\begin{figure*}
\includegraphics[width=0.9\textwidth]{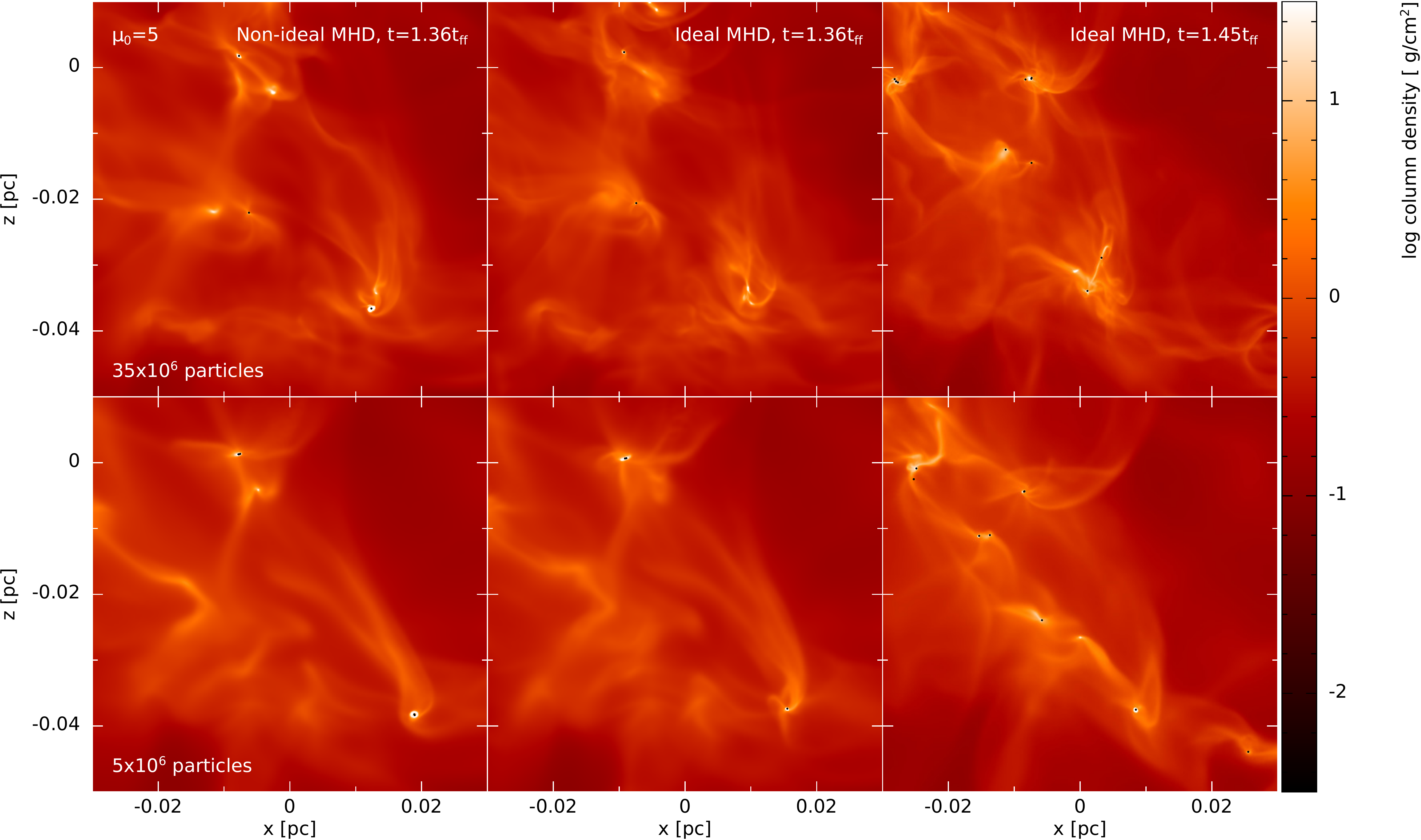}  
\caption{Gas column density of a small region of the cluster for models with \mueq{5}.  Each black dot represent the location of a sink particle (not to scale).  The resolution effects are highlighted on these scales, showing denser clumps and more well-defined structures in the high resolution models.  The non-ideal processes have a larger effect at higher resolutions, such that the qualitative differences between  \emph{N05}$^h$ and \emph{I05}$^h$ are greater than those between  \emph{N05} and \emph{I05} at \tendNh. As time evolves, the differences between resolutions increases, as shown by the difference structures shown in the final column. }
\label{fig:res:columndensity:local}
\end{figure*} 

\begin{figure}
\centering
\includegraphics[width=\columnwidth]{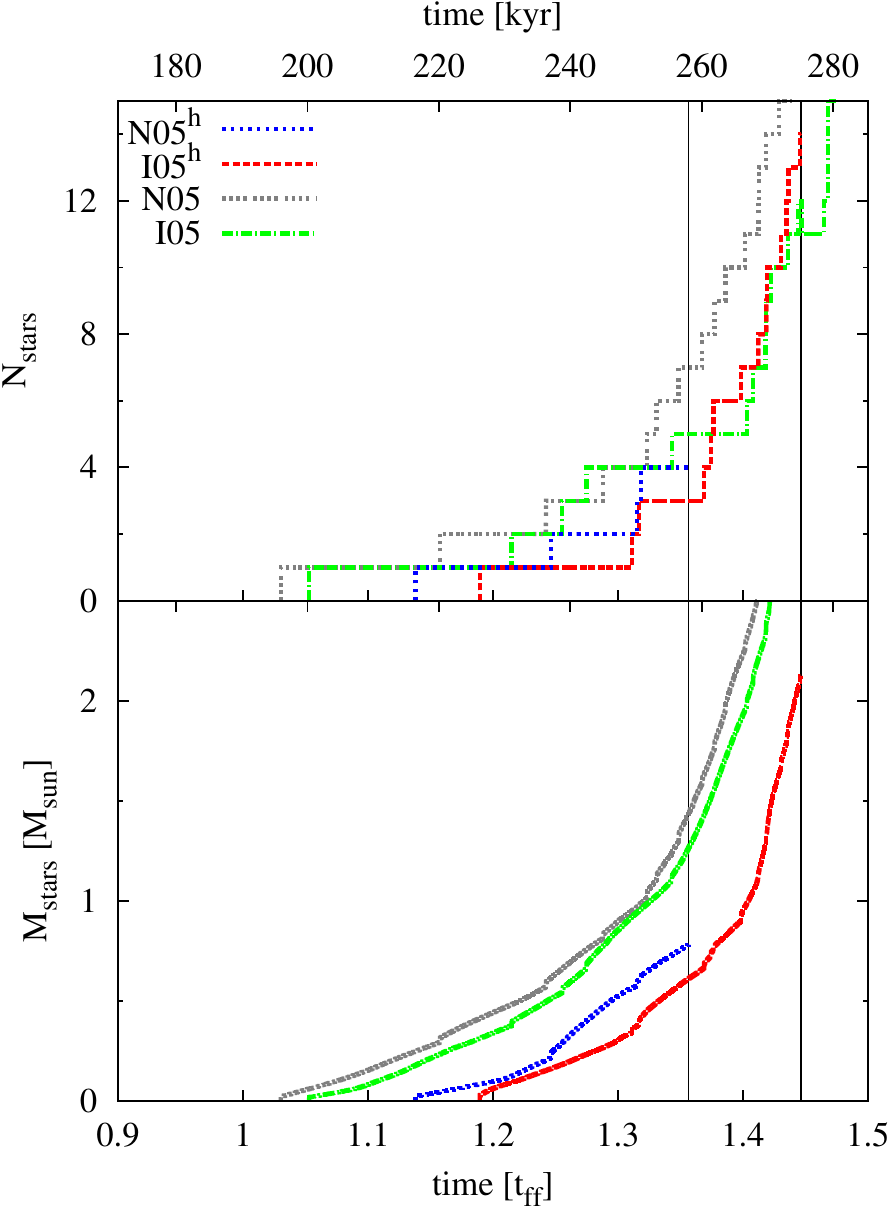}  
\caption{The total number of stars (top) and the total mass in stars (bottom) for the models with \mueq{5}.  The vertical axes are truncated for a better comparison between the high and fiducial resolution models.  The vertical lines represent \tend{} of \emph{N05}$^h$ and \emph{I05}$^h$.  Star formation begins later in the high resolution models, but  in all the models the star formation rate is accelerating as the calculations proceed so that high and fiducial resolution models have similar numbers of stars at later times.  At a given time, there is alway less stellar mass in the higher resolution simulations.}
\label{fig:res:stars}
\end{figure} 
\begin{table*}
\begin{center}
\begin{tabular}{c c c c c c c c c c c c c}
\hline
Name   & $M^\text{c}_\text{stars}$ & $N^\text{c}_\text{stars}$  & $N^\text{c}_\text{mergers}$  & $N^\text{c}_\text{systems}$  & $N^\text{c}_\text{discs}$   
             & $M^\text{f}_\text{stars}$ & $N^\text{f}_\text{stars}$    & $N^\text{f}_\text{mergers}$  &  $N^\text{f}_\text{systems}$  & $N^\text{f}_\text{discs}$  \\
             &  [\Msun] & & &  1, 2, 3, 4 & 1, 2, 3, 4  & [\Msun] & & &  1, 2, 3, 4 & 1, 2, 3, 4 &\\
\hline 
\emph{N05}$^h$ & 0.78 &  4 &  0 &  4, 0, 0, 0 & 4 (4), 0 (0), 0, 0  \\ 
\emph{N05}         & 1.43 &  7 &  0 &  0, 0, 1, 1 &  0 (7), 0 (2), 1, 0   \\ 
\emph{I05}$^h$   & 0.61 &  3 &  0 & 3, 0, 0, 0 & 3 (3), 0 (0), 0, 0   & 2.13 & 14 & 0 & 5, 1, 1, 1 & 5 (14), 0 (2), 1, 1     \\ 
\emph{I05}          &  1.28 &  5 &  0 & 0, 1, 1, 0 & 0 (5), 1 (1), 0, 0  & 3.25 & 12 & 0 & 5, 2, 1, 0 &   4 (9), 1 (1), 1, 0   \\ 
\hline
\end{tabular}
\caption{Summary of the stellar population properties for the models with \mueq{5}.  The models are compared at \ttendNh \ (for all models, superscript c) and at \ttendIh \ (for the ideal models, superscript f).  The columns are the same as in \tabref{table:stars}.}
\label{table:res:stars} 
\end{center}
\end{table*}

We perform two high-resolution versions of our models with \mueq{5}.  For these models, we employ a mass resolution seven times higher than our fiducial resolution models, yielding an $\approx 2$ times improvement in spatial resolution.  These models are named \emph{N05}$^h$ and \emph{I05}$^h$ for the non-ideal and ideal MHD versions, respectively, and are computed to $t = 1.356 \approx 1.36$~\tff{} and $t = 1.446 \approx 1.45$~\tff{}, respectively.

 \subsection{Structure}

\fig{fig:res:columndensity:global} shows the global structure of the models in our resolution comparison at the final times of the high resolution models.  We find qualitative agreement between the two resolutions, and as expected, the small scale filamentary structure appears better resolved in the high resolution models.

The differences can be better seen on smaller scales, as shown in \fig{fig:res:columndensity:local}.  This figure shows the better defined structures, with narrower and less wispy filaments in the high resolution models.  In several regions, there are denser, more well-defined clumps in the high resolution models compared to similar regions in fiducial resolution counterparts.  Given the chaotic nature of the evolution, as time progresses, the differences between the resolutions becomes more apparent.  This also suggests that the high resolution clusters evolve slightly faster, since their higher densities results in higher gravitational accelerations. 

At \tendNh, the qualitative structure is similar between \emph{N05} and \emph{I05}.  Although \emph{N05}$^h$ and \emph{I05}$^h$ are also similar at this time, there are a few more differences, including two high-density clumps in \emph{N05}$^h$ and one in  \emph{I05}$^h$ that do not appear in the other model.  This suggests that non-ideal effects are more important at higher resolutions, which is reasonable since artificial resistivity is resolution-dependent;  there is less resistivity in \emph{I05}$^h$ compared to \emph{I05}, while  \emph{N05}$^h$ and \emph{N05} should have similar amounts in dense regions since physical resistivity is expected to dominate over artificial resistivity in these regions, and should be resolution-independent. 

\subsection{Stellar populations}
Our star formation prescription does not depend on resolution --- we still insert a sink of radius 0.5~au when \rhoxeq{-5} is reached.  \fig{fig:res:stars} shows the total number of stars (top panel) and the total mass in stars (bottom panel) as a function of time; \tabref{table:res:stars} lists the stellar population properties at $t \approx 1.36$ and $1.45$~\tff.

Star formation begins later in the high resolution models, with higher-resolution models collapsing slower than their lower-resolution counterparts \citep[as previously discussed in][]{WursterBatePrice2018ff}.   Once the star formation begins in \emph{I05}$^h$, the star formation rate is similar to that of  \emph{I05}, and both models have accelerating star formation rates. Thus, although initial star formation may be delayed in the high resolution models, similar numbers of stars ultimately form.

In the non-ideal models, there is a similar delay of the onset of star formation at high resolution.  At \tendNh, the high resolution model has fewer stars and less total stellar mass but this is simply caused by the later start.  Starting at \ttendNh,  \emph{N05} undergoes a rapid star formation epoch.  Given the high-density clumps in \fig{fig:res:stars}, it is likely that \emph{N05}$^h$ is also about to undergo a rapid star formation epoch.  

For all time, the total stellar mass is lower in the high resolution model, which is mainly a result of the delay in the collapse due to better resolved pressure gradients.  High-density circumstellar material is also accreted by sink particles at a lower rate due to the lower numerical viscosity \citep[see the Appendix of][]{Bate2018}. 

For the high resolution non-ideal MHD model at \tendNh, all stars are single stars, but this is simply because only four stars have formed by this point.  In the ideal MHD model at \tendIh, both ideal models have a similar number of stars and higher order systems (Table \ref{table:res:stars}).

 \subsection{Protostellar discs}
\begin{figure}
\includegraphics[width=\columnwidth]{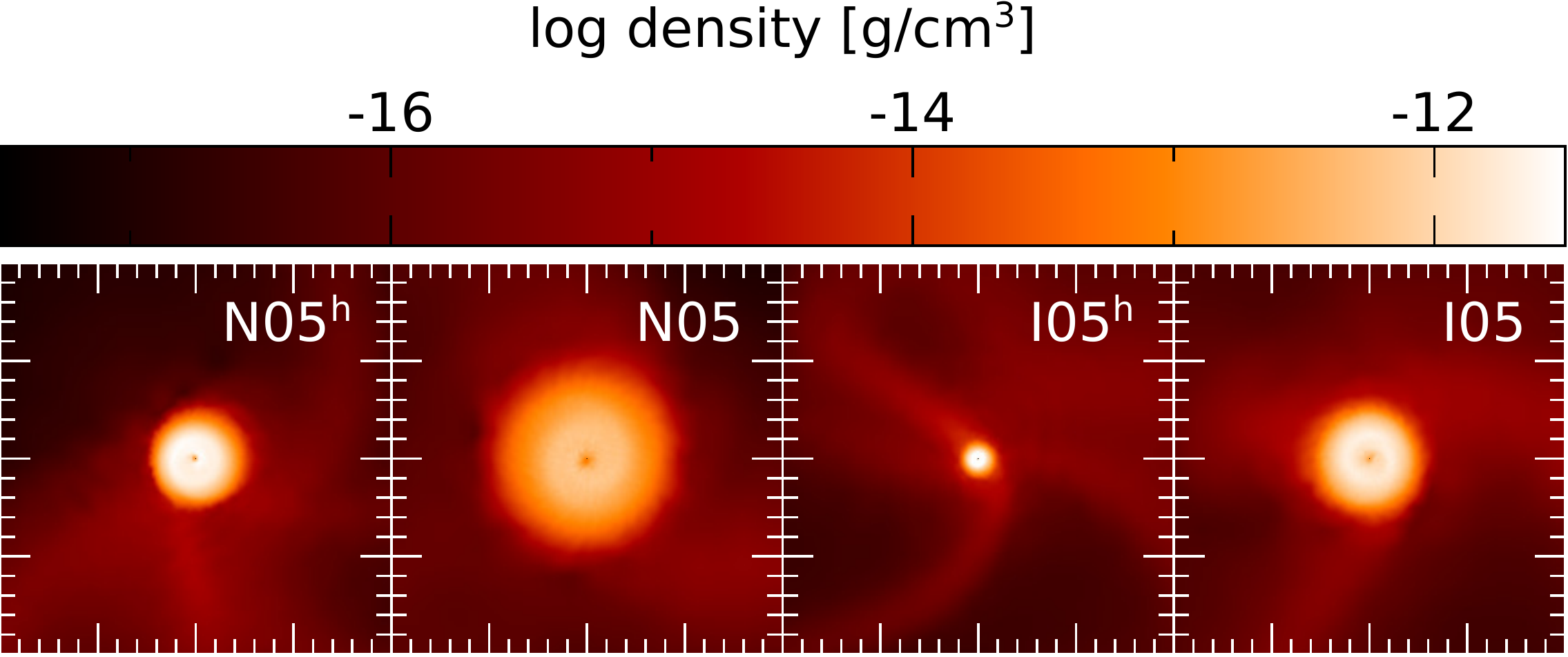}  
\includegraphics[width=\columnwidth]{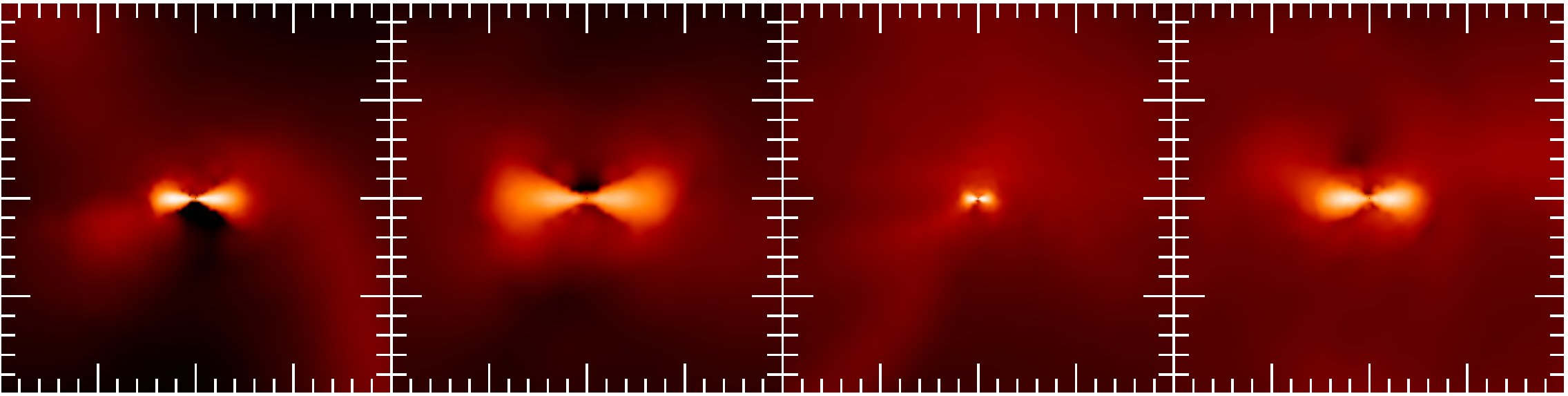}  
\includegraphics[width=\columnwidth]{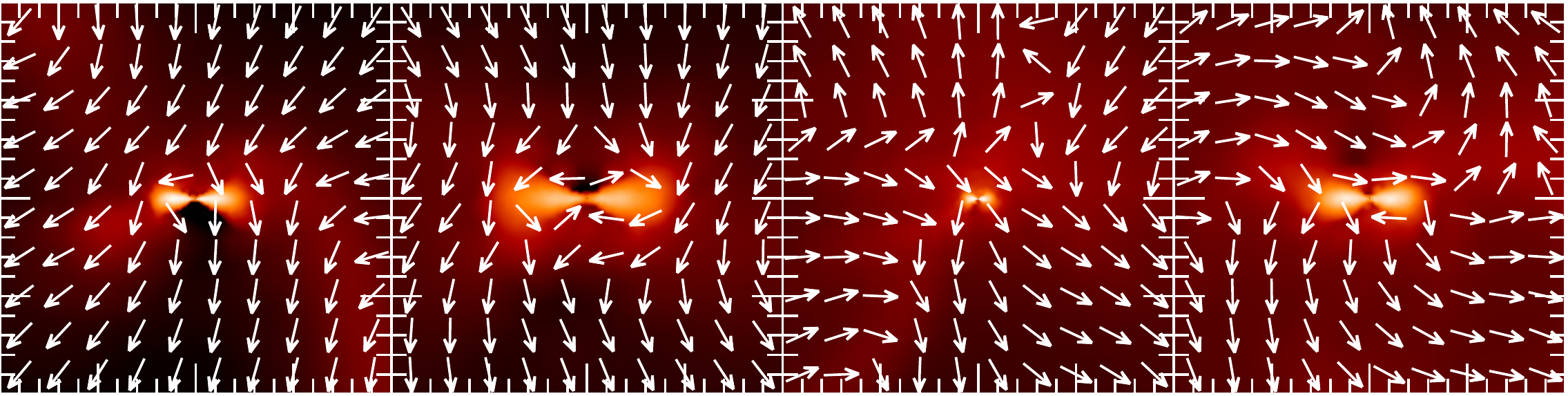}  
\includegraphics[width=\columnwidth]{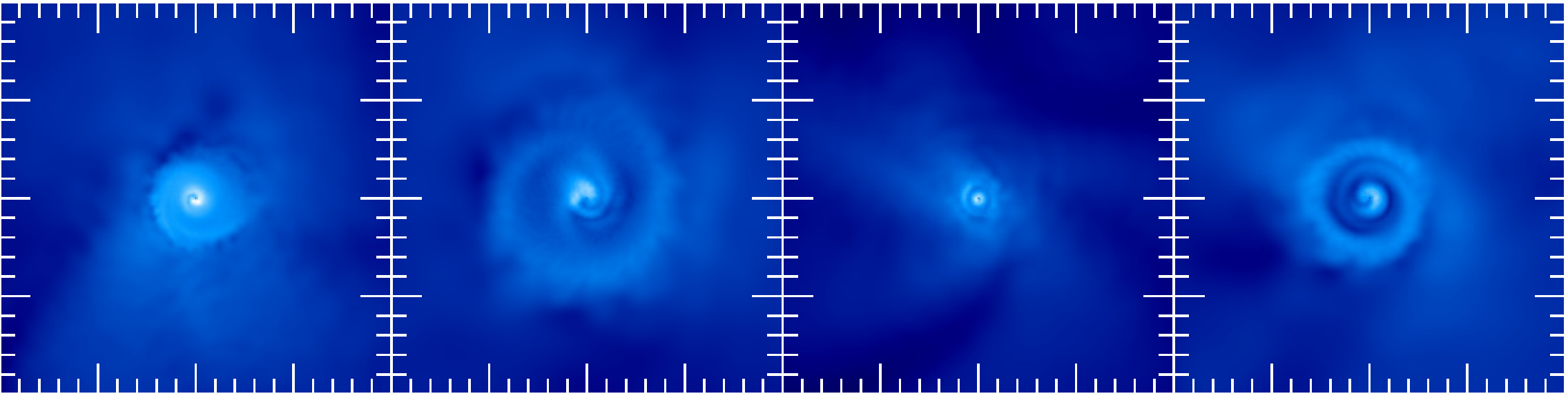}  
\includegraphics[width=\columnwidth]{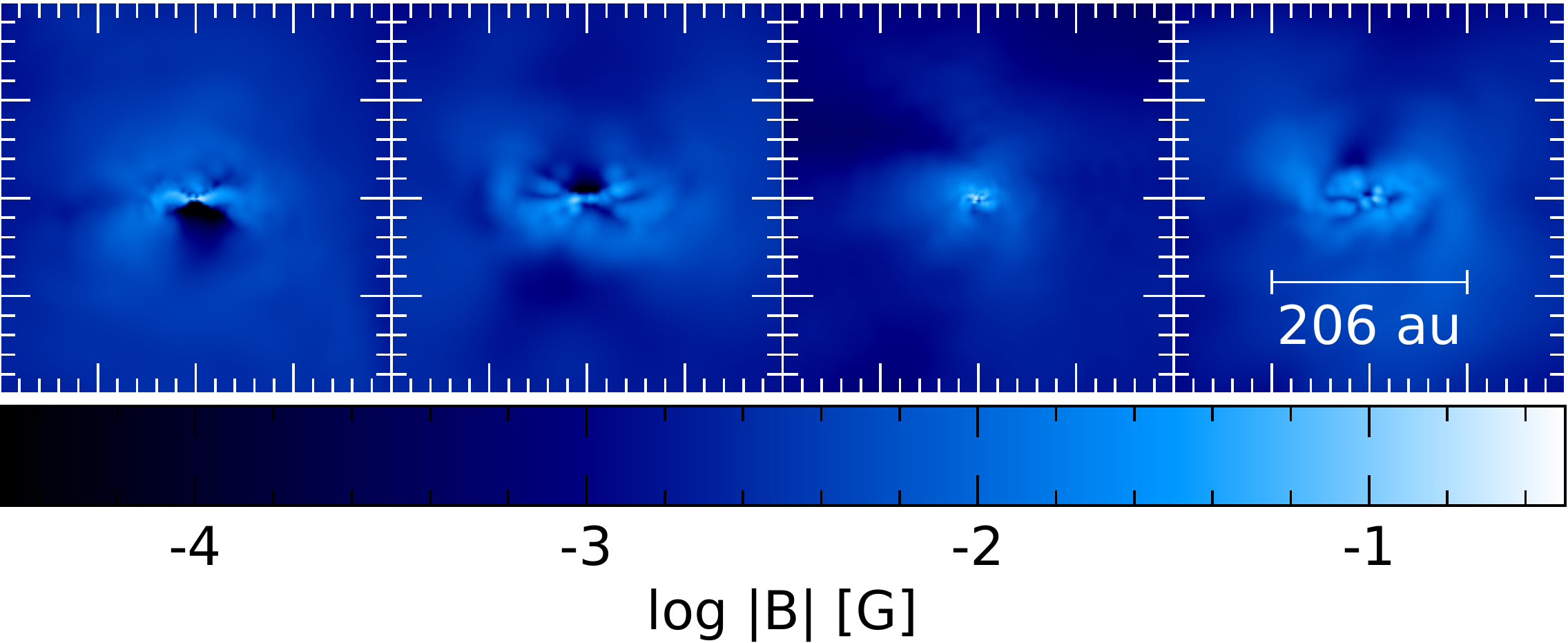}  
\caption{A cross-section of the largest, most well-defined circumstellar disc from each \mueq{5} model at \tendNh.  As in \fig{fig:disc:One}, from top to bottom the rows are face-on gas density, edge-on gas density, edge-on gas density over-plotted with magnetic field vectors representing direction only, face-on magnetic field strength, and edge-on magnetic field strength.   All panels are slices through the host star.  Discs are generally smaller with stronger magnetic fields in the high resolution models.}
\label{fig:disc:Res:One}
\end{figure}

At \tendNh, every star has a circumstellar disc in the high resolution models, and higher order discs form at later times in \emph{I05}$^h$.  Given the chaotic nature of the simulations, each fiducial resolution disc does not have a high resolution counterpart, thus we are unable to perform a direct disc-to-disc comparison.  \fig{fig:disc:Res:One} shows the gas density and magnetic field strength and direction in a slice through the centre of the the most well-defined circumstellar disc in each model.  At both resolutions the magnetic field retains a spiral structure in the discs (third row of \fig{fig:disc:Res:One}), and the azimuthal component remains the largest component.

The high resolution discs generally have smaller radii and less mass than those in the fiducial resolution models.  These results are to be expected in the ideal MHD calculations as a consequence of the factor of \sm4 decrease in the artificial resistivity by increasing the resolution.  The higher artificial resistivity in the fiducial resolution models will artificially promote disc formation \citep{WursterPriceBate2016}, increasing their size and decreasing their magnetic field strength.  With non-ideal MHD, the non-ideal processes should dominate over artificial resistivity.  However, because circumstellar discs tend to grow in both mass and size with time \citep[see also][]{Bate2018}, the difference in size and mass may simply be due to the delay in the onset of the star formation (i.e., the discs have had less time to grow in the high resolution simulations). 

Large protostellar discs still form in both of our high resolution simulations, again suggesting that the disc formation is promoted by the turbulent velocity field and that the magnetic braking catastrophe is a result of idealised initial conditions.  However, given that we do see a tentative resolution dependence on disc size and mass, we cannot completely rule out the catastrophe.  It is possible that, given enough resolution, strong magnetic fields may again hinder disc formation even in turbulent gas.  To study this properly would require even higher resolution simulations, which are currently computationally infeasible.   

\section{Protostellar disc populations}
\label{app:discs}
\begin{figure*}
\includegraphics[width=0.9\textheight, angle=90]{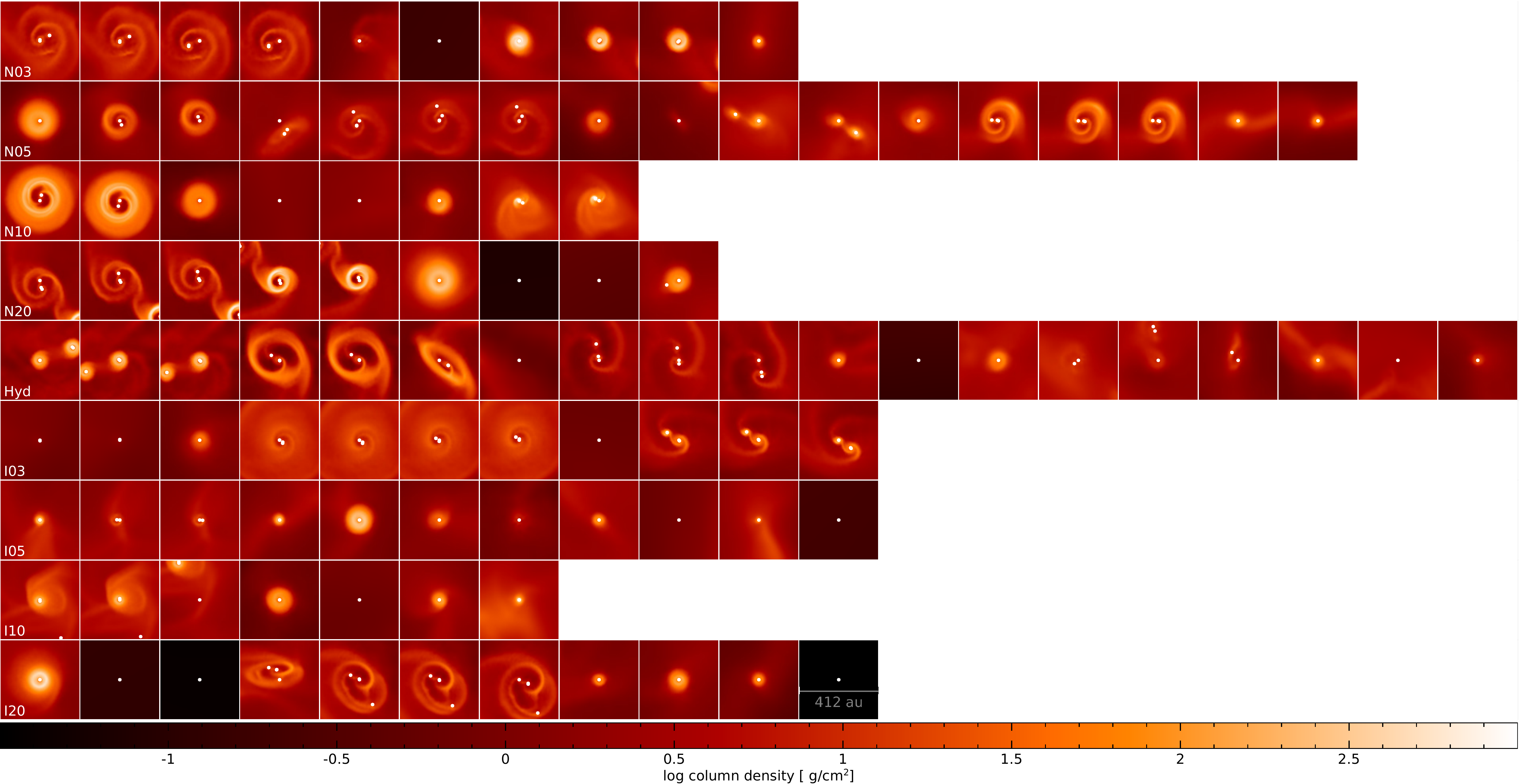}  
\includegraphics[width=0.9\textheight, angle=90]{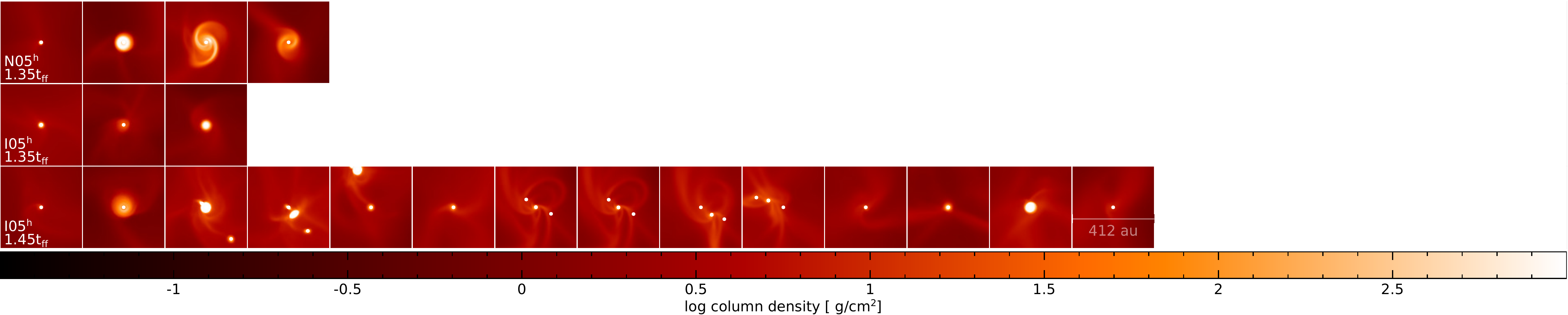}  
\caption{Gas column density around every star in our fiducial resolution simulations at \tnow{} (top panel) and at the times discussed in Appendix~\ref{app:resolution} for the high-resolution models (bottom panel).  All the discs are orientated to be face-on, and the target star is placed at the centre of the frame.  Each white dot represents the location of a sink particle (not to scale), and stars are shown as long as they are in the frame, independent of vertical distance.  The stars in the \tendNh{} row of \emph{I05}$^h$ correspond to the panels directly below at \tendIh{}. Not all stars have discs, but a variety of circumstellar, circumbinary and circumsystem discs form. }
\label{fig:discs:rho}
\end{figure*}

\begin{figure*}
\includegraphics[width=0.9\textheight, angle=90]{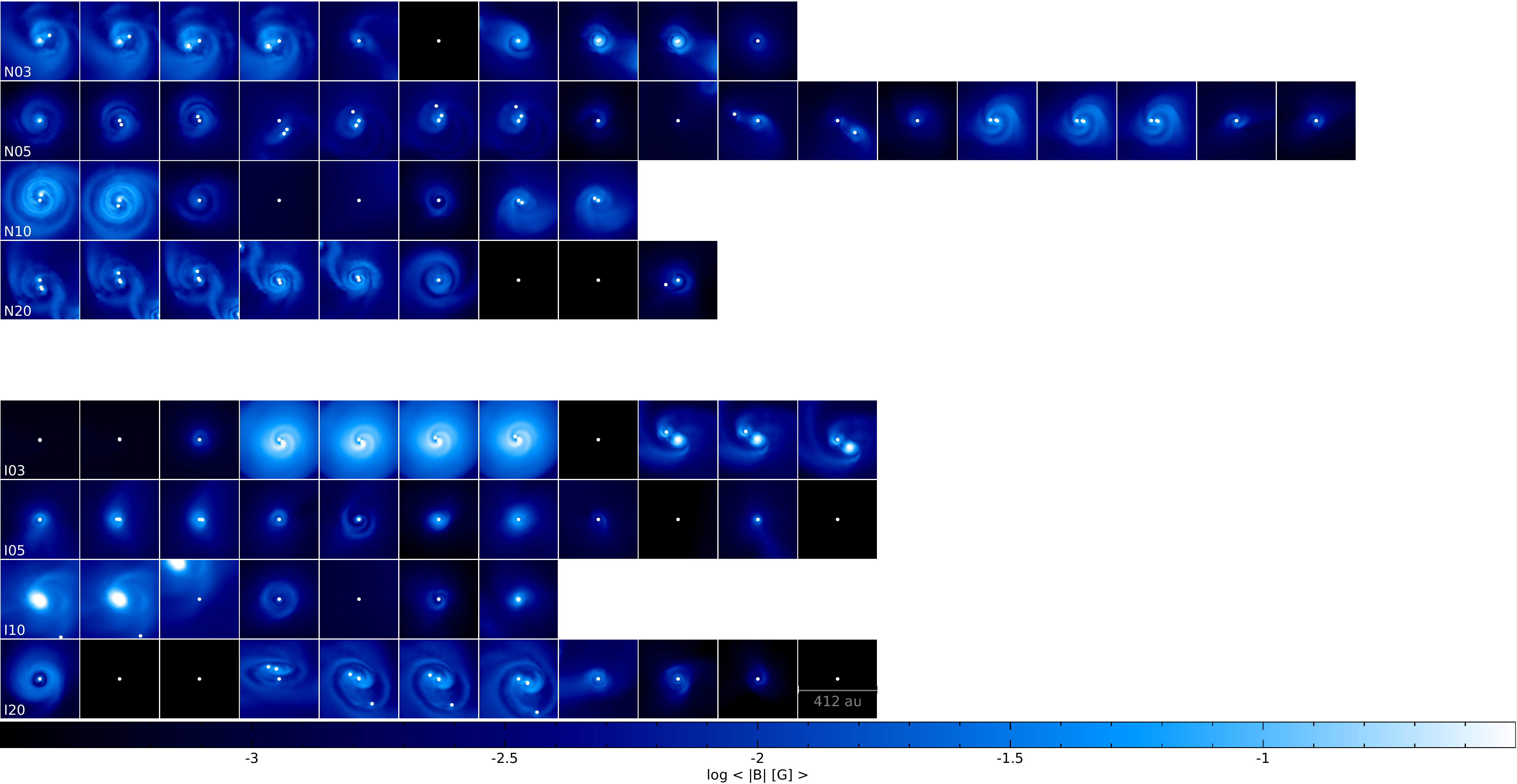}  
\includegraphics[width=0.9\textheight, angle=90]{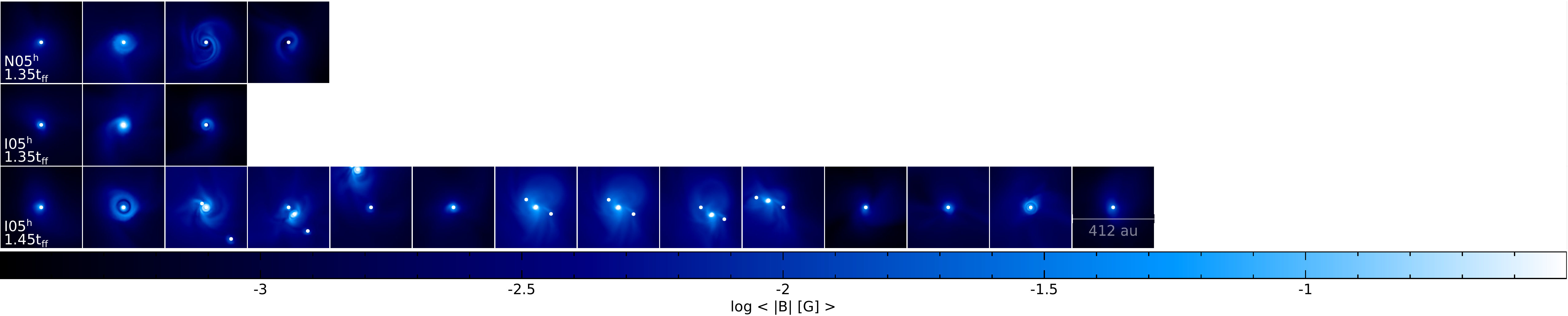}  
\caption{Magnetic field strength in the gas surrounding every star, as in \fig{fig:discs:rho}.}
\label{fig:discs:B}
\end{figure*}

Figs.~\ref{fig:discs:rho} and \ref{fig:discs:B} show a 412 $\times$ 412 au$^2$ region around every star in our fiducial resolution simulations at \tnow, and at the times discussed in Appendix~\ref{app:resolution} for the high-resolution models; the images are rotated such that the disc is face-on and show the gas column density and density-weighted line-of-sight-averaged magnetic field strength, respectively (i.e. $\left<|B|\right> = \int |B| \rho \text{d}z' / \int \rho \text{d}z'$).  These figures differ slightly from the slice through the centre of the star as in Figs.~\ref{fig:disc:One} and \ref{fig:disc:Res:One} to better show how the discs interact with their surroundings, by also showing non-coplanar, nearby objects.  Since we plot every star, one may observe the wide variety of disc structures, multiplicities, and the stars that do not form discs.

\label{lastpage}
\end{document}